\documentclass[12pt]{revtex4-1}
\pdfoutput=1
\usepackage{hyperref, graphicx, multirow, chngpage}
\usepackage{amsmath,amssymb,amsfonts}
\usepackage{caption}
\usepackage{float}
\usepackage{xcolor}
\usepackage[makeroom]{cancel} 
\usepackage{ulem}

\begin{document}

\renewcommand\tablename{{\small Table}}

\renewcommand{\vec}[1]{{\boldsymbol{#1}}} 
\newcommand\Wtilde{\tilde{W}}
\newcommand\bfd{\overset\leftrightarrow{D}}
\newcommand\htilde{\tilde{H}}


\title{Primary Observables for\\ Top Quark Collider Signals}

\author{Layne Bradshaw, Spencer Chang}

\address{Department of Physics and Institute for Fundamental Science\\ 
University of Oregon, Eugene, Oregon 97403}

\begin{abstract}
At the HL-LHC and future high energy colliders, a sample of a billion top quarks will be produced, allowing precision searches for new physics in top quark decay and production.  To aid in this endeavor, we characterize the independent three and four point on-shell amplitudes involving top quarks, under the assumption of $SU(3)_c\times U(1)_{em}$ invariance.  The four point amplitudes factorize into primary and descendent amplitudes, where descendants are primaries multiplied by Mandelstam variables.  By enumerating the allowed amplitudes, we can check for amplitude redundancies to find the number of independent terms and convert those into a Lagrangian which parameterizes these amplitudes.  These results are then cross checked by utilizing the Hilbert series to count the number of independent Lagrangian operators.  Interestingly, we find situations where the Hilbert series has cancellations which, if na\"{i}vely interpreted,  would lead to the incorrect conclusion that that there are no primary operators at a given mass dimension.  We characterize the four fermion ($ffff$) and two fermion, two gauge boson ($ffVV$) operators respectively up to  dimension 12 and 13.  Finally, by combining unitarity bounds on the coupling strengths and simple estimates of the branching ratio sensitivities, we highlight interesting amplitudes for top quark decay that should be studied more closely at the HL-LHC.  Of those highlighted, there are both new charge current and flavor changing neutral current decays that occur at dimension 8 and 10 in SMEFT.      

\end{abstract}
\maketitle

\noindent
\section{Introduction}
\label{sec:intro}
The search for new physics beyond the Standard Model, at the LHC and beyond, has been led by the well established methods of effective field theory (EFT).   To  parameterize the indirect effects of new physics there are the two main paradigms of SMEFT \cite{Buchmuller:1985jz,Grzadkowski:2010es} and HEFT \cite{Feruglio:1992wf}.  These two approaches have differing assumptions about the physics at high energy scales and the relative importance of different effects.  

There are however a variety of issues that can obfuscate the connection between EFTs and experimental signals.   There is the large number of allowed interactions and also the complication of redundant (or incomplete) bases from equivalences due to equations of motion and integration by parts.  These issues have motivated work to understand the direct connection between dimension 6 SMEFT terms and the physical observables they parameterize \cite{Gupta:2014rxa,Gonzalez-Alonso:2014eva,Greljo:2015sla, Falkowski:2001958}.  

These redundancies on the Lagrangian side do not affect the predictions of physical amplitudes where external particles are on-shell.   Since these amplitudes are the direct observables accessible to experiment,  they provide a useful intermediary between theory and experiment.  Recent work in the study of amplitudes has allowed greater insight into the independent amplitudes for a given process.  In particular, the general structure of beyond the Standard Model amplitudes, given just $SU(3)_c \times U(1)_{em}$ invariance, has been analyzed, using both spinor helicity variables \cite{Shadmi:2018xan, Durieux:2019eor, Durieux:2020gip,Dong:2022jru,Liu:2023jbq} as well as standard variables \cite{Chang:2022crb}.    

Ref.~\cite{Chang:2022crb} was able to characterize the  structure of on-shell 3 and 4 point amplitudes involving the Higgs.  To complete this procedure, a set of potential on-shell amplitudes was constructed out of Lorentz invariant combinations of momenta and polarizations.  By studying their Taylor expansion in the kinematic variables, a set of independent amplitudes was determined.  These could then be converted into a basis of Lagrangian operators.  As a cross check, the number of independent operators at each mass dimension could be determined using the Hilbert Series approach \cite{Lehman:2015via,Henning:2015daa,Lehman:2015coa,Henning:2015alf,Henning:2017fpj, Graf:2020yxt, Graf:2022rco}.  For the four point couplings, this lead to a number of primary amplitudes/operators whose multiplication by Mandelstam variables gave descendant amplitudes/operators.  If these new interactions are mediated by the exchange of a massive particle, the lowest order primary amplitude would be a first approximation to the relevant phenomenology.  Finally, by requiring unitarity up to an energy $E_\text{max}$, one can place upper bounds on their coupling strength.  These results, when combined with simple estimates, suggested that there are new amplitudes in Higgs decays into $Z\bar{f}f, W\bar{f}f , \gamma \bar{f}f,$ and $Z\gamma\gamma$ that could be searched for at the HL-LHC. 

In this paper, we extend this study to amplitudes involving the top quark.  At the HL-LHC and future TeV colliders, over a billion top quarks will be produced, allowing the study for rare decays as well as new production mechanisms.  This requires understanding the general structure of four fermion operators and two fermion operators with two gauge bosons, which can result in primaries up to dimension 11.  This vector space of amplitudes is spanned by these primary and descendant amplitudes, which in a model agnostic analysis can be taken to be independent \footnote{One can certainly model build situations where high dimension operators can be the leading correction to the Standard Model.  For example, a theory having multiple massive mediators, which all need to be integrated out to generate an operator with just Standard Model particles, could be a potential realization.}.    Interestingly, in this classification, we find interactions (e.g.~$\gamma\gamma\bar{f}f$) whose Hilbert series numerator has a complete cancellation in the coefficient for one of the terms, where a na\"{i}ve inspection incorrectly concludes that there are no primary operators at a certain mass dimension.   
In our analysis, we have also checked that the primary and descendant structure up to at least dimension 12, going beyond the existing dimension 8 results using spinor-helicity variables \cite{Dong:2022jru,Liu:2023jbq}.  As an initial look at the phenomenology of these operators, we give simple estimates that top quark decays for which FCNC  modes (e.g.~ $t\to c(\bar{\ell}\ell, h\gamma, hg, Z\gamma, Zg, \gamma\gamma, \gamma g)$)  and charged current decay modes could be interesting to search for at the HL-LHC.  These simple estimates indicate that there are some decay modes that appear at dimension 8 and 10 in SMEFT that are worth studying in more detail.       

The rest of this paper is organized as follows:  Section~\ref{sec:indptamps} describes what amplitudes we will explore and how to determine independent amplitudes.  Section~\ref{sec:HilbertSeries} discusses the Hilbert series results for our top quark operators.  In Section~\ref{sec:pheno}, we discuss some relevant phenomenological issues, such as unitarity bounds on coupling strengths and also rough estimates for top quark decays at the HL-LHC.  Section~\ref{sec:amplitudes} is the main body of results, where we list the operators for the primary amplitudes.  In Section~\ref{sec:interestingdecays}, we estimate which top decay amplitudes are interesting for exploration at HL-LHC. Finally in Section~\ref{sec:conclusions}, we conclude.

\section{Finding Independent Amplitudes/Couplings for Top Quarks}
\label{sec:indptamps}
The general on-shell amplitudes needed for top quark phenomenology are invariant under $SU(3)_c\times U(1)_{em}$ and Lorentz symmetry.  For 3 and 4 point interactions, imposing $SU(3)_c$ and Lorentz symmetry gives the following list:  
\begin{align}
\text{3pt}:  \bar{q}qV, \bar{q}qh, \quad \text{4pt}:  \bar{q}q\bar{\ell}\ell, qqq \ell, \bar{q}q\bar{q}q, \bar{q}qhh, \bar{q}qhV, \bar{q}qVV
\end{align}
where $q$ is a quark, $\ell$ is a lepton (charged or neutral), $h$ is a Higgs boson, and $V$ is any gauge boson.  To fully characterize these 4 point interactions, we also need additional 3 point interactions for exchange diagrams, which add 
\begin{align}
\text{3pt additional}:  VVV, hVV, hhh, \bar{\ell}\ell h, \bar{\ell}\ell V.
\end{align}
Of these couplings,  the three point couplings and $\bar{q}qhh, \bar{q}qhV$ have been fully characterized (e.g~~\cite{Chang:2022crb}), so in this paper this leaves the following four point couplings to determine:
\begin{align}
\bar{q}qVV:& \quad  WW\bar{q}q, WZ \bar{q}q', ZZ\bar{q}q, Z\gamma\bar{q}q, Zg\bar{q}q,W\gamma\bar{q}q', Wg\bar{q}q', g\gamma\bar{q}q, \gamma\gamma \bar{q}q, gg \bar{q}q, \\
\text{Four fermion}:& \quad  \bar{q}q \bar{\ell}\ell,\bar{q}q'\bar{e}\nu, qqq\ell, \bar{q}q\bar{q}q.
\end{align}
When there are identical particles involved, the form of the amplitude must respect the relevant exchange symmetry and for these, there are no amplitudes with 3 or more identical particles (note that, if we were characterizing down quark interactions, we would have to consider $ddd\bar{e}$).

In \cite{Chang:2022crb}, a general approach for finding independent amplitudes for 3 and 4 point on-shell amplitudes was presented.  Here, we give a brief overview of the process and refer to that paper for further details, but will also note where changes in that approach need to be made.  To characterize four point on-shell amplitudes, we form Lorentz invariants out of particle momenta, fermion wavefunctions, and gauge boson polarizations.  For massless gauge bosons, we use the field strength contribution $\epsilon_\mu p_\nu - \epsilon_\nu p_\mu$, so that the amplitude is manifestly gauge invariant.  Three point interactions with a covariant derivative can also give a four point contact interaction with a gauge boson; for our cases, the only one that will be relevant  is $\bar{q} \sigma_{\mu\nu} q' W^{\mu\nu}$, which generates a $\bar{q}q' W \gamma$ interaction.   This results in a set of amplitudes ${\cal M}_a$, giving a linear parameterization of the general amplitudes ${\cal M} = \sum_a C_a {\cal M}_a$.  For each on-shell amplitude ${\cal M}_a$, we can associate a local Lagrangian operator, which we choose to have the lowest mass dimension possible, $\frac{c_a}{v^{d_{\cal O}-4}} {\cal O}_a$, where we've normalized its coefficient with factors of the Higgs vev to give a dimensionless coupling $c_a$, resulting in a Lagrangian which parameterizes the on-shell amplitudes
\begin{align}
{\cal L_\text{amp}} = \sum_a  \frac{c_a}{v^{d_{\cal O}-4}} {\cal O}_a.
\end{align}

By connecting these amplitudes to Lagrangian operators, we can work in increasing mass dimension of the corresponding operators.  For example, $\bar{q}qWW$ starts at dimension 5, since the lowest local operator needs two fermions and two gauge bosons, while $\bar{q}q\gamma\gamma$ will start at dimension 7.  At a given mass dimension, we write out all of the amplitudes for the allowed particle helicities.  In cases where there are two particles that are identical, we symmetrize and anti-symmetrize with respect to those two particles.  After finding the allowed primary amplitudes for the distinguishable case, we can achieve the indistinguishable case by imposing the Bose/Fermi symmetry.  We'll have more to say on that later, when we have the Hilbert series results.  

For our four point amplitudes, we consider $1+ 2\to 3+4$ scattering in the center of mass frame, where $p_1=(E_1,0,0,p_i), p_2=(E_2,0,0,-p_i), p_3=(E_3,0,p_f \sin\theta,p_f\cos\theta), p_4=(E_4,0,-p_f \sin\theta,-p_f \cos\theta)$.  On-shell these have the constraints
\begin{align}
E_1= \frac{E_{com}^2+m_1^2-m_2^2}{2 E_{com}}, E_2= \frac{E_{com}^2+m_2^2-m_1^2}{2 E_{com}}, E_3= \frac{E_{com}^2+m_3^2-m_4^2}{2 E_{com}}, E_4= \frac{E_{com}^2+m_4^2-m_3^2}{2 E_{com}} 
\end{align} 
A general kinematic configuration is determined by the two continuous parameters $E_{com}$ and $\cos \theta$ as well as the choice of helicities.  However, treating $p_i, p_f,$ and $\sin \theta$ as independent is advantageous for finding amplitude redundancies.  On-shell, one can replace even powers of these variables as $\sin^2 \theta = (1-\cos^2\theta), p_i^2 = \frac{(E_{com}^2-(m_1+m_2)^2)(E_{com}^2-(m_1-m_2)^2)}{4E_{com}^2}, p_f^2 = \frac{(E_{com}^2-(m_3+m_4)^2)(E_{com}^2-(m_3-m_4)^2)}{4E_{com}^2}$.    After doing this, as shown in detail in \cite{Chang:2022crb}, the Taylor series coefficients of the amplitudes expansion in $E_{com}, p_i,p_f,\cos\theta, \sin\theta$ must all vanish if there is an amplitude redundancy.  Schematically, if there are Taylor series coefficients $B_\alpha$, we then form the matrix $\frac{\partial B_\alpha}{\partial C_a}$, evaluate it for random numerical values for the particle masses, and numerically evaluate its singular value decomposition.  The number of nonzero values in that decomposition is the number of independent amplitudes and one can find the independent ones by removing $C_a$'s one at a time. 

There are a few modifications to \cite{Chang:2022crb} needed to address the amplitudes of this paper.  First of all, for four fermion amplitudes, we are required to have fermions in the final state.  Similar to that paper, we can choose a mass configuration, either $m_3 =0, m_4\neq 0$ or $m_3=m_4$, to constrain the variable dependence of the kinematic variables in the fermion wavefunctions.  We have checked that this mass assumption doesn't affect the basis of independent amplitudes.  Having final state fermions also results in dependence on $\cos \frac{\theta}{2}, \sin \frac{\theta}{2}$, which can  be treated by replacing $\cos \theta = 2\cos^2 \frac{\theta}{2} -1$ and $\sin \theta = 2 \cos\frac{\theta}{2} \sin\frac{\theta}{2}$ and using $\cos \frac{\theta}{2}$ and $\sin\frac{\theta}{2}$ as our variables.  Another complication is that the allowed $SU(3)$ gauge invariant contractions are more diverse than before.  This issue interplays with the Bose/Fermi symmetries of the amplitudes.  As an example, for $\bar{q}qgg$, interchange of the gluons must result in the same amplitude.  If the gluons are contracted with an $f_{ABC}$ then the amplitude must also be odd under exchange of the momenta and polarizations of the gluons.  On the other hand if the gluons are contracted with a $d_{ABC}$ then    
the amplitude must also be even under exchange of the momenta and polarizations of the gluons.  

\section{Hilbert Series}
\label{sec:HilbertSeries}
The Hilbert series gives a systematic way to count the number of gauge invariant independent operators, up to equation of motion and integration by part redundancies \cite{Lehman:2015via,Henning:2015daa,Lehman:2015coa,Henning:2015alf,Henning:2017fpj, Graf:2020yxt, Graf:2022rco}, which provides a useful cross check on our amplitude counting.  It gives a function, whose Taylor series expansion in a parameter $q$ gives the number of independent operators at each mass dimension \footnote{How to treat massive gauge bosons has only recently been worked out and is best explained in \cite{Graf:2022rco}.}.  In Eqn.~\ref{eqn:Hilbert4pt}, we list the Hilbert series for each of the four point operators that we will characterize.  The three point and the other four point operator results can be found in \cite{Chang:2022crb}.  
\begin{align}
\begin{split}
& H_{WW\bar{f}f}=H_{WZ\bar{f}f'} = \frac{4q^5+12q^6+16q^7+6q^8-2q^9}{(1-q^2)^2},\\
& H_{ZZ\bar{f}f} = \frac{2q^5+6q^6+12q^7+6q^8+6q^9+6q^{10}-2q^{11}}{(1-q^2)(1-q^4)},\\
& H_{Z\gamma\bar{f}f}=H_{Zg\bar{f}f}=H_{W\gamma\bar{f}f'}=H_{Wg\bar{f}f'} = \frac{4q^6+12q^7+8q^8+(2-2)q^9}{(1-q^2)^2},\\
& H_{g \gamma\bar{f}f} = \frac{6q^7+8q^8+(4-2)q^9}{(1-q^2)^2}, \quad H_{\gamma \gamma\bar{f}f} = \frac{4q^7+2q^8+4q^9+6q^{10}+(2-2)q^{11}}{(1-q^2)(1-q^4)},\\
& H_{gg\bar{f}f} = \frac{10q^7+10q^8+(14-2)q^9+14q^{10}+(6-4)q^{11}}{(1-q^2)(1-q^4)},\\
& H_{\bar{q}q \bar{\ell}\ell} = H_{\bar{q}q' \bar{e}\nu} = H_{q_1 q_2 q_3 \ell} = \frac{10q^6+8q^7-2q^8}{(1-q^2)^2}, \\
& H_{q q q' \ell} = \frac{4q^6+6q^7+(6-2)q^8+2q^9}{(1-q^2)(1-q^4)}, \quad H_{\bar{q}\bar{q}' q q'} = \frac{2(10q^6+8q^7-2q^8)}{(1-q^2)^2}, \\
& H_{\bar{q}\bar{q}' q q}= H_{\bar{q}\bar{q} q q'} = \frac{10q^6+8q^7+(10-2)q^8+8q^9-2q^{10}}{(1-q^2)(1-q^4)}, \\
&  H_{\bar{q}\bar{q} q q} = \frac{8q^6+4q^7+(8-2)q^8+4q^9-2q^{10}}{(1-q^2)(1-q^4)}. 
\end{split}
\label{eqn:Hilbert4pt}
\end{align}
These fractional forms are interpretable in the following way:  the numerator counts the number of primary operators and the denominator allows for the dressing of these operators with Mandelstam factors.

For example, looking at $H_{\bar{q}q\bar{\ell}\ell} =  \frac{10q^6+8q^7-2q^8}{(1-q^2)^2}$, the numerator says that there are 10 dimension 6 primary operators and 8 dimension 7 primary operators.  Ignore for now the $-2q^8$, which we'll see denotes two constraints that appear at dimension 8.  The denominator of $1/(1-q^2)^2$ has an expansion of $(1+q^2 +q^4 + \cdots)^2$ which is just counting the number of operators from multiplying the primaries by Mandelstam factors of $s,t$ ($u$ is redundant to the on-shell condition).  As we will see when we analyze the amplitudes of this interaction, two primary amplitudes at dimension $6$, say $M_a, M_b$ (with respective operators ${\cal O}_a, {\cal O}_b$), when multiplied by a factor of $s$ are redundant to a linear combination of other amplitudes, so are no longer independent at dimension 8.  This explains the $-2q^8$ since treating this as the loss of the two related operators $s {\cal O}_a$ and $s {\cal O}_b$ and all of their descendants gives the correct counting of the number of independent terms.  Such negative coefficients in the Hilbert series often occur when the particles have nonzero spin \cite{Lehman:2015via,Henning:2015daa,Lehman:2015coa,Henning:2015alf,Henning:2017fpj, Graf:2020yxt, Graf:2022rco}, as identities relate operators of different tensor structures when combined with derivatives.  For four point functions, there is an argument from counting conformal correlators that the number of primary operators is equal  to the product of the spin degrees of freedom of the participating particles \cite{Henning:2017fpj,  Schomerus:2016epl, Kravchuk:2016qvl}.  In our results, this is correct for all cases except $\bar{q}\bar{q}qq$, if one includes the negative coefficients and takes into account possible $SU(3)_c$ contractions.  For example, for $\bar{q}q\bar{\ell}\ell$, the sum of the numerator coefficients $10+8-2 = 16$ is equal to the spin counting of $2^4$.  On the other hand, the case of $\bar{q}\bar{q}qq$ has further constraints from the crossing symmetry of the $\bar{q}$ and $q$, resulting in fewer operators.

We also  note that for some denominators, the factors are $(1-q^2)(1-q^4)$.  This results for situations where there are two identical particles in the amplitude.  Assuming the two initial state particles are the identical pair,  $s$ and $(t-u)^2$ are the Mandelstam factors that have the correct exchange symmetry between the two particles, so we are allowed to multiply the primary by an arbitrary set of $s$ and $(t-u)^2$ factors (note that the primary already has a factor of $+/-$ when exchanging bosons/fermions).    

As you'll notice in the Hilbert series list, some of the numerator coefficients are written in an unusual way, for example the $(14-2)q^9$ and $(6-4)q^{11}$ in $H_{gg\bar{f}f}$.  When we evaluated the Hilbert series, these would of course have been $12q^9$ and $2q^{11}$.  However, when examining the number of independent amplitudes at dimension 9, we found 14 new primaries and 2 redundancies when 2 of the dimension 7 amplitudes were multiplied by $s$.  In this way, the Hilbert series must be interpreted with care, as there can be hidden cancellations.  In some case, there is even a complete cancellation like the $(2-2)q^{11}$ term for $\gamma\gamma\bar{f}f$, where a na\"{i}ve interpretation would have  missed the new primaries at dimension 11.  

The Hilbert series also allows for understanding of the constraints of Bose/Fermi symmetry.  For example, for $gg\bar{f}f$ there are two symmetric contractions for the gluon $SU(3)$ indices ($\delta_{AB}, d_{ABC}$) and one antisymmetric contraction ($f_{ABC}$), then  swapping the kinematic variables of the two gluons would result respectively in a $+$ sign for the first two and a $-$ sign for the last one.  Now, if we calculated the Hilbert series assuming photons were odd under interchange, then $H^{asym}_{\gamma \gamma\bar{f}f} = \frac{2q^7+6q^8+(6-2)q^9+2q^{10}+2q^{11}}{(1-q^2)(1-q^4)}$.  One can then check that $H_{gg\bar{f}f} = 2 H_{\gamma \gamma\bar{f}f} + H^{asym}_{\gamma \gamma\bar{f}f}$ as expected from the behavior under kinematic variable exchange and the allowed $SU(3)$ contractions.   

Note that unlike in \cite{Chang:2022crb}, due to complications of enumerating all of the terms, we do not claim to have examined the full, allowed tensor structures of the amplitudes.  Instead, we have checked that we agree with the Hilbert series up to dimension 13 for $\bar{q}q VV$ amplitudes and dimension 12 for four fermion amplitudes.  Up to those dimensions, the numerator of these Hilbert series do not have any additional cancellations.  As the Hilbert series shows, the redundancies that appear at higher dimension appear in pairs so it seems unlikely there are more, but still we cannot guarantee that others do not appear at higher dimension.

\section{Phenomenology}
\label{sec:pheno}
\subsection{Unitarity}
As in \cite{Chang:2022crb}, we utilize unitarity to constrain the coupling strengths of these operators.  Since these are new couplings beyond the Standard Model, they violate unitarity at high energies.  Requiring the amplitudes to satisfy perturbative unitarity up to a scale $E_\text{max}$, gives an upper bound on the couplings.  The technique follows the work \cite{Chang:2019vez,Abu-Ajamieh:2020yqi, Abu-Ajamieh:2021egq, Abu-Ajamieh:2022ppp}, where the unitarity bounds due to high multiplicity scattering was developed (see also \cite{Maltoni:2001dc,Dicus:2004rg,Dror:2015nkp, Falkowski:2019tft, Maltoni:2019aot}).  

To stand in for a more detailed calculation of each amplitude, we utilize a SMEFT operator realization of the amplitude to act as a proxy.  As an example, consider the case of $\frac{c}{v}\bar{q}qWW.$  This is realized by the dimension 8 SMEFT operator $\frac{1}{\Lambda^4}(\bar{Q_L} \tilde{H} u_R +\text{h.c.})|D^\mu H|^2$ \footnote{Note that the dimension 6 operator $\frac{1}{\Lambda^2}(\bar{Q_L} (D^2 \tilde{H}) u_R +\text{h.c.})$ can be reduced by equations of motion and does not result in the correct high energy behavior of the $\bar{q}qWW$ interaction.}.  Since we are only looking for an approximate bound, we ignore $O(1)$ factors like $\sqrt{2}, g, g', \sin\theta_W, \cos\theta_W$ and only take into account factors of $v$.  Under this approximation, $c \approx v^4/\Lambda^4$.    The SMEFT operator has many contact interactions that violate unitarity, but we find that either the lowest and highest multiplicity give the best bound as a function of $E_\text{max}$, so we will calculate these for all interactions and include them in our tables.  For this example, the lowest multiplicity amplitude is for two quarks and two Goldstones, with a matrix element that goes as $M_{2\to2} \approx \frac{v E_\text{max}^3}{\Lambda^4}$, where one factor of $E_\text{max}$ comes from the fermion bilinear and the other two come from the two derivatives acting on the Goldstones.  This is bounded by phase space factors $M_{2\to 2} \leq 8\pi$ \cite{Chang:2019vez}, which translates into $c\leq (8\pi)v^3/E_\text{max}^3 \approx \frac{0.4}{E_\text{TeV}^3}$ where $E_\text{TeV} = E_\text{max}/\text{TeV}$.  The highest multiplicity amplitude is for two quarks and 3 Goldstones, with $M_{2\to3} \approx  \frac{E_\text{max}^3}{\Lambda^4} \leq \frac{32\pi^2}{E_\text{max}}$, where the bound again depends on the phase space.  This gives the bound $c\leq (32\pi^2)v^4/E_\text{max}^4 \approx \frac{1.2}{E_\text{TeV}^4}.$   As this example illustrates, we generally find that the low multiplicity constraint is stronger for $E_\text{max} < 4\pi v$ and the high multiplicity one is stronger for energies above that.  

\subsection{Top Quark Decays}
The HL-LHC will produce about 5 billion top quarks, allowing searches for rare decays as well as new production modes.  Here we will consider decay modifications due to our amplitudes.  The on-shell 2 and 3 body decay modes of the top quark allowed by the Standard Model quantum numbers are
\begin{align}
t \to dW, u(Z, h), d(e \nu, \bar{d}u, WZ, W\gamma, Wg), u(\bar{\ell}\ell, \bar{q}q,WW,Z\gamma, Zg, \gamma\gamma, \gamma g, gg)  
\end{align}  
along with changes in flavors of quarks and leptons.  

Searches for the flavor changing two body decays are actively being pursued at the LHC (e.g.~\cite{CMS:2021hug, CMS:2021gfa, ATLAS:2021amo, ATLAS:2022per, ATLAS:2022gzn,CMS-PAS-TOP-21-013, ATLAS:2023qzr}), where theoretical analyses are often performed in SMEFT (e.g.~\cite{Aguilar-Saavedra:2008nuh, Durieux:2014xla, Aguilar-Saavedra:2018ksv, Altmannshofer:2023bfk}).  Some of the three body decays are higher order decays that exist in the Standard Model at tree level (e.g.~$dW(Z, \gamma, g), uWW$), while the others require flavor changing neutral current interactions which should be suppressed in the Standard Model.   Searches for new decay modes can be triggered by requiring one of the tops decays in the standard leptonic channel and then looking for the new decay mode for the other top quark.  

For this simple analysis of the phenomenology, we will approximate top decay amplitudes as a constant, assuming the top quark mass is the only relevant mass scale
\begin{align}
\mathcal{M}_{\mathcal {O}}(t \to 2) 
&\simeq \frac{c_{\mathcal {O}}}{v^{d_{\mathcal {O}} - 4}} 
m_t^{d_{\mathcal {O}} - 3} \approx c_{\mathcal {O}} \left(\frac{m_t}{v}\right)^{d_{\mathcal {O}} - 4}m_t \approx c_{\mathcal {O}} 2^{2-d_{\mathcal {O}}/2}m_t,
\\
\mathcal{M}_{\mathcal {O}}(t \to 3) &\simeq \frac{c_{\mathcal{O}}}{v^{d_{\mathcal{O}} - 4}}
m_t^{d_{\mathcal{O}} - 4} \approx c_{\mathcal {O}} \left(\frac{m_t}{v}\right)^{d_{\mathcal {O}} - 4}\approx c_{\mathcal {O}} 2^{2-d_{\mathcal {O}}/2},
\end{align}
where we've approximated $v \approx \sqrt{2} m_t$.  Note that this ignores $O(1)$ enhancements of the form $(m_t/m_W)$ that can come from longitudinal polarizations, but is sufficient for our estimates.  

Let's first consider non-FCNC top decays that are not suppressed in the Standard Model, such as $t\to b(W, \ell \nu, W\gamma, Wg)$.  In such cases, one has at least the Standard Model top background to contend with.  For new amplitudes which are CP even, they will interfere with the Standard Model amplitude and have enhanced sensitivities (unless one designs CP violating observables).  In this case, we want to compare the number of new decays to the fluctuation in the Standard Model top background.  Under our approximation the branching ratios in the Standard Model and the modification due to interference are  
\begin{align}
& Br(t\to 2)_{SM} \approx \frac{1}{16\pi m_t \Gamma_t} |\mathcal{M}(t\to 2)_{SM}|^2, \\
& \delta Br(t\to 2) \approx \frac{1}{16\pi m_t \Gamma_t} |\mathcal{M}(t\to 2)_{SM}||\mathcal{M}(t\to 2)_{BSM}|.
\end{align}
To estimate sensitivity, we require that the new top decays must be as large as a one sigma deviation in the Standard Model top background, which for a sample of $N_t$ top quarks gives $N_t \delta Br(t\to 2) \gtrsim \sqrt{N_t Br(t\to 2)_{SM}}$.  Such a calculation gives for two and three body decays the constraints  
\begin{align}
\begin{split}
\text{2 Body Decays}:& \quad  c\gtrsim 5\times10^{-6} \left(\frac{10^9}{N_t}\right)^{1/2} 2^{d_\mathcal{O}/2}, \\
\text{3 Body Decays}: & \quad c\gtrsim 6\times 10^{-5} \left(\frac{10^9}{N_t}\right)^{1/2} 2^{d_\mathcal{O}/2} 
\end{split}\label{eqn:cdecaybound}
\end{align}
where we've normalized to a total sample of a billion top quarks.  

For FCNC decays, such as $t\to c (Z, \gamma, g, WW, Z\gamma, Zg, \gamma\gamma, \gamma g, gg)$, the branching ratios predicted in the Standard Model ($10^{-12}$ to $10^{-17}$) are too small to occur at the HL-LHC (e.g.~\cite{Diaz-Cruz:1989tem, Eilam:1990zc, Jenkins:1996zd, Mele:1998ag, Aguilar-Saavedra:2004mfd}).  Thus, for these decays we can ignore interference and give an estimate that works for both CP even and odd interactions.  If we make an optimistic assumption that other backgrounds can be neglected, this requires that the new branching ratios $Br_{BSM}$ give a few events at the HL-LHC or $N_t Br_{BSM} \gtrsim 1$. Under our approximation, this gives the same bounds as Eqn.~\ref{eqn:cdecaybound}. 

To get some sense of how well this approximation works, we've checked in a few existing FCNC searches, whether the background free assumption works at the $O(1)$ level.  As one might expect, one finds that for final states with a single gluon or photon, where hadronic backgrounds and fakes are relevant, that this is a poor assumption and gives a branching ratio bound that is too strong by two and three orders of magnitude  for photon  and gluon decays, respectively.  Thus, estimates for these final states should be viewed as very optimistic.  However, we found that the searches with a Higgs decaying into two photons agree roughly with our bounds.  Similarly, the final states with $e, \mu$'s give bounds that are correct to a factor of $2-3$ as long as one takes into account tagging efficiencies for $b$ ($\sim 0.5$), $e/\mu$ ($\sim 0.8$) and, when relevant, $Z$ and $W$ leptonic branching ratios ($\sim 0.06$ and $0.2$).   Thus, as long as one take these factors into account, these final states should be more reliable.  Later, when combined with our upper bounds from perturbative unitarity, these calculations will enable us to give a simple estimate of which decay amplitudes that are worth exploring further at the HL-LHC. 

\section{Independent Amplitudes for Top Quark Physics}
\label{sec:amplitudes}
In the following subsections, we will list operators corresponding to the primary amplitudes for  $ffVV$ and $ffff$ interactions involving the top quark.  We will make comparisons to the Hilbert series to show consistency with the number of independent operators, including discussions of redundancies that occur at certain mass dimensions.  We will also give $CP$ properties of the operators and unitarity bounds on the coupling constants for these interactions.  

\subsection{$ffVV$ Amplitudes}

Tables~\ref{tab:qqww1} and~\ref{tab:qqww2} list the primary operators for $\bar{q}qWW$ interactions. Note that for the primary operators, covariant derivatives are with respect to $SU(3)_c\times U(1)_{em}$ and thus only involve the photon and gluon.  From the Hilbert series, we expect that there should be 4 operators at dimension 5, 12 operators at dimension 6, 16 operators at dimension 7, 6 operators at dimension 8, and at least two redundancies at dimension 9. This is precisely what we find, with the 38 listed operators and at dimension 9, $s\mathcal{O}_{26}$ and $s\mathcal{O}_{27}$, where $s=(p_q+p_{\bar{q}})^2$, become redundant to other operators.  To be concrete, one can replace these two operators with an operator of the following form 
\begin{align}
& \sum_{i=1\cdots 4} (c_i + c_{i,s} s + c_{i,t} t +c_{i,ss} s^2 +c_{i,st} st+ c_{i,tt} t^2) \mathcal{O}_i  \\ &+ \sum_{i=5\cdots 25,\, 28\cdots 32} (c_i + c_{i,s} s + c_{i,t} t) \mathcal{O}_i + \sum_{i=26, 27} (c_i +c_{i,t} t) \mathcal{O}_i,
\end{align}
where the coefficients $c_i$'s only depend on the particle masses and predict the same on-shell amplitudes as $s\mathcal{O}_{26}$ and $s\mathcal{O}_{27}.$
To generate an independent set of operators, one needs to add descendants of the primaries, which involve multiplying by arbitrary powers of $s$ and $t$.  However, because of the redundancies at dimension 9 for $s\mathcal{O}_{26}$ and $s\mathcal{O}_{27}$, one only needs the descendants $t^n \mathcal{O}_{26}$ and $t^n \mathcal{O}_{27}$ for $\mathcal{O}_{26}$ and $\mathcal{O}_{27}$.  Note that this explains the $\frac{-2q^9}{(1-q^2)^2}$ part of the Hilbert series for $H_{WW\bar{f}f}$, since operators of the form $s^n t^m \mathcal{O}_{26}$ and $s^n t^m \mathcal{O}_{27}$ (with $n\geq 1$) are redundant, so one needs this term in the Hilbert series to correct the counting of independent operators.  We've also listed the lowest dimensional SMEFT-like operator (that we could find) which realizes each operator, where the covariant derivatives are with respect to $SU(3)_c\times SU(2)_L \times U(1)_Y$.  We also list the unitarity bounds for each SMEFT operator, assuming the lowest and highest particle multiplicity. These operators can also be reworked to account for $\bar{q}q^{\prime}WZ$ amplitudes provided we take $q\to q^{\prime}$ and $W\to Z$. Here, we use $q^{\prime}$ to denote a different quark flavor of the correct charge. 

\begin{table}[p]
\begin{adjustwidth}{-.5in}{-.5in}  
\begin{center}
\footnotesize
\centering
\renewcommand{\arraystretch}{0.9}
\setlength{\tabcolsep}{6pt}
\begin{tabular}{|c|c|c|c|c|c|}

\hline
\multirow{2}{*}{$i$} & \multirow{2}{*}{$\mathcal{O}_i^{\bar{q}qW^{+}W^{-}}$}   & \multirow{2}{*}{CP} & \multirow{2}{*}{$d_{\mathcal{O}_i}$}& SMEFT & $c$ Unitarity  \\
 & & & & Operator & Bound \\

\hline 
1 & \scriptsize{$\left(\bar{q}q\right) \left(W^{+}_{\mu}W^{-\mu}\right)$} & $+$ & \multirow{4}{*}{5} & \scriptsize{$\left(\bar{Q}_{L}\htilde u_{R}+\textrm{h.c.}\right)\left(|D^{\mu}H|^{2}\right)$} &  \multirow{4}{*}{  $\frac{0.4}{E_\text{TeV}^3}, \frac{1.2}{E_\text{TeV}^4}$ }  \\

2 & \scriptsize{$\left(i\bar{q}\gamma_{5}q\right) \left(W^{+}_{\mu}W^{-\mu}\right)$} & $-$ &  & \scriptsize{$\left(i\bar{Q}_{L}\htilde u_{R}+\textrm{h.c.}\right)\left(|D^{\mu}H|^{2}\right)$} &  \\

3 & \scriptsize{$\left(\bar{q}\sigma^{\mu\nu}q\right) \left(iW^{+}_{\mu}W^{-}_{\nu}\right)$} & $+$ &  & \scriptsize{$\left(\bar{Q}_{L}\sigma^{\mu\nu}\htilde u_{R}+\textrm{h.c.}\right)\left(i\left[D_{\mu}H\right]^{\dagger}\left[D_{\nu}H\right]+\textrm{h.c.}\right)$} &  \\

4 & \scriptsize{$\left(i\bar{q}\sigma^{\mu\nu}\gamma_{5}q\right) \left(iW^{+}_{\mu}W^{-}_{\nu}\right)$} &$-$&  & \scriptsize{$\left(i\bar{Q}_{L}\sigma^{\mu\nu}\gamma_{5}\htilde u_{R}+\textrm{h.c.}\right)\left(i\left[D_{\mu}H\right]^{\dagger}\left[D_{\nu}H\right]+\textrm{h.c.}\right)$} &  \\
\hline

5 & \scriptsize{$\left(\bar{q}\gamma^{\nu}q\right) \left(iW^{+\mu}\bfd_{\nu}W^{-}_{\mu}\right)$} & + & \multirow{12}{*}{6} & \scriptsize{$\left(\bar{Q}_{L}\gamma^{\nu}Q_{L}+\bar{u}_{R}\gamma^{\nu}u_{R}\right)\left(i\left[D^{\mu}H\right]^{\dagger}\bfd_{\nu}\left[D_{\mu}H\right]+\textrm{h.c.}\right)$} & \multirow{12}{*}{  $\frac{0.09}{E_\text{TeV}^4}$ }  \\

6 & \scriptsize{$\left(\bar{q}\gamma^{\nu}\gamma_{5}q\right) \left(iW^{+\mu}\bfd_{\nu}W^{-}_{\mu}\right)$} & + &  & \scriptsize{$\left(\bar{Q}_{L}\gamma^{\nu}Q_{L}-\bar{u}_{R}\gamma^{\nu}u_{R}\right)\left(i\left[D^{\mu}H\right]^{\dagger}\bfd_{\nu}\left[D_{\mu}H\right]+\textrm{h.c.}\right)$} & \\

7 & \scriptsize{$\left(i\bar{q}\gamma^{\nu}\bfd_{\mu}q\right) \left(W^{+\mu}W^{-}_{\nu}+\textrm{h.c.}\right)$} & + &  & \scriptsize{$\left(i\bar{Q}_{L}\gamma^{\nu}\bfd_{\mu}Q_{L}+i\bar{u}_{R}\gamma^{\nu}\bfd_{\mu}u_{R}\right)\left(\left[D^{\mu}H\right]^{\dagger}\left[D_{\nu}H\right]+\textrm{h.c.}\right)$} & \\

8 & \scriptsize{$\left(i\bar{q}\gamma^{\nu}\gamma_{5}\bfd_{\mu}q\right) \left(W^{+\mu}W^{-}_{\nu}+\textrm{h.c.}\right)$} & + &  & \scriptsize{$\left(i\bar{Q}_{L}\gamma^{\nu}\bfd_{\mu}Q_{L}-i\bar{u}_{R}\gamma^{\nu}\bfd_{\mu}u_{R}\right)\left(\left[D^{\mu}H\right]^{\dagger}\left[D_{\nu}H\right]+\textrm{h.c.}\right)$} & \\

9 & \scriptsize{$\left(\bar{q}\gamma^{\nu}q\right) \left(iW^{+\mu}D_{\mu}W^{-}_{\nu}+\textrm{h.c.}\right)$} & + &  & \scriptsize{$\left(\bar{Q}_{L}\gamma^{\nu}Q_{L}+\bar{u}_{R}\gamma^{\nu}u_{R}\right)\left(i\left[D^{\mu}H\right]^{\dagger}\left[D_{\mu\nu}H\right]+\textrm{h.c.}\right)$} & \\

10 & \scriptsize{$\left(\bar{q}\gamma^{\nu}\gamma_{5}q\right) \left(iW^{+\mu}D_{\mu}W^{-}_{\nu}+\textrm{h.c.}\right)$} & + &  & \scriptsize{$\left(\bar{Q}_{L}\gamma^{\nu}Q_{L}-\bar{u}_{R}\gamma^{\nu}u_{R}\right)\left(i\left[D^{\mu}H\right]^{\dagger}\left[D_{\mu\nu}H\right]+\textrm{h.c.}\right)$} & \\

11 & \scriptsize{$\left(i\bar{q}\gamma^{\mu}\bfd_{\nu}q\right) \left(iW^{+}_{\mu}W^{-\nu}+\textrm{h.c.}\right)$} & $-$ &  & \scriptsize{$\left(i\bar{Q}_{L}\gamma^{\mu}\bfd_{\nu}Q_{L}+i\bar{u}_{R}\gamma^{\mu}\bfd_{\nu}u_{R}\right)\left(i\left[D_{\mu}H\right]^{\dagger}\left[D^{\nu}H\right]+\textrm{h.c.}\right)$} & \\

12 & \scriptsize{$\left(i\bar{q}\gamma^{\mu}\gamma_{5}\bfd_{\nu}q\right) \left(iW^{+}_{\mu}W^{-\nu}+\textrm{h.c.}\right)$} & $-$ &  & \scriptsize{$\left(i\bar{Q}_{L}\gamma^{\mu}\bfd_{\nu}Q_{L}-i\bar{u}_{R}\gamma^{\mu}\bfd_{\nu}u_{R}\right)\left(i\left[D_{\mu}H\right]^{\dagger}\left[D^{\nu}H\right]+\textrm{h.c.}\right)$} & \\

13 & \scriptsize{$\left(\bar{q}\gamma^{\nu}q\right) \left(W^{+\mu}D_{\mu}W^{-}_{\nu}+\textrm{h.c.}\right)$} & $-$ &  & \scriptsize{$\left(\bar{Q}_{L}\gamma^{\nu}Q_{L}+\bar{u}_{R}\gamma^{\nu}u_{R}\right)\left(\left[D^{\mu}H\right]^{\dagger}\left[D_{\mu\nu}H\right]+\textrm{h.c.}\right)$} & \\

14 & \scriptsize{$\left(\bar{q}\gamma^{\nu}\gamma_{5}q\right) \left(W^{+\mu}D_{\mu}W^{-}_{\nu}+\textrm{h.c.}\right)$} & $-$ &  & \scriptsize{$\left(\bar{Q}_{L}\gamma^{\nu}Q_{L}-\bar{u}_{R}\gamma^{\nu}u_{R}\right)\left(\left[D^{\mu}H\right]^{\dagger}\left[D_{\mu\nu}H\right]+\textrm{h.c.}\right)$} & \\

15 & \scriptsize{$\epsilon_{\mu\nu\rho\sigma}\left(\bar{q}\gamma^{\nu}q\right) \left(W^{+\rho}\overset\leftrightarrow{D^{\mu}}W^{-\sigma}\right)$} & + &  & \scriptsize{$\epsilon_{\mu\nu\rho\sigma}\left(\bar{Q}_{L}\gamma^{\nu}Q_{L}+\bar{u}_{R}\gamma^{\nu}u_{R}\right)\left(\left[D^{\rho}H\right]^{\dagger}\overset\leftrightarrow{D^{\mu}}\left[D^{\sigma}H\right]+\textrm{h.c.}\right)$} & \\

16 & \scriptsize{$\epsilon_{\mu\nu\rho\sigma}\left(\bar{q}\gamma^{\nu}\gamma_{5}q\right) \left(W^{+\rho}\overset\leftrightarrow{D^{\mu}}W^{-\sigma}\right)$} & + &  &  \scriptsize{$\epsilon_{\mu\nu\rho\sigma}\left(\bar{Q}_{L}\gamma^{\nu}Q_{L}-\bar{u}_{R}\gamma^{\nu}u_{R}\right)\left(\left[D^{\rho}H\right]^{\dagger}\overset\leftrightarrow{D^{\mu}}\left[D^{\sigma}H\right]+\textrm{h.c.}\right)$} & \\
\hline

\end{tabular}
\begin{minipage}{5.7in}
\medskip
\caption{\label{tab:qqww1} \footnotesize Primary 5- and 6-dimension operators for $\bar{q}qW^{+}W^{-}$ interactions.  As outlined in the text, these operators can be modified to yield the operators for $\bar{q}q^{\prime}WZ$ interactions. Under the assumption that $\bar{q}$ and $q$ are each other's anti-particles, the operators are Hermitean and have the listed CP properties. If they are not, each of these operators has a Hermitean conjugate, which can be used to create a CP even and a CP odd operator.  To simplify the expressions, we use the shorthand $\bfd_{\mu\nu}=\bfd_{\mu}\bfd_{\nu}$, and similarly, $D_{\mu\nu}=D_{\mu}D_{\nu}$. To get the descendant operators, once can add contracted derivatives to get arbitrary Mandelstam factors of $s, t$.}
\end{minipage}
\end{center}
\end{adjustwidth}
\end{table}

\begin{table}[p]
\begin{adjustwidth}{-.75in}{-.75in} 
\begin{center}
\footnotesize
\centering
\renewcommand{\arraystretch}{0.9}
\setlength{\tabcolsep}{6pt}
\begin{tabular}{|c|c|c|c|c|c|}

\hline
\multirow{2}{*}{$i$} & \multirow{2}{*}{$\mathcal{O}_i^{\bar{q}qW^{+}W^{-}}$}   & \multirow{2}{*}{CP} & \multirow{2}{*}{$d_{\mathcal{O}_i}$}& SMEFT & $c$ Unitarity  \\
 & & & & Operator & Bound \\
 \hline

17 & \scriptsize{$\left(\bar{q}\bfd_{\mu\nu}q\right) \left(W^{+\mu}W^{-\nu}\right)$} & + & \multirow{13}{*}{7} & \scriptsize{$\left(\bar{Q}_{L}\bfd_{\mu\nu}\htilde u_{R}+\textrm{h.c.}\right)\left(\left[D^{\mu}H\right]^{\dagger}\left[D^{\nu}H\right]+\textrm{h.c.}\right)$} & \multirow{13}{*}{  $\frac{0.02}{E_\text{TeV}^5}, \frac{0.07}{E_\text{TeV}^6}$ }  \\

18 & \scriptsize{$\left(i\bar{q}\gamma_{5}\bfd_{\mu\nu}q\right) \left(W^{+\mu}W^{-\nu}\right)$} & $-$ &  & \scriptsize{$\left(i\bar{Q}_{L}\bfd_{\mu\nu}\htilde u_{R}+\textrm{h.c.}\right)\left(\left[D^{\mu}H\right]^{\dagger}\left[D^{\nu}H\right]+\textrm{h.c.}\right)$} & \\

19 & \scriptsize{$\left(i\bar{q}\bfd_{\mu}q\right) \left(W^{+\nu}D_{\nu}W^{-\mu}+\textrm{h.c.}\right)$} & $-$ &  & \scriptsize{$\left(i\bar{Q}_{L}\bfd_{\mu}\htilde u_{R}+\textrm{h.c.} \right)\left(\left[D_{\nu}H\right]^{\dagger}\left[D^{\nu\mu}H\right]+\textrm{h.c.}\right)$} & \\

20 & \scriptsize{$\left(\bar{q}\gamma_{5}\bfd_{\mu}q\right) \left(W^{+\nu}D_{\nu}W^{-\mu}+\textrm{h.c.}\right)$} & + &  & \scriptsize{$\left(\bar{Q}_{L}\bfd_{\mu}\htilde u_{R}+\textrm{h.c.} \right)\left(\left[D_{\nu}H\right]^{\dagger}\left[D^{\nu\mu}H\right]+\textrm{h.c.}\right)$} & \\

21 & \scriptsize{$\left(i\bar{q}\bfd_{\mu}q\right) \left(iW^{+\nu}D_{\nu}W^{-\mu}+\textrm{h.c.}\right)$} & + &  & \scriptsize{$\left(i\bar{Q}_{L}\bfd_{\mu}\htilde u_{R}+\textrm{h.c.} \right)\left(i\left[D_{\nu}H\right]^{\dagger}\left[D^{\nu\mu}H\right]+\textrm{h.c.}\right)$} & \\

22 & \scriptsize{$\left(\bar{q}\gamma_{5}\bfd_{\mu}q\right) \left(iW^{+\nu}D_{\nu}W^{-\mu}+\textrm{h.c.}\right)$} & $-$ &  &\scriptsize{$\left(\bar{Q}_{L}\bfd_{\mu}\htilde u_{R}+\textrm{h.c.} \right)\left(i\left[D_{\nu}H\right]^{\dagger}\left[D^{\nu\mu}H\right]+\textrm{h.c.}\right)$} & \\

23 & \scriptsize{$\epsilon_{\mu\nu\rho\sigma}\left(i\bar{q}\overset\leftrightarrow{D^{\mu}}q\right)\left(W^{+\rho}\overset\leftrightarrow{D^{\nu}}W^{-\sigma}\right)$} & $+$ & & \scriptsize{$\epsilon_{\mu\nu\rho\sigma}\left(i\bar{Q}_{L}\overset\leftrightarrow{D^{\mu}}\htilde u_{R}+\textrm{h.c.}\right)\left(\left[D^{\rho}H\right]^{\dagger}\overset\leftrightarrow{D^{\nu}}\left[D^{\sigma}H\right]+\textrm{h.c.}\right)$} & \\

24 & \scriptsize{$\epsilon_{\mu\nu\rho\sigma}\left(\bar{q}\gamma_{5}\overset\leftrightarrow{D^{\mu}}q\right)\left(W^{+\rho}\overset\leftrightarrow{D^{\nu}}W^{-\sigma}\right)$} & $-$ & & \scriptsize{$\epsilon_{\mu\nu\rho\sigma}\left(\bar{Q}_{L}\overset\leftrightarrow{D^{\mu}}\htilde u_{R}+\textrm{h.c.}\right)\left(\left[D^{\rho}H\right]^{\dagger}\overset\leftrightarrow{D^{\nu}}\left[D^{\sigma}H\right]+\textrm{h.c.}\right)$} & \\

25 & \scriptsize{$\left(i\bar{q}\sigma^{\mu\nu}\bfd_{\rho}q\right)\left(iW^{+}_{\nu}\bfd_{\mu}W^{-\rho}+\textrm{h.c.}\right)$} & $-$ &  & \scriptsize{$\left(i\bar{Q}_{L}\sigma^{\mu\nu}\bfd_{\rho}\htilde u_{R}+\textrm{h.c.}\right)\left(i\left[D_{\nu}H\right]^{\dagger}\bfd_{\mu}\left[D^{\rho}H\right]+\textrm{h.c.}\right)$} & \\

26 & \scriptsize{$\left(\bar{q}\sigma^{\mu\nu}q\right)\left(i[D_{\rho}W^{+}_{\nu}]\bfd_{\mu}W^{-\rho}+\textrm{h.c.}\right)$} & $+$ & & \scriptsize{$\left(\bar{Q}_{L}\sigma^{\mu\nu}\htilde u_{R}+\textrm{h.c.}\right)\left(i\left[D_{\rho\nu}H\right]^{\dagger}\bfd_{\mu}\left[D^{\rho}H\right]+\textrm{h.c.}\right)$} & \\

27 & \scriptsize{$\left(i\bar{q}\sigma^{\mu\nu}\gamma_{5}q\right)\left(i[D_{\rho}W^{+}_{\nu}]\bfd_{\mu}W^{-\rho}+\textrm{h.c.}\right)$} & $-$ & & \scriptsize{$\left(i\bar{Q}_{L}\sigma^{\mu\nu}\htilde u_{R}+\textrm{h.c.}\right)\left(i\left[D_{\rho\nu}H\right]^{\dagger}\bfd_{\mu}\left[D^{\rho}H\right]+\textrm{h.c.}\right)$} & \\

28 & \scriptsize{$\left(\bar{q}\sigma^{\mu\nu}\gamma_{5}\bfd_{\rho}q\right)\left(iW^{+}_{\nu}\overset\leftrightarrow{D_{\mu}}W^{-\rho}+\textrm{h.c.}\right)$} & $+$ & & \scriptsize{$\left(\bar{Q}_{L}\sigma^{\mu\nu}\bfd_{\rho}\htilde u_{R}+\textrm{h.c.}\right)\left(i\left[D_{\nu}H\right]^{\dagger}\overset\leftrightarrow{D_{\mu}}\left[D^{\rho}H\right]+\textrm{h.c.}\right)$} & \\
\hline

29 & \scriptsize{$\left(\bar{q}q\right)\left(W^{\mu\nu}W_{\mu\nu}\right)$} & + & \multirow{4}{*}{7} & \scriptsize{$\left(\bar{Q}_{L}\htilde u_{R}+\textrm{h.c.}\right)\left(W^{a\mu\nu}W^{a}_{\mu\nu}\right)$} & \multirow{4}{*}{  $\frac{0.4}{E_\text{TeV}^3}, \frac{1.2}{E_\text{TeV}^4}$ } \\

30 & \scriptsize{$\left(i\bar{q}\gamma_{5}q\right)\left(W^{\mu\nu}W_{\mu\nu}\right)$} & $-$ & & \scriptsize{$\left(i\bar{Q}_{L}\htilde u_{R}+\textrm{h.c.}\right)\left(W^{a\mu\nu}W^{a}_{\mu\nu}\right)$} & \\

31 & \scriptsize{$\left(\bar{q}q\right)\left(W^{+\mu\nu} \Wtilde{\vphantom{W}}^{-}_{\mu\nu}\right)$} & $-$ & & \scriptsize{$\left(\bar{Q}_{L}\htilde u_{R}+\textrm{h.c.}\right)\left(W^{a\, \mu\nu}\Wtilde{\vphantom{W}}^{a}_{\mu\nu}+\textrm{h.c.}\right)$} &  \\

32 & \scriptsize{$\left(i\bar{q}\gamma_{5}q\right)\left(W^{+\mu\nu}\Wtilde{\vphantom{W}}^{-}_{\!\mu\nu}\right)$} & + & & \scriptsize{$\left(i\bar{Q}_{L}\htilde u_{R}+\textrm{h.c.}\right)\left(W^{a\, \mu\nu}\Wtilde{\vphantom{W}}^{a}_{\!\mu\nu}+\textrm{h.c.}\right)$} & \\
\hline

33 & \scriptsize{$\left(\bar{q}\gamma^{\mu}\bfd_{\nu\rho}q\right)\left(i W^{+\nu}\overset\leftrightarrow{D_{\mu}}W^{-\rho}\right)$} & + & \multirow{6}{*}{8} & \scriptsize{$\left(\bar{Q}_{L}\gamma^{\mu}\bfd_{\nu\rho}Q_{L}+\bar{u}_{R}\gamma^{\mu}\bfd_{\nu\rho}u_{R}\right)\left(i\left[D^{\nu}H\right]^{\dagger}\bfd_{\mu}\left[D^{\rho}H\right]+\textrm{h.c.}\right)$} & \multirow{6}{*}{  $\frac{0.006}{E_\text{TeV}^6}$ }  \\

34 & \scriptsize{$\left(\bar{q}\gamma^{\mu}\gamma_{5}\bfd_{\nu\rho}q\right)\left(i W^{+\nu}\overset\leftrightarrow{D_{\mu}}W^{-\rho}\right)$} & + & & \scriptsize{$\left(\bar{Q}_{L}\gamma^{\mu}\bfd_{\nu\rho}Q_{L}-\bar{u}_{R}\gamma^{\mu}\bfd_{\nu\rho}u_{R}\right)\left(i\left[D^{\nu}H\right]^{\dagger}\bfd_{\mu}\left[D^{\rho}H\right]+\textrm{h.c.}\right)$} & \\

35 & \scriptsize{$\left(i\bar{q}\gamma^{\mu}\bfd_{\rho}q\right)\left(iW^{+\nu}\bfd_{\mu}D_{\nu}W^{-\rho}+\textrm{h.c.}\right)$} & $-$ & & \scriptsize{$\left(i\bar{Q}_{L}\gamma^{\mu}\bfd_{\rho}Q_{L}+i\bar{u}_{R}\gamma^{\mu}\bfd_{\rho}u_{R}\right)\left(i\left[D_{\nu}H\right]^{\dagger}\bfd_{\mu}\left[D^{\rho\nu}H\right]+\textrm{h.c.}\right)$} & \\

36 & \scriptsize{$\left(i\bar{q}\gamma^{\mu}\gamma_{5}\bfd_{\rho}q\right)\left(iW^{+\nu}\bfd_{\mu}D_{\nu}W^{-\rho}+\textrm{h.c.}\right)$} & $-$ & & \scriptsize{$\left(i\bar{Q}_{L}\gamma^{\mu}\bfd_{\rho}Q_{L}-i\bar{u}_{R}\gamma^{\mu}\bfd_{\rho}u_{R}\right)\left(i\left[D_{\nu}H\right]^{\dagger}\bfd_{\mu}\left[D^{\rho \nu}H\right]+\textrm{h.c.}\right)$} & \\

37 & \scriptsize{$\left(\bar{q}\gamma^{\mu}q\right)\left(i[D_{\rho}W^{+}_{\nu}]\bfd_{\mu}D^{\nu}W^{-\rho}\right)$} & + & & \scriptsize{$\left(\bar{Q}_{L}\gamma^{\mu}Q_{L}+\bar{u}_{R}\gamma^{\mu}u_{R}\right)\left(i\left[D_{\nu \rho}H\right]^{\dagger}\bfd_{\mu}\left[D^{\rho  \nu}H\right]+\textrm{h.c.}\right)$} & \\

38 & \scriptsize{$\left(\bar{q}\gamma^{\mu}\gamma_{5}q\right)\left(i[D_{\rho}W^{+}_{\nu}]\bfd_{\mu}D^{\nu}W^{-\rho}\right)$} & + & & \scriptsize{$\left(\bar{Q}_{L}\gamma^{\mu}Q_{L}-\bar{u}_{R}\gamma^{\mu}u_{R}\right)\left(\left[D_{\nu  \rho}H\right]^{\dagger}\bfd_{\mu}\left[D^{\rho \nu}H\right]+\textrm{h.c.}\right)$} & \\
\hline

\end{tabular}
\begin{minipage}{7in}
\medskip
\caption{\label{tab:qqww2} \footnotesize Primary 7- and 8-dimension operators for $\bar{q}qW^{+}W^{-}$ interactions, where $\tilde{W}_{\mu\nu} = \frac{1}{2}\epsilon_{\mu\nu\rho\sigma}W^{\rho\sigma}$.  As outlined in the text, these operators can be modified to yield the operators for $\bar{q}q^{\prime}WZ$ interactions. Under the assumption that $\bar{q}$ and $q$ are each other's anti-particles, the operators are Hermitean and have the listed CP properties. If they are not, each of these operators has a Hermitean conjugate, which can be used to create a CP even and a CP odd operator. To simplify the expressions, we use the shorthand $\bfd_{\mu\nu}=\bfd_{\mu}\bfd_{\nu}$, and similarly, $D_{\mu\nu}=D_{\mu}D_{\nu}$. To get the descendant operators, once can add contracted derivatives to get arbitrary Mandelstam factors of $s, t$. At dimension 9, $s\mathcal{O}_{26}$ and $s\mathcal{O}_{27}$ become redundant to other operators and thus one only needs their descendants $t^{n}\mathcal{O}_{26}$ and $t^{n}\mathcal{O}_{27}$ for an independent set of operators.}
\end{minipage}
\end{center}
\end{adjustwidth}
\end{table}

In Tables~\ref{tab:qqzz1} and~\ref{tab:qqzz2}, we list the primary operators for $\bar{q}qZZ$ interactions. Reading off from the Hilbert series, we expect to see 2 operators at dimension 5, 6 operators at dimension 6, 12 operators at dimension 7, 6 operators at dimensions 8, 9, and 10, and at least 2 constraints at dimension 11. We do indeed find that there are 38 primary operators, as well as two redundancies at dimension 11, for $s\mathcal{O}_{31}$ and $s\mathcal{O}_{32}$.  To generate an independent set of operators, one needs to add descendants of the primaries, which involve multiplying by arbitrary powers of $s$ and $(t-u)^2$ (note that $(t-u)^2$ respects the exchange symmetry of the $Z$'s).   
However because of the redundancies at dimension 11, for $\mathcal{O}_{31}$ and $\mathcal{O}_{32}$, one only needs their descendants $(t-u)^{2n} \mathcal{O}_{31}$ and $(t-u)^{2n}\mathcal{O}_{32}$.

\begin{table}[p]
\begin{adjustwidth}{-.5in}{-.5in}  
\begin{center}
\footnotesize
\centering
\renewcommand{\arraystretch}{0.9}
\setlength{\tabcolsep}{6pt}
\begin{tabular}{|c|c|c|c|c|c|}

\hline
\multirow{2}{*}{$i$} & \multirow{2}{*}{$\mathcal{O}_i^{\bar{q}qZZ}$}   & \multirow{2}{*}{CP} & \multirow{2}{*}{$d_{\mathcal{O}_i}$}& SMEFT & $c$ Unitarity  \\
 & & & & Operator & Bound \\

\hline 
1 & \scriptsize{$\left(\bar{q}q\right) \left(Z^{\mu}Z_{\mu}\right)$} & $+$ & \multirow{2}{*}{5} & \scriptsize{$\left(\bar{Q}_{L}\htilde u_{R}+\textrm{h.c.}\right)|D_{\mu}H|^{2}$} &  \multirow{2}{*}{  $\frac{0.4}{E_\text{TeV}^3}, \frac{1.2}{E_\text{TeV}^4}$ }  \\

2 & \scriptsize{$\left(\bar{q}i\gamma_{5}q\right) \left(Z^{\mu}Z_{\mu}\right)$} & $-$ & & \scriptsize{$\left(i\bar{Q}_{L}\htilde u_{R}+\textrm{h.c.}\right)|D_{\mu}H|^{2}$} &  \\
\hline

3 & \scriptsize{$\left(i\bar{q}\gamma^{\nu}\bfd_{\mu}q\right) \left(Z^{\mu}Z_{\nu}\right)$} & + & \multirow{4}{*}{6} & \scriptsize{$\left(i\bar{Q}_{L}\gamma^{\nu}\bfd_{\mu}Q_{L}+i\bar{u}_{R}\gamma^{\nu}\bfd_{\mu}u_{R}\right)\left(\left[D^{\mu}H\right]^{\dagger}\left[D_{\nu}H\right]+\textrm{h.c.}\right)$} & \multirow{4}{*}{  $\frac{0.09}{E_\text{TeV}^4}$ }  \\

4 & \scriptsize{$\left(i\bar{q}\gamma^{\nu}\gamma_{5}\bfd_{\mu}q\right) \left(Z^{\mu}Z_{\nu}\right)$} & + &  & \scriptsize{$\left(i\bar{Q}_{L}\gamma^{\nu}\bfd_{\mu}Q_{L}-i\bar{u}_{R}\gamma^{\nu}\bfd_{\mu}u_{R}\right)\left(\left[D^{\mu}H\right]^{\dagger}\left[D_{\nu}H\right]+\textrm{h.c.}\right)$} & \\

5 & \scriptsize{$\left(\bar{q}\gamma^{\nu}q\right) \left(Z^{\mu}\partial_{\mu}Z_{\nu}\right)$} & $-$ &  & \scriptsize{$\left(\bar{Q}_{L}\gamma^{\nu}Q_{L}+\bar{u}_{R}\gamma^{\nu}u_{R}\right)\left(\left[D^{\mu}H\right]^{\dagger}\left[D_{\mu\nu}H\right]+\textrm{h.c.}\right)$} & \\

6 & \scriptsize{$\left(\bar{q}\gamma^{\nu}\gamma_{5}q\right) \left(Z^{\mu}\partial_{\mu}Z_{\nu}\right)$} & $-$ &  & \scriptsize{$\left(\bar{Q}_{L}\gamma^{\nu}Q_{L}-\bar{u}_{R}\gamma^{\nu}u_{R}\right)\left(\left[D^{\mu}H\right]^{\dagger}\left[D_{\mu\nu}H\right]+\textrm{h.c.}\right)$} & \\
\hline

7 & \scriptsize{$\left(\bar{q}\gamma^{\nu}q\right) \left(\tilde{Z}_{\nu\sigma}Z^{\sigma}\right)$} & + & \multirow{2}{*}{6}& \scriptsize{$\left(\bar{Q}_{L}\gamma^{\nu}Q_{L}+\bar{u}_{R}\gamma^{\nu}u_{R}\right)\left(\tilde{B}_{\nu\sigma}H^{\dagger}D^{\sigma}H+\textrm{h.c.}\right)$} &\multirow{2}{*}{  $\frac{0.4}{E_\text{TeV}^3}, \frac{1.2}{E_\text{TeV}^4}$ } \\

8 & \scriptsize{$\left(\bar{q}\gamma^{\nu}\gamma_{5}q\right) \left(\tilde{Z}_{\nu\sigma}Z^{\sigma}\right)$} & + &  & \scriptsize{$\left(\bar{Q}_{L}\gamma^{\nu}Q_{L}-\bar{u}_{R}\gamma^{\nu}u_{R}\right)\left(\tilde{B}_{\nu\sigma}H^{\dagger}D^{\sigma}H+\textrm{h.c.}\right)$} & \\
\hline

9 & \scriptsize{$\left(\bar{q}q\right)\left(Z_{\mu\nu}Z^{\mu\nu}\right)$} & + &\multirow{4}{*}{7}& \scriptsize{$\left(\bar{Q}_{L}\htilde u_{R}+\textrm{h.c.}\right)\left(B^{\mu\nu}B_{\mu\nu}\right)$} &  \multirow{4}{*}{  $\frac{0.4}{E_\text{TeV}^3}, \frac{1.2}{E_\text{TeV}^4}$ }\\

10 & \scriptsize{$\left(i\bar{q}\gamma_{5}q\right)\left(Z^{\mu\nu}Z_{\mu\nu}\right)$} & $-$ & & \scriptsize{$\left(i\bar{Q}_{L}\htilde u_{R}+\textrm{h.c.}\right)\left(B^{\mu\nu}B_{\mu\nu}\right)$} & \\

11 & \scriptsize{$\left(\bar{q}q\right)\left(Z^{\mu\nu}\tilde{Z}_{\mu\nu}\right)$} & $-$ & & \scriptsize{$\left(\bar{Q}_{L}\htilde u_{R}+\textrm{h.c.}\right)\left(B^{\mu\nu}\tilde{B}_{\mu\nu}\right)$} & \\

12 & \scriptsize{$\left(i\bar{q}\gamma_{5}q\right)\left(Z^{\mu\nu}\tilde{Z}_{\mu\nu}\right)$} & + & & \scriptsize{$\left(i\bar{Q}_{L}\htilde u_{R}+\textrm{h.c.}\right)\left(B^{\mu\nu}\tilde{B}_{\mu\nu}\right)$} & \\
\hline

13 & \scriptsize{$\left(i\bar{q}\sigma_{\mu\nu}\bfd_{\rho}q\right)\left(Z^{\mu}\partial^{\rho}Z^{\nu}\right)$}& + & \multirow{6}{*}{7} & \scriptsize{$\left(i\bar{Q}_{L}\sigma_{\mu\nu}\bfd_{\rho}\htilde u_{R}+\textrm{h.c.}\right)\left(\left[D^{\mu}H\right]^{\dagger}\left[D^{\rho\nu}H\right]+\textrm{h.c.}\right)$} & \multirow{6}{*}{  $\frac{0.02}{E_\text{TeV}^5}, \frac{0.07}{E_\text{TeV}^6}$ }  \\

14 & \scriptsize{$\left(\bar{q}\sigma_{\mu\nu}\gamma_{5}\bfd_{\rho}q\right)\left(Z^{\mu}\partial^{\rho}Z^{\nu}\right)$} & $-$ & & \scriptsize{$\left(\bar{Q}_{L}\sigma_{\mu\nu}\bfd_{\rho}\htilde u_{R}+\textrm{h.c.}\right)\left(\left[D^{\mu}H\right]^{\dagger}\left[D^{\rho\nu}H\right]+\textrm{h.c.}\right)$} & \\

15  & \scriptsize{$\left(\bar{q}\bfd_{\mu\nu}q\right) \left(Z^{\mu}Z^{\nu}\right)$} & + & & \scriptsize{$\left(\bar{Q}_{L}\bfd_{\mu\nu}\htilde u_{R}+\textrm{h.c.}\right)\left(\left[D^{\mu}H\right]^{\dagger}\left[D^{\nu}H\right]+\textrm{h.c.}\right)$ }& \\

16 & \scriptsize{$\left(i\bar{q}\gamma_{5}\bfd_{\mu\nu}q\right) \left(Z^{\mu}Z^{\nu}\right)$} & $-$ & & \scriptsize{$\left(i\bar{Q}_{L}\bfd_{\mu\nu}\htilde u_{R}+\textrm{h.c.}\right)\left(\left[D^{\mu}H\right]^{\dagger}\left[D^{\nu}H\right]+\textrm{h.c.}\right)$} & \\

17 & \scriptsize{$\left(i\bar{q}\bfd_{\nu}q\right)\left(Z^{\mu}\partial_{\mu}Z^{\nu}\right)$} & $-$ & & \scriptsize{$\left(i\bar{Q}_{L}\bfd_{\nu}\htilde u_{R}+\textrm{h.c.}\right)\left(\left[D^{\mu}H\right]^{\dagger}\left[D_{\mu}^{\ \nu}H\right]+\textrm{h.c.}\right)$} & \\

18 & \scriptsize{$\left(\bar{q}\gamma_{5}\bfd_{\nu}q\right)\left(Z^{\mu}\partial_{\mu}Z^{\nu}\right)$} & + & & \scriptsize{$\left(\bar{Q}_{L}\bfd_{\nu}\htilde u_{R}+\textrm{h.c.}\right)\left(\left[D^{\mu}H\right]^{\dagger}\left[D_{\mu}^{\ \nu}H\right]+\textrm{h.c.}\right)$} & \\
\hline

19 & \scriptsize{$\left(i\bar{q}\overset\leftrightarrow{D^{\mu}}q\right)\left(\tilde{Z}_{\mu\sigma}Z^{\sigma}\right)$} & + &\multirow{2}{*}{7} & \scriptsize{$\left(i\bar{Q}_{L}\overset\leftrightarrow{D^{\mu}}\htilde u_{R}+\textrm{h.c.}\right)\left(\tilde{B}_{\mu\sigma}H^{\dagger}D^{\sigma}H+\textrm{h.c.}\right)$} &\multirow{2}{*}{  $\frac{0.09}{E_\text{TeV}^4}, \frac{0.9}{E_\text{TeV}^6}$ } \\

20 & \scriptsize{$\left(\bar{q}\gamma_{5}\overset\leftrightarrow{D^{\mu}}q\right)\left(\tilde{Z}_{\mu\sigma}Z^{\sigma}\right)$} & $-$ & & \scriptsize{$\left(\bar{Q}_{L}\overset\leftrightarrow{D^{\mu}}\htilde u_{R}+\textrm{h.c.}\right)\left(\tilde{B}_{\mu\sigma}H^{\dagger}D^{\sigma}H+\textrm{h.c.}\right)$} & \\
\hline

\end{tabular}
\begin{minipage}{5.7in}
\medskip
\caption{\label{tab:qqzz1} \footnotesize Primary 5-, 6-, and 7-dimension operators for $\bar{q}qZZ$ interactions. Under the assumption that $\bar{q}$ and $q$ are each other's anti-particles, the operators are Hermitean and have the listed CP properties. If they are not, each of these operators has a Hermitean conjugate, which can be used to create a CP even and a CP odd operator.  To simplify the expressions, we use the shorthand $\bfd_{\mu\nu}=\bfd_{\mu}\bfd_{\nu}$, and similarly, $\partial_{\mu\nu}=\partial_{\mu}\partial_{\nu}$. To get the descendant operators, once can add contracted derivatives to get arbitrary Mandelstam factors of $s, (t-u)^2$.}
\end{minipage}
\end{center}
\end{adjustwidth}
\end{table}

\begin{table}[p]
\begin{adjustwidth}{-.5in}{-.5in}  
\begin{center}
\footnotesize
\renewcommand{\arraystretch}{0.9}
\setlength{\tabcolsep}{6pt}
\begin{tabular}{|c|c|c|c|c|c|}

\hline
\multirow{2}{*}{$i$} & \multirow{2}{*}{$\mathcal{O}_i^{\bar{q}qZZ}$}   & \multirow{2}{*}{CP} & \multirow{2}{*}{$d_{\mathcal{O}_i}$}& SMEFT & $c$ Unitarity  \\
 & & & & Operator & Bound \\

\hline 
21 & \scriptsize{$\left(i\bar{q}\gamma^{\nu}\bfd_{\rho}q\right) \left([\partial_{\nu}Z^{\mu}]\partial^{\rho}Z_{\mu}\right)$} & + & \multirow{6}{*}{8} & \scriptsize{$\left(i\bar{Q}_{L}\gamma^{\nu}\bfd_{\rho}Q_{L}+i\bar{u}_{R}\gamma^{\nu}\bfd_{\rho}u_{R}\right)\left(\left[D^{\ \mu}_{\nu}H\right]^{\dagger}\left[D^{\rho}_{\ \mu}H\right]+\textrm{h.c.}\right)$} &  \multirow{6}{*}{  $\frac{0.006}{E_\text{TeV}^6}$ }  \\

22 & \scriptsize{$\left(i\bar{q}\gamma^{\nu}\gamma_{5}\bfd_{\rho}q\right) \left([\partial_{\nu}Z^{\mu}]\partial^{\rho}Z^{\mu}\right)$} & $+$ &  & \scriptsize{$\left(i\bar{Q}_{L}\gamma^{\nu}\bfd_{\rho}Q_{L}-i\bar{u}_{R}\gamma^{\nu}\bfd_{\rho}u_{R}\right)\left(\left[D^{\ \mu}_{\nu}H\right]^{\dagger} \left[D^{\rho}_{\ \mu}H\right]+\textrm{h.c.}\right)$} &  \\

23 & \scriptsize{$\left(\bar{q}\gamma^{\nu}\bfd_{\mu\rho}q\right)\left(Z^{\mu}\partial^{\rho}Z_{\nu}\right)$} & $-$ &  & \scriptsize{$\left(\bar{Q}_{L}\gamma^{\nu}\bfd_{\mu\rho}Q_{L}+\bar{u}_{R}\gamma^{\nu}\bfd_{\mu\rho}u_{R}\right)\left(\left[D^{\mu}H\right]^{\dagger} \left[D^{\rho}_{\ \nu}H\right]+\textrm{h.c.}\right)$} &  \\

24 & \scriptsize{$\left(\bar{q}\gamma^{\nu}\gamma_{5}\bfd_{\mu\rho}q\right)\left(Z^{\mu}\partial^{\rho}Z_{\nu}\right)$} & $-$ &  & \scriptsize{$\left(\bar{Q}_{L}\gamma^{\nu}\bfd_{\mu\rho}Q_{L}-\bar{u}_{R}\gamma^{\nu}\bfd_{\mu\rho}u_{R}\right)\left(\left[D^{\mu}H\right]^{\dagger} \left[D^{\rho}_{\ \nu}H\right]+\textrm{h.c.}\right)$} &  \\

25 & \scriptsize{$\left(i\bar{q}\gamma^{\nu}\bfd_{\rho}q\right)\left(Z_{\mu}\partial^{\rho\mu}Z_{\nu}\right)$} & $+$ &  & \scriptsize{$\left(i\bar{Q}_{L}\gamma^{\nu}\bfd_{\rho}Q_{L}+i\bar{u}_{R}\gamma^{\nu}\bfd_{\rho}u_{R}\right)\left(\left[D_{\mu}H\right]^{\dagger}\left[D^{\rho\mu}_{\ \ \nu}H\right]+\textrm{h.c.}\right)$} &  \\

26 & \scriptsize{$\left(i\bar{q}\gamma^{\nu}\gamma_{5}\bfd_{\rho}q\right)\left(Z_{\mu}\partial^{\rho\mu}Z_{\nu}\right)$} & $+$ &  & \scriptsize{$\left(i\bar{Q}_{L}\gamma^{\nu}\bfd_{\rho}Q_{L}-i\bar{u}_{R}\gamma^{\nu}\bfd_{\rho}u_{R}\right)\left(\left[D_{\mu}H\right]^{\dagger}\left[D^{\rho\mu}_{\ \ \nu}H\right]+\textrm{h.c.}\right)$} &  \\
\hline

27 & \scriptsize{$\left(\bar{q}\bfd_{\nu\alpha}q\right)\left(Z_{\mu}\partial^{\alpha\mu}Z^{\nu}\right)$} & + & \multirow{3}{*}{9} & \scriptsize{$\left(\bar{Q}_{L}\bfd_{\nu\alpha}\htilde u_{R}+\textrm{h.c.}\right)\left(\left[D_{\mu}H\right]^{\dagger}\left[D^{\alpha\mu\nu}H\right]+\textrm{h.c.}\right)$} & \multirow{3}{*}{  $\frac{0.001}{E_\text{TeV}^7}, \frac{0.004}{E_\text{TeV}^8}$ }  \\

28 & \scriptsize{$\left(i\bar{q}\gamma_{5}\bfd_{\nu\alpha}q\right)\left(Z_{\mu}\partial^{\alpha\mu}Z^{\nu}\right)$} & $-$ &   & \scriptsize{$\left(i\bar{Q}_{L}\bfd_{\nu\alpha}\htilde u_{R}+\textrm{h.c.}\right)\left(\left[D_{\mu}H\right]^{\dagger} \left[D^{\alpha\mu\nu}H\right]+\textrm{h.c.}\right)$} & \\

29 & \scriptsize{$\left(i\bar{q}\sigma_{\mu\nu}\gamma_{5}\bfd_{\rho\sigma}q\right)\left([\partial^{\mu}Z^{\rho}]\partial^{\sigma}Z^{\nu}\right)$} & + &  & \scriptsize{$\left(i\bar{Q}_{L}\sigma_{\mu\nu}\bfd_{\rho\sigma}\htilde u_{R}+\textrm{h.c.}\right)\left(\left[D^{\mu\rho}H\right]^{\dagger}\left[D^{\sigma\nu}H\right]+\textrm{h.c.}\right)$} & \\
\hline

30 & \scriptsize{$\left(\bar{q}\sigma^{\mu\nu}\bfd_{\rho\sigma}q\right)\left(Z_{\mu\nu}\partial^{\sigma}Z^{\rho}\right)$} & $-$ &  \multirow{3}{*}{9} & \scriptsize{$\left(\bar{Q}_{L}\sigma^{\mu\nu}\bfd_{\rho\sigma}\htilde u_{R}+\textrm{h.c.}\right)\left(B_{\mu\nu}H^{\dagger}D^{\sigma\rho}H+\textrm{h.c.}\right)$} & \multirow{3}{*}{  $\frac{0.006}{E_\text{TeV}^6}, \frac{0.05}{E_\text{TeV}^8}$ } \\

31 & \scriptsize{$\left(i\bar{q}\sigma^{\mu\nu}\bfd_{\sigma}q\right)\left(\left[\partial_{\rho}Z_{\mu\nu}\right]\partial^{\sigma}Z^{\rho}\right)$} & + &  & \scriptsize{$\left(i\bar{Q}_{L}\sigma^{\mu\nu}\bfd_{\sigma}\htilde u_{R}+\textrm{h.c.}\right)\left(\left[\partial_{\rho}B_{\mu\nu}\right]H^{\dagger}D^{\sigma\rho}H+\textrm{h.c.}\right)$} & \\

32 & \scriptsize{$\left(i\bar{q}\sigma^{\mu\nu}\bfd_{\rho}q\right)\left(\left[\partial_{\mu}\tilde{Z}_{\nu\sigma}\right]\partial^{\rho}Z^{\sigma}\right)$} & $-$ &   & \scriptsize{$\left(i\bar{Q}_{L}\sigma^{\mu\nu}\bfd_{\rho}\htilde u_{R}+\textrm{h.c.}\right)\left(\left[\partial_{\mu}\tilde{B}_{\nu\sigma}\right]H^{\dagger}D^{\rho\sigma}H+\textrm{h.c.}\right)$} & \\
\hline

33 & \scriptsize{$\left(\bar{q}\gamma^{\mu}\bfd_{\nu\sigma}q\right)\left([\partial_{\mu\rho}Z^{\nu}]\partial^{\sigma}Z^{\rho}\right)$} & $-$ &  \multirow{3}{*}{10} & \scriptsize{$\left(\bar{Q}_{L}\gamma^{\mu}\bfd_{\nu\sigma}Q_{L}+\bar{u}_{R}\gamma^{\mu}\bfd_{\nu\sigma}u_{R}\right)\left(\left[D_{\ \mu\rho}^{\nu}H\right]^{\dagger}\left[D^{\sigma\rho}H\right]+\textrm{h.c.}\right)$} & \multirow{3}{*}{  $\frac{3\times10^{-4}}{E_\text{TeV}^8}$ }  \\

34 & \scriptsize{$\left(i\bar{q}\gamma^{\mu}\bfd_{\sigma}q\right)\left([\partial_{\mu\rho}Z_{\nu}]\partial^{\sigma\nu}Z^{\rho}\right)$} & + & & \scriptsize{$\left(i\bar{Q}_{L}\gamma^{\mu}\bfd_{\sigma}Q_{L}+i\bar{u}_{R}\gamma^{\mu}\bfd_{\sigma}u_{R}\right)\left(\left[D_{\mu\rho\nu}H\right]^{\dagger} \left[D^{\sigma\nu\rho}H\right]+\textrm{h.c.}\right)$} & \\

35 & \scriptsize{$\left(i\bar{q}\gamma^{\mu}\gamma_{5}\bfd_{\sigma}q\right)\left([\partial_{\mu\rho}Z_{\nu}]\partial^{\sigma\nu}Z^{\rho}\right)$} & + & & \scriptsize{$\left(i\bar{Q}_{L}\gamma^{\mu}\bfd_{\sigma}Q_{L}-i\bar{u}_{R}\gamma^{\mu}\bfd_{\sigma}u_{R}\right)\left(\left[D_{\mu\rho\nu}H\right]^{\dagger} \left[D^{\sigma\nu\rho}H\right]+\textrm{h.c.}\right)$} & \\
\hline

36 & \scriptsize{$\left(\bar{q}\gamma^{\alpha}\overset\leftrightarrow{D^{\mu}}_{\beta}q\right)\left(\tilde{Z}_{\mu\rho}\partial^{\rho\beta}Z_{\alpha}\right)$} & + &\multirow{3}{*}{10} & \scriptsize{$\left(\bar{Q}_{L}\gamma^{\alpha}\overset\leftrightarrow{D^{\mu}}_{\beta}Q_{L}+\bar{u}_{R}\gamma^{\alpha}\overset\leftrightarrow{D^{\mu}}_{\beta}u_{R}\right)\left(\tilde{B}_{\mu\rho}H^{\dagger}D^{\rho\beta}_{\ \ \alpha}H+\textrm{h.c.}\right)$} & \multirow{3}{*}{  $\frac{0.001}{E_\text{TeV}^7}, \frac{0.004}{E_\text{TeV}^8}$ } \\

37 & \scriptsize{$\left(\bar{q}\gamma^{\alpha}\gamma_{5}\overset\leftrightarrow{D^{\mu}}_{\beta}q\right)\left(\tilde{Z}_{\mu\rho}\partial^{\rho\beta}Z_{\alpha}\right)$} & + & & \scriptsize{$\left(\bar{Q}_{L}\gamma^{\alpha}\overset\leftrightarrow{D^{\mu}}_{\beta}Q_{L}-\bar{u}_{R}\gamma^{\alpha}\overset\leftrightarrow{D^{\mu}}_{\beta}u_{R}\right)\left(\tilde{B}_{\mu\rho}H^{\dagger}D^{\rho\beta}_{\ \ \alpha}H+\textrm{h.c.}\right)$} & \\

38 & \scriptsize{$\left(i\bar{q}\gamma^{\rho}\gamma_{5}\overset\leftrightarrow{D^{\mu}}_{\alpha\beta}q\right)\left(\tilde{Z}_{\mu\rho}\partial^{\beta}Z^{\alpha}\right)$} & $-$ &  & \scriptsize{$\left(i\bar{Q}_{L}\gamma^{\rho}\overset\leftrightarrow{D^{\mu}}_{\alpha\beta}Q_{L}-i\bar{u}_{R}\gamma^{\rho}\overset\leftrightarrow{D^{\mu}}_{\beta\alpha}u_{R}\right)\left(\tilde{B}_{\mu\rho}H^{\dagger}D^{\alpha\beta}H+\textrm{h.c.}\right)$} & \\
\hline

\end{tabular}
\begin{minipage}{5.7in}
\medskip
\caption{\label{tab:qqzz2} \footnotesize Primary 8-, 9-, and 10-dimension operators for $\bar{q}qZZ$ interactions. Under the assumption that $\bar{q}$ and $q$ are each other's anti-particles, the operators are Hermitean and have the listed CP properties. If they are not, each of these operators has a Hermitean conjugate, which can be used to create a CP even and a CP odd operator.  To simplify the expressions, we use the shorthand $\bfd_{\mu\nu}=\bfd_{\mu}\bfd_{\nu}$, and similarly, $\partial_{\mu\nu}=\partial_{\mu}\partial_{\nu}$. To get the descendant operators, once can add contracted derivatives to get arbitrary Mandelstam factors of $s, (t-u)^2$. At dimension 11, $s\mathcal{O}_{31}$ and $s\mathcal{O}_{32}$ become redundant to other operators. Thus, for these two, we need only their $(t-u)^{2n}\mathcal{O}_{31}$ and $(t-u)^{2n}\mathcal{O}_{32}$ descendants.}
\end{minipage}
\end{center}
\end{adjustwidth}
\end{table}

We have listed all of the primary operators for $\bar{q}qZ\gamma$ interactions in Table~\ref{tab:qqza1}. The Hilbert series tells us to expect 4 operators at dimension 6, 12 new operators at dimension 7, 8 operators at dimension 8, and 2 new operators and 2 new redundancies at dimension 9. We note that a na\"ive interpretation of the Hilbert series would have missed the 2 new primary operators that appear at dimension 9. We find that there are 26 primary operators, in agreement with the Hilbert series, as well as two constraints at dimension 9---$s\mathcal{O}_{7}$ and $s\mathcal{O}_{8}$.  Thus for those two operators, one only needs their descendant operators $t^n \mathcal{O}_{7}$ and $t^n \mathcal{O}_{8}$. These operators can also be adapted to account for $\bar{q}q^{\prime}W\gamma$, $\bar{q}qZg$, and $\bar{q}q^{\prime}Wg$ where we use a prime to denote a different quark flavor. To get $\bar{q}qZg$ operators, one replaces $F^{\mu\nu}\to G^{\mu\nu}$, to get $\bar{q}q^{\prime}W\gamma$ operators, one should make the replacement $q\to q^{\prime}$ and $Z\to W$, and to get $\bar{q}q^{\prime}Wg$ operators one needs to make the replacements $q\to q^{\prime}$, $F^{\mu\nu}\to G^{\mu\nu}$, and $Z\to W$. 

\begin{table}[p]
\begin{adjustwidth}{-.5in}{-.5in}  
\begin{center}
\footnotesize
\renewcommand{\arraystretch}{0.9}
\setlength{\tabcolsep}{6pt}
\begin{tabular}{|c|c|c|c|c|c|}

\hline
\multirow{2}{*}{$i$} & \multirow{2}{*}{$\mathcal{O}_i^{\bar{q}qZ\gamma}$}   & \multirow{2}{*}{CP} & \multirow{2}{*}{$d_{\mathcal{O}_i}$}& SMEFT & $c$ Unitarity  \\
 & & & & Operator & Bound \\
 \hline
 
 1 & \scriptsize{$\left(\bar{q}\gamma^{\nu}q\right)\left(F_{\nu\mu}Z^{\mu}\right)$} & $-$ & \multirow{4}{*}{6} & \scriptsize{$\left(\bar{Q}_{L}\gamma^{\nu}Q_{L}+\bar{u}_{R}\gamma^{\nu}u_{R}\right)\left(B_{\nu\mu}H^{\dagger}D^{\mu}H+\textrm{h.c.}\right)$} & \multirow{4}{*}{  $\frac{0.4}{E_\text{TeV}^3}, \frac{1.2}{E_\text{TeV}^4}$ } \\

2 & \scriptsize{$\left(\bar{q}\gamma^{\nu}\gamma_{5}q\right)\left(F_{\nu\mu}Z^{\mu}\right)$} & $-$ & & \scriptsize{$\left(\bar{Q}_{L}\gamma^{\nu}Q_{L}-\bar{u}_{R}\gamma^{\nu}u_{R}\right)\left(B_{\nu\mu}H^{\dagger}D^{\mu}H+\textrm{h.c.}\right)$} & \\

3 & \scriptsize{$\left(\bar{q}\gamma^{\nu}q\right)\left(\tilde{F}_{\nu\sigma}Z^{\sigma}\right)$} & $+$ & & \scriptsize{$\left(\bar{Q}_{L}\gamma^{\nu}Q_{L}+\bar{u}_{R}\gamma^{\nu}u_{R}\right)\left(\tilde{B}_{\nu\sigma}H^{\dagger}D^{\sigma}H+\textrm{h.c.}\right)$} &  \\

4 & \scriptsize{$\left(\bar{q}\gamma^{\nu}\gamma_{5}q\right)\left(\tilde{F}_{\nu\sigma}Z^{\sigma}\right)$} & $+$ & & \scriptsize{$\left(\bar{Q}_{L}\gamma^{\nu}Q_{L}-\bar{u}_{R}\gamma^{\nu}u_{R}\right)\left(\tilde{B}_{\nu\sigma}H^{\dagger}D^{\sigma}H+\textrm{h.c.}\right)$} & \\
\hline 

5 & \scriptsize{$\left(\bar{q}q\right)\left(F_{\mu\nu}Z^{\mu\nu}\right)$} & $+$ &\multirow{4}{*}{7} & \scriptsize{$\left(\bar{Q}_{L}\htilde u_{R}+\textrm{h.c.}\right)\left(B_{\mu\nu}B^{\mu\nu}\right)$} &\multirow{4}{*}{  $\frac{0.4}{E_\text{TeV}^3}, \frac{1.2}{E_\text{TeV}^4}$ } \\

6 & \scriptsize{$\left(i\bar{q}\gamma_{5}q\right)\left(F_{\mu\nu}Z^{\mu\nu}\right)$} & $-$ & & \scriptsize{$\left(i\bar{Q}_{L}\htilde u_{R}+\textrm{h.c.}\right)\left(B_{\mu\nu}B^{\mu\nu}\right)$} & \\

7 & \scriptsize{$\left(\bar{q}q\right)\left(\tilde{F}_{\mu\nu}Z^{\mu\nu}\right)$} & $-$ & & \scriptsize{$\left(\bar{Q}_{L}\htilde u_{R}+\textrm{h.c.}\right)\left(B^{\mu\nu}\tilde{B}_{\mu\nu}\right)$} & \\

8 & \scriptsize{$\left(i\bar{q}\gamma_{5}q\right)\left(\tilde{F}_{\mu\nu}Z^{\mu\nu}\right)$} & $+$ & & \scriptsize{$\left(i\bar{Q}_{L}\htilde u_{R}+\textrm{h.c.}\right)\left(B^{\mu\nu}\tilde{B}_{\mu\nu}\right)$} & \\
\hline

9 & \scriptsize{$\left(i\bar{q}\bfd_{\nu}q\right)\left(F^{\nu\mu}Z_{\mu}\right)$} & $-$ & \multirow{8}{*}{7} & \scriptsize{$\left(i\bar{Q}_{L}\bfd_{\nu}\htilde u_{R}+\textrm{h.c.}\right)\left(B^{\nu\mu}H^{\dagger}D_{\mu}H+\textrm{h.c.}\right)$} & \multirow{8}{*}{  $\frac{0.09}{E_\text{TeV}^4}, \frac{0.9}{E_\text{TeV}^6}$ } \\

10 & \scriptsize{$\left(\bar{q}\bfd_{\nu}\gamma_{5}q\right)\left(F^{\nu\mu}Z_{\mu}\right)$} & $+$ & & \scriptsize{$\left(\bar{Q}_{L}\bfd_{\nu}\htilde u_{R}+\textrm{h.c.}\right)\left(B^{\nu\mu}H^{\dagger}D_{\mu}H+\textrm{h.c.}\right)$} & \\

11 & \scriptsize{$\left(i\bar{q}\sigma_{\mu\nu}\bfd_{\rho}q\right)\left(F^{\mu\rho}Z^{\nu}\right)$} & $+$ & & \scriptsize{$\left(i\bar{Q}_{L}\sigma_{\mu\nu}\bfd_{\rho}\htilde u_{R}+\textrm{h.c.}\right)\left(B^{\rho\mu}H^{\dagger}D^{\nu}H+\textrm{h.c.}\right)$} & \\

12 & \scriptsize{$\left(\bar{q}\sigma_{\mu\nu}q\right)\left(F^{\mu\rho}\partial_{\rho}Z^{\nu}\right)$} & $-$ & & \scriptsize{$\left(\bar{Q}_{L}\sigma_{\mu\nu}\htilde u_{R}+\textrm{h.c.}\right)\left(B^{\mu\rho}H^{\dagger}D^{\nu}_{\ \rho}H+\textrm{h.c.}\right)$} & \\

13 & \scriptsize{$\left(\bar{q}\sigma_{\mu\nu}\gamma_{5}\bfd_{\rho}q\right)\left(F^{\mu\rho}Z^{\nu}\right)$} & $-$ & & \scriptsize{$\left(\bar{Q}_{L}\sigma_{\mu\nu}\bfd_{\rho}\htilde u_{R}+\textrm{h.c.}\right)\left(B^{\mu\rho}H^{\dagger}D^{\nu}H+\textrm{h.c.}\right)$} & \\

14 & \scriptsize{$\left(i\bar{q}\sigma_{\mu\nu}\gamma_{5}q\right)\left(F^{\mu\rho}\partial_{\rho}Z^{\nu}\right)$} & $+$ & & \scriptsize{$\left(i\bar{Q}_{L}\sigma_{\mu\nu}\htilde u_{R}+\textrm{h.c.}\right)\left(B^{\mu\rho}H^{\dagger}D^{\nu}_{\ \rho}H+\textrm{h.c.}\right)$} & \\

15 & \scriptsize{$\left(i\bar{q}\overset\leftrightarrow{D^{\mu}}q\right)\left(\tilde{F}_{\mu\sigma}Z^{\sigma}\right)$} & $+$ & & \scriptsize{$\left(i\bar{Q}_{L}\overset\leftrightarrow{D^{\mu}}\htilde u_{R}+\textrm{h.c.}\right)\left(\tilde{B}_{\mu\sigma}H^{\dagger}D^{\sigma}H+\textrm{h.c.}\right)$} & \\

16 & \scriptsize{$\left(\bar{q}\gamma_{5}\overset\leftrightarrow{D^{\mu}}q\right)\left(\tilde{F}_{\mu\sigma}Z^{\sigma}\right)$} & $-$ & & \scriptsize{$\left(\bar{Q}_{L}\overset\leftrightarrow{D^{\mu}}\htilde u_{R}+\textrm{h.c.}\right)\left(\tilde{B}_{\mu\sigma}H^{\dagger}D^{\sigma}H+\textrm{h.c.}\right)$} & \\
\hline

17 & \scriptsize{$\left(\bar{q}\gamma^{\nu}q\right)\left(\left[\partial_{\nu}F^{\mu\rho}\right]Z_{\mu\rho}\right)$} & $-$ &  \multirow{2}{*}{8} & \scriptsize{$\left(\bar{Q}_{L}\gamma^{\nu}Q_{L}+\bar{u}_{R}\gamma^{\nu}u_{R}\right)\left(\left[\partial_{\nu}B^{\mu\rho}\right]B_{\mu\rho}\right)$} & \multirow{2}{*}{  $\frac{0.09}{E_\text{TeV}^4}$ } \\

18 & \scriptsize{$\left(\bar{q}\gamma^{\nu}\gamma_{5}q\right)\left(\left[\partial_{\nu}F^{\mu\rho}\right]Z_{\mu\rho}\right)$} & $-$ & & \scriptsize{$\left(\bar{Q}_{L}\gamma^{\nu}Q_{L}-\bar{u}_{R}\gamma^{\nu}u_{R}\right)\left(\left[\partial_{\nu}B^{\mu\rho}\right]B_{\mu\rho}\right)$} & \\
\hline

19 & \scriptsize{$\left(i\bar{q}\gamma^{\nu}\bfd_{\rho}q\right)\left(\left[\partial_{\nu}F^{\mu\rho}\right]Z_{\mu}\right)$} & $+$ & \multirow{6}{*}{8} & \scriptsize{$\left(i\bar{Q}_{L}\gamma^{\nu}\bfd_{\rho}Q_{L}+i\bar{u}_{R}\gamma^{\nu}\bfd_{\rho}u_{R}\right)\left(\left[\partial_{\nu}B^{\mu\rho}\right]H^{\dagger}D_{\mu}H+\textrm{h.c.}\right)$} & \multirow{6}{*}{  $\frac{0.02}{E_\text{TeV}^5}, \frac{0.07}{E_\text{TeV}^6}$ } \\

20 & \scriptsize{$\left(i\bar{q}\gamma^{\nu}\gamma_{5}\bfd_{\rho}q\right)\left(\left[\partial_{\nu}F^{\mu\rho}\right]Z_{\mu}\right)$} & $+$ & & \scriptsize{$\left(i\bar{Q}_{L}\gamma^{\nu}\bfd_{\rho}Q_{L}-\bar{u}_{R}\gamma^{\nu}\bfd_{\rho}u_{R}\right)\left(\left[\partial_{\nu}B^{\mu\rho}\right]H^{\dagger}D_{\mu}H+\textrm{h.c.}\right)$} & \\

21 & \scriptsize{$\left(i\bar{q}\gamma^{\nu}\bfd_{\mu}q\right)\left(F^{\mu\rho}\partial_{\rho}Z_{\nu}\right)$} & $+$ & & \scriptsize{$\left(i\bar{Q}_{L}\gamma^{\nu}\bfd_{\mu}Q_{L}+i\bar{u}_{R}\gamma^{\nu}\bfd_{\mu}u_{R}\right)\left(B^{\mu\rho}H^{\dagger}D_{\nu\rho}H+\textrm{h.c.}\right)$} & \\

22 & \scriptsize{$\left(i\bar{q}\gamma^{\nu}\gamma_{5}\bfd_{\mu}q\right)\left(F^{\mu\rho}\partial_{\rho}Z_{\nu}\right)$} & $+$ & & \scriptsize{$\left(i\bar{Q}_{L}\gamma^{\nu}\bfd_{\mu}Q_{L}-i\bar{u}_{R}\gamma^{\nu}\bfd_{\mu}u_{R}\right)\left(B^{\mu\rho}H^{\dagger}D_{\nu\rho}H+\textrm{h.c.}\right)$} & \\

23 & \scriptsize{$\left(\bar{q}\gamma_{\mu}\bfd_{\nu\rho}q\right)\left(F^{\mu\rho}Z^{\nu}\right)$} & $-$ & & \scriptsize{$\left(\bar{Q}_{L}\gamma_{\mu}\bfd_{\nu\rho}Q_{L}+\bar{u}_{R}\gamma_{\mu}\bfd_{\nu\rho}u_{R}\right)\left(B^{\mu\rho}H^{\dagger}D^{\nu}H+\textrm{h.c.}\right)$} & \\

24 & \scriptsize{$\left(\bar{q}\gamma_{\mu}\gamma_{5}\bfd_{\nu\rho}q\right)\left(F^{\mu\rho}Z^{\nu}\right)$} & $-$ & & \scriptsize{$\left(\bar{Q}_{L}\gamma_{\mu}\bfd_{\nu\rho}Q_{L}-\bar{u}_{R}\gamma_{\mu}\bfd_{\nu\rho}u_{R}\right)\left(B^{\mu\rho}H^{\dagger}D^{\nu}H+\textrm{h.c.}\right)$} & \\
\hline

25 & \scriptsize{$\left(\bar{q}\bfd_{\mu\nu}q\right)\left(F^{\mu\rho}\partial_{\rho}Z^{\nu}\right)$} & $+$ & \multirow{2}{*}{9} & \scriptsize{$\left(\bar{Q}_{L}\bfd_{\mu\nu}\htilde u_{R}+\textrm{h.c.}\right)\left(B^{\mu\rho}H^{\dagger}D^{\nu}_{\ \rho}H+\textrm{h.c.}\right)$} & \multirow{2}{*}{  $\frac{0.006}{E_\text{TeV}^6}, \frac{0.05}{E_\text{TeV}^8}$ } \\

26 & \scriptsize{$\left(i\bar{q}\gamma_{5}\bfd_{\mu\nu}q\right)\left(F^{\mu\rho}\partial_{\rho}Z^{\nu}\right)$} & $-$ & & \scriptsize{$\left(i\bar{Q}_{L}\bfd_{\mu\nu}\htilde u_{R}+\textrm{h.c.}\right)\left(B^{\mu\rho}H^{\dagger}D^{\nu}_{\ \rho}H+\textrm{h.c.}\right)$} & \\
\hline

\end{tabular}
\begin{minipage}{6.5in}
\medskip
\caption{\label{tab:qqza1} \scriptsize Primary operators for $\bar{q}qZ\gamma$ interactions. As outlined in the text, these operators can be modified to yield the operators for $\bar{q}qZg$, $\bar{q}q^{\prime}W\gamma$, and $\bar{q}q^{\prime}Wg$ interactions. Under the assumption that $\bar{q}$ and $q$ are each other's anti-particles, the operators are Hermitean and have the listed CP properties. If they are not, each of these operators has a Hermitean conjugate, which can be used to create a CP even and a CP odd operator. To simplify the expressions, we use the shorthand $\bfd_{\mu\nu}=\bfd_{\mu}\bfd_{\nu}$, and similarly, $D_{\mu\nu}=D_{\mu}D_{\nu}$. To get the descendant operators, once can add contracted derivatives to get arbitrary Mandelstam factors of $s, t$. At dimension 9, $s\mathcal{O}_{7}$ and $s\mathcal{O}_{8}$ become redundant to other operators. For these two, one only needs their $t^{n}\mathcal{O}_{7}$ and $t^{n}\mathcal{O}_{8}$ descendants.}
\end{minipage}
\end{center}
\end{adjustwidth}
\end{table}

Table~\ref{tab:qqga1} lists the primary operators for $\bar{q}qg\gamma$ interactions. Reading the appropriate Hilbert series, we expect to find 6 dimension 7 operators, 8 dimension 8 operators, and 4 dimension 9 operators, as well as 2 operators that become redundant at dimension 9, so the analysis again finds 2 additional dimension 9 primary operators that a quick interpretation of the Hilbert series would have missed. We indeed find the 18 operators we expect from the Hilbert series analysis, as well as two operators that become redundant at dimension 9---$s\mathcal{O}_{5}$ and $s\mathcal{O}_{6}$. Thus, for those two operators, we can  just add their descendants $t^{n}\mathcal{O}_{5}$ and $t^{n}\mathcal{O}_{6}$. 

\begin{table}[p]
\begin{center}
\footnotesize
\renewcommand{\arraystretch}{0.9}
\setlength{\tabcolsep}{6pt}
\begin{tabular}{|c|c|c|c|c|c|}

\hline
\multirow{2}{*}{$i$} & \multirow{2}{*}{$\mathcal{O}_i^{\bar{q}qg\gamma}$}   & \multirow{2}{*}{CP} & \multirow{2}{*}{$d_{\mathcal{O}_i}$}& SMEFT & $c$ Unitarity  \\
 & & & & Operator & Bound \\
 \hline
 
1 & \scriptsize{$\left(\bar{q}q\right)\left(F^{\mu\nu}G_{\mu\nu}\right)$} & $+$ & \multirow{6}{*}{7} & \scriptsize{$\left(\bar{Q}_{L}\htilde u_{R}+\textrm{h.c.}\right)\left(B^{\mu\nu}G_{\mu\nu}\right)$} & \multirow{6}{*}{  $\frac{0.4}{E_\text{TeV}^3}, \frac{1.2}{E_\text{TeV}^4}$ } \\

2 & \scriptsize{$\left(i\bar{q}\gamma_{5}q\right)\left(F^{\mu\nu}G_{\mu\nu}\right)$} & $-$ & & \scriptsize{$\left(i\bar{Q}_{L}\htilde u_{R}+\textrm{h.c.}\right)\left(B^{\mu\nu}G_{\mu\nu}\right)$} & \\

3 & \scriptsize{$\left(\bar{q}\sigma_{\mu\nu}q\right)\left(F^{\mu\rho}G^{\nu}_{\ \rho}\right)$} & $-$ & & \scriptsize{$\left(\bar{Q}_{L}\sigma_{\mu\nu}\htilde u_{R}+\textrm{h.c.}\right)\left(B^{\mu\rho}G^{\nu}_{\ \rho}\right)$} & \\

4 & \scriptsize{$\left(i\bar{q}\sigma_{\mu\nu}\gamma_{5}q\right)\left(F^{\mu\rho}G^{\nu}_{\ \rho}\right)$} & $+$ & & \scriptsize{$\left(i\bar{Q}_{L}\sigma_{\mu\nu}\gamma_{5}\htilde u_{R}+\textrm{h.c.}\right)\left(B^{\mu\rho}G^{\nu}_{\ \rho}\right)$} & \\

5 & \scriptsize{$\left(\bar{q}q\right)\left(F^{\mu\nu}\tilde{G}_{\mu\nu}\right)$} & $-$ & & \scriptsize{$\left(\bar{Q}_{L}\htilde u_{R}+\textrm{h.c.}\right)\left(B^{\mu\nu}\tilde{G}_{\mu\nu}\right)$} & \\

6 & \scriptsize{$\left(i\bar{q}\gamma_{5}q\right)\left(F^{\mu\nu}\tilde{G}_{\mu\nu}\right)$} & $+$ & & \scriptsize{$\left(\bar{Q}_{R}\htilde u_{R}+\textrm{h.c.}\right)\left(B^{\mu\nu}\tilde{G}_{\mu\nu}\right)$} & \\
\hline

7 & \scriptsize{$\left(\bar{q}\gamma^{\nu}q\right)\left(\left[\partial_{\nu}F^{\mu\rho}\right]G_{\mu\rho}\right)$} & $-$ & \multirow{8}{*}{8} & \scriptsize{$\left(\bar{Q}_{L}\gamma^{\nu}Q_{L}+\bar{u}_{R}\gamma^{\nu}u_{R}\right) \left(\left[\partial_{\nu}B^{\mu\rho}\right]G_{\mu\rho}\right)$} & \multirow{8}{*}{  $\frac{0.09}{E_\text{TeV}^4}$ } \\

8 & \scriptsize{$\left(\bar{q}\gamma^{\nu}\gamma_{5}q\right)\left(\left[\partial_{\nu}F^{\mu\rho}\right]G_{\mu\rho}\right)$} & $-$ & & \scriptsize{$\left(\bar{Q}_{L}\gamma^{\nu}Q_{L}-\bar{u}_{R}\gamma^{\nu}u_{R}\right)\left(\left[\partial_{\nu}B^{\mu\rho}\right]G_{\mu\rho}\right)$} & \\

9 & \scriptsize{$\left(i\bar{q}\gamma^{\nu}\bfd_{\mu}q\right)\left(F^{\mu\rho}G_{\nu\rho}\right)$} & $+$ & & \scriptsize{$\left(i\bar{Q}_{L}\gamma^{\nu}\bfd_{\mu}Q_{L}+i\bar{u}_{R}\gamma^{\nu}\bfd_{\mu}u_{R}\right)\left(B^{\mu\rho}G_{\nu\rho}\right)$} & \\

10 & \scriptsize{$\left(i\bar{q}\gamma^{\nu}\gamma_{5}\bfd_{\mu}q\right)\left(F^{\mu\rho}G_{\nu\rho}\right)$} & $+$ & & \scriptsize{$\left(i\bar{Q}_{L}\gamma^{\nu}\bfd_{\mu}Q_{L}-i\bar{u}_{R}\gamma^{\nu}\bfd_{\mu}u_{R}\right)\left(B^{\mu\rho}G_{\nu\rho}\right)$} & \\

11 & \scriptsize{$\left(i\bar{q}\gamma^{\nu}\bfd_{\rho}q\right)\left(F_{\nu\mu}G^{\rho\mu}\right)$} & $+$ & & \scriptsize{$\left(i\bar{Q}_{L}\gamma^{\nu}\bfd_{\rho}Q_{L}+i\bar{u}_{R}\gamma^{\nu}\bfd_{\rho}u_{R}\right)\left(B_{\nu\mu}G^{\rho\mu}\right)$} & \\

12 & \scriptsize{$\left(i\bar{q}\gamma^{\nu}\gamma_{5}\bfd_{\rho}q\right)\left(F_{\nu\mu}G^{\rho\mu}\right)$} & $+$ & & \scriptsize{$\left(i\bar{Q}_{L}\gamma^{\nu}\bfd_{\rho}Q_{L}-i\bar{u}_{R}\gamma^{\nu}\bfd_{\rho}u_{R}\right)\left(B_{\nu\mu}G^{\rho\mu}\right)$} & \\

13 & \scriptsize{$\left(i\bar{q}\gamma^{\nu}\bfd_{\rho}q\right)\left(\tilde{F}_{\mu\nu}G^{\mu\rho}\right)$} & $-$ & & \scriptsize{$\left(i\bar{Q}_{L}\gamma^{\nu}\bfd_{\rho}Q_{L}+i\bar{u}_{R}\gamma^{\nu}\bfd_{\rho}u_{R}\right)\left(\tilde{B}_{\mu\nu}G^{\mu\rho}\right)$} & \\

14 & \scriptsize{$\left(i\bar{q}\gamma^{\nu}\gamma_{5}\bfd_{\rho}q\right)\left(\tilde{F}_{\mu\nu}G^{\mu\rho}\right)$} & $-$ & & \scriptsize{$\left(i\bar{Q}_{L}\gamma^{\nu}\bfd_{\rho}Q_{L}-i\bar{u}_{R}\gamma^{\nu}\bfd_{\rho}u_{R}\right)\left(\tilde{B}_{\mu\nu}G^{\mu\rho}\right)$} & \\
\hline

15 & \scriptsize{$\left(\bar{q}\bfd_{\mu\nu}q\right)\left(F^{\mu\rho}G^{\nu}_{\ \rho}\right)$} & $+$ & \multirow{4}{*}{9} & \scriptsize{$\left(\bar{Q}_{L}\bfd_{\mu\nu}\htilde u_{R}+\textrm{h.c.}\right)\left(B^{\mu\rho}G^{\nu}_{\ \rho}\right)$} & \multirow{4}{*}{  $\frac{0.02}{E_\text{TeV}^5}, \frac{0.07}{E_\text{TeV}^6}$ } \\

16 & \scriptsize{$\left(i\bar{q}\gamma_{5}\bfd_{\mu\nu}q\right)\left(F^{\mu\rho}G^{\nu}_{\ \rho}\right)$} & $-$ & & \scriptsize{$\left(i\bar{Q}_{L}\bfd_{\mu\nu} \htilde u_{R}+\textrm{h.c.}\right)\left(B^{\mu\rho}G^{\nu}_{\ \rho}\right)$} & \\

17 & \scriptsize{$\left(i\bar{q}\sigma_{\mu\nu}\bfd_{\sigma}q\right)\left(F^{\mu\rho}D_{\rho}G^{\nu\sigma}\right)$} & $+$ & & \scriptsize{$\left(i\bar{Q}_{L}\sigma_{\mu\nu}\bfd_{\sigma}\htilde u_{R}+\textrm{h.c.}\right)\left(B^{\mu\rho}D_{\rho}G^{\nu\sigma}\right)$} & \\

18 & \scriptsize{$\left(\bar{q}\sigma_{\mu\nu}\gamma_{5}\bfd_{\rho}q\right )\left(F^{\mu\sigma}D_{\sigma}G^{\nu\rho}\right)$} & $-$ & & \scriptsize{$\left(\bar{Q}_{L}\sigma_{\mu\nu}\bfd_{\rho}u_{R}+\textrm{h.c.}\right) \left(B^{\mu\sigma}D_{\sigma}G^{\nu\rho}\right)$} & \\
\hline
 \end{tabular}
\begin{minipage}{5.7in}
\medskip
\caption{\label{tab:qqga1} \footnotesize Primary operators for $\bar{q}qg\gamma$ interactions. Under the assumption that $\bar{q}$ and $q$ are each other's anti-particles, the operators are Hermitean and have the listed CP properties. If they are not, each of these operators has a Hermitean conjugate, which can be used to create a CP even and a CP odd operator. To simplify the expressions, we use the shorthand $\bfd_{\mu\nu}=\bfd_{\mu}\bfd_{\nu}$. To get the descendant operators, once can add contracted derivatives to get arbitrary Mandelstam factors of $s, t$. At dimension 9, $s\mathcal{O}_{5}$ and $s\mathcal{O}_{6}$ become redundant to other operators. For these two, one only needs their $t^{n}\mathcal{O}_{5}$ and $t^{n}\mathcal{O}_{6}$ descendants.}
\end{minipage}
\end{center}
\end{table}

We list the primary operators for $\bar{q}q\gamma\gamma$ interactions in Table~\ref{tab:qqaa1}. From the Hilbert series, we expect that there should be 4 operators at dimension 7, 2 operators at dimension 8, 4 operators at dimension 9, 6 operators at dimension 10, and 2 operators at dimension 11,  giving 18 total primary operators in agreement with the Hilbert series.   We also find that there are two new redundancies at dimension 11 for $s\mathcal{O}_{7}$ and $s\mathcal{O}_{8}$. This gives rise to a  complete cancellation in the Hilbert series at dimension 11 between the two new operators $\mathcal{O}_{17}, \mathcal{O}_{18}$ and the two redundancies for  $s\mathcal{O}_{7}$ and $s\mathcal{O}_{8}$.  Given the redundancies, for $\mathcal{O}_{7}$ and $\mathcal{O}_{8}$, we only need the descendant operators $(t-u)^{2n}\mathcal{O}_{7}$ and $(t-u)^{2n}\mathcal{O}_{8}$.  

\begin{table}[p]
\begin{adjustwidth}{-.5in}{-.5in}  
\begin{center}
\footnotesize
\renewcommand{\arraystretch}{0.9}
\setlength{\tabcolsep}{6pt}
\begin{tabular}{|c|c|c|c|c|c|}

\hline
\multirow{2}{*}{$i$} & \multirow{2}{*}{$\mathcal{O}_i^{\bar{q}q\gamma\gamma}$}   & \multirow{2}{*}{CP} & \multirow{2}{*}{$d_{\mathcal{O}_i}$}& SMEFT & $c$ Unitarity  \\
 & & & & Operator & Bound \\
 \hline
 
1 & \scriptsize{$\left(\bar{q}q\right)\left(F^{\mu\nu}F_{\mu\nu}\right)$} & $+$ & \multirow{4}{*}{7} & \scriptsize{$\left(\bar{Q}_{L}\htilde u_{R}+\textrm{h.c.}\right)\left(B^{\mu\nu}B_{\mu\nu}\right)$} & \multirow{4}{*}{  $\frac{0.4}{E_\text{TeV}^3}, \frac{1.2}{E_\text{TeV}^4}$ } \\

2 & \scriptsize{$\left(\bar{q}i\gamma_{5}q\right)\left(F^{\mu\nu}F_{\mu\nu}\right)$} & $-$ & & \scriptsize{$\left(i\bar{Q}_{L}\htilde u_{R}+\textrm{h.c}\right)\left(B^{\mu\nu}B_{\mu\nu}\right)$} & \\

3 & \scriptsize{$\left(\bar{q}q\right)\left(F^{\mu\nu}\tilde{F}_{\mu\nu}\right)$} & $-$ & & \scriptsize{$\left(\bar{Q}_{L}\htilde u_{R}+\textrm{h.c.}\right)\left(B^{\mu\nu}\tilde{B}_{\mu\nu}\right)$} & \\

4 & \scriptsize{$\left(i\bar{q}\gamma_{5}q\right)\left(F^{\mu\nu}\tilde{F}_{\mu\nu}\right)$} & $+$ & & \scriptsize{$\left(i\bar{Q}_{L}\htilde u_{R}+\textrm{h.c.}\right)\left(B^{\mu\nu}\tilde{B}_{\mu\nu}\right)$} & \\
\hline
 
5 & \scriptsize{$\left(i\bar{q}\gamma^{\nu}\bfd_{\mu}q\right)\left(F^{\mu\rho}F_{\rho\nu}\right)$} & $+$ & \multirow{2}{*}{8} & \scriptsize{$\left(i\bar{Q}_{L}\bfd_{\mu}\gamma^{\nu}Q_{L}+i\bar{u}_{R}\bfd_{\mu}\gamma^{\nu}u_{R}\right)\left(B^{\mu\rho}B_{\rho\nu}\right)$} & \multirow{2}{*}{  $\frac{0.09}{E_\text{TeV}^4}$ } \\

6 & \scriptsize{$\left(i\bar{q}\gamma^{\nu}\gamma_{5}\bfd_{\mu}q\right)\left(F^{\mu\rho}F_{\rho\nu}\right)$} & $+$ & & \scriptsize{$\left(i\bar{Q}_{L}\bfd_{\mu}\gamma^{\nu}Q_{L}-i\bar{u}_{R}\bfd_{\mu}\gamma^{\nu}u_{R}\right)\left(B^{\mu\rho}B_{\rho\nu}\right)$} & \\
\hline
 
7 & \scriptsize{$\left(i\bar{q}\sigma_{\mu\nu}\bfd_{\rho}q\right)\left(F^{\mu\sigma}\partial^{\rho}F^{\nu}_{\ \sigma}\right)$} & $+$ & \multirow{4}{*}{9} & \scriptsize{$\left(i\bar{Q}_{L}\sigma_{\mu\nu}\bfd_{\rho}\htilde u_{R}+\textrm{h.c.}\right)\left(B^{\mu\sigma}\partial^{\rho}B^{\nu}_{\ \sigma}\right)$} & \multirow{4}{*}{  $\frac{0.02}{E_\text{TeV}^5}, \frac{0.07}{E_\text{TeV}^6}$ } \\

8 & \scriptsize{$\left(\bar{q}\sigma_{\mu\nu}\gamma_{5}\bfd_{\rho}q\right)\left(F^{\mu\sigma}\partial^{\rho}F^{\nu}_{\ \sigma}\right)$} & $-$ & & \scriptsize{$\left(\bar{Q}_{L}\sigma_{\mu\nu}\bfd_{\rho}\htilde u_{R}+\textrm{h.c.}\right)\left(B^{\mu\sigma}\partial^{\rho}B^{\nu}_{\ \sigma}\right)$} & \\

9 & \scriptsize{$\left(\bar{q}\bfd_{\mu\nu}q\right)\left(F^{\mu\rho}F^{\nu}_{\ \rho}\right)$} & $+$ & & \scriptsize{$\left(\bar{Q}_{L}\bfd_{\mu\nu}\htilde u_{R}+\textrm{h.c.}\right)\left(B^{\mu\rho}B^{\nu}_{\ \rho}\right)$} & \\

10 & \scriptsize{$\left(i\bar{q}\gamma_{5}\bfd_{\mu\nu}q\right)\left(F^{\mu\rho}F^{\nu}_{\ \rho}\right)$} & $-$ & & \scriptsize{$\left(i\bar{Q}_{L}\bfd_{\mu\nu}\htilde u_{R}+\textrm{h.c.}\right)\left(B^{\mu\rho}B^{\nu}_{\ \rho}\right)$} & \\
\hline
 
11 & \scriptsize{$\left(i\bar{q}\gamma^{\nu}\bfd_{\rho}q\right)\left(\left[\partial_{\nu}F^{\mu\sigma}\right]\partial^{\rho}F_{\mu\sigma}\right)$} & $+$ & \multirow{6}{*}{10} & \scriptsize{$\left(i\bar{Q}_{L}\gamma^{\nu}\bfd_{\rho}Q_{L}+i\bar{u}_{R}\gamma^{\nu}\bfd_{\rho}u_{R}\right)\left(\left[\partial_{\nu}B^{\mu\sigma}\right]\partial^{\rho}B_{\mu\sigma}\right)$} & \multirow{6}{*}{  $\frac{0.006}{E_\text{TeV}^6}$ } \\

12 & \scriptsize{$\left(i\bar{q}\gamma^{\nu}\gamma_{5}\bfd_{\rho}q\right)\left(\left[\partial_{\nu}F^{\mu\sigma}\right]\partial^{\rho}F_{\mu\sigma}\right)$} & $+$ & & \scriptsize{$\left(i\bar{Q}_{L}\gamma^{\nu}\bfd_{\rho}Q_{L}-i\bar{u}_{R}\gamma^{\nu}\bfd_{\rho}u_{R}\right)\left(\left[\partial_{\nu}B^{\mu\sigma}\right]\partial^{\rho}B_{\mu\sigma}\right)$} & \\

13 & \scriptsize{$\left(\bar{q}\gamma^{\nu}\bfd_{\mu\sigma}q\right)\left(F^{\mu\rho}\partial^{\sigma}F_{\nu\rho}\right)$} & $-$ & & \scriptsize{$\left(\bar{Q}_{L}\gamma^{\nu}\bfd_{\mu\sigma}Q_{L}+\bar{u}_{R}\gamma^{\nu}\bfd_{\mu\sigma}u_{R}\right)\left(B^{\mu\rho}\partial^{\sigma}B_{\nu\rho}\right)$} & \\

14 & \scriptsize{$\left(\bar{q}\gamma^{\nu}\gamma_{5}\bfd_{\mu\sigma}q\right)\left(F^{\mu\rho}\partial^{\sigma}F_{\nu\rho}\right)$} & $-$ & & \scriptsize{$\left(\bar{Q}_{L}\gamma^{\nu}\bfd_{\mu\sigma}Q_{L}-\bar{u}_{R}\gamma^{\nu}\bfd_{\mu\sigma}u_{R}\right)\left(B^{\mu\rho}\partial^{\sigma}B_{\nu\rho}\right)$} & \\

15 & \scriptsize{$\left(\bar{q}\gamma^{\nu}\bfd_{\alpha\beta}q\right)\left(\tilde{F}_{\nu\sigma}\partial^{\beta}F^{\sigma\alpha}\right)$} & $+$ & & \scriptsize{$\left(\bar{Q}_{L}\gamma^{\nu}\bfd_{\alpha\beta}Q_{L}+\bar{u}_{R}\gamma^{\nu}\bfd_{\alpha\beta}u_{R}\right)\left(\tilde{B}_{\nu\sigma}\partial^{\beta}B^{\sigma\alpha}\right)$} & \\

16 & \scriptsize{$\left(\bar{q}\gamma^{\nu}\gamma_{5}\bfd_{\alpha\beta}q\right)\left(\tilde{F}_{\nu\sigma}\partial^{\beta}F^{\sigma\alpha}\right)$} & $+$ & & \scriptsize{$\left(\bar{Q}_{L}\gamma^{\nu}\bfd_{\alpha\beta}Q_{L}-\bar{u}_{R}\gamma^{\nu}\bfd_{\alpha\beta}u_{R}\right)\left(\tilde{B}_{\nu\sigma}\partial^{\beta}B^{\sigma\alpha}\right)$} & \\
\hline
 
17 & \scriptsize{$\left(\bar{q}\sigma_{\mu\nu}\bfd_{\sigma\alpha}q\right)\left(F^{\mu\rho}\partial^{\alpha}_{\ \rho}F^{\nu\sigma}\right)$} & $-$ & \multirow{2}{*}{11} & \scriptsize{$\left(\bar{Q}_{L}\sigma_{\mu\nu}\bfd_{\sigma\alpha}\htilde u_{R}+\textrm{h.c.}\right)\left(B^{\mu\rho}\partial^{\alpha}_{\ \rho}B^{\nu\sigma}\right)$} & \multirow{2}{*}{  $\frac{0.001}{E_\text{TeV}^7}, \frac{0.004}{E_\text{TeV}^8}$ } \\

18 & \scriptsize{$\left(i\bar{q}\sigma_{\mu\nu}\gamma_{5}\bfd_{\sigma\alpha}q\right)\left(F^{\mu\rho}\partial^{\alpha}_{\ \rho}F^{\nu\sigma}\right)$} & $+$ & & \scriptsize{$\left(i\bar{Q}_{L}\sigma_{\mu\nu}\bfd_{\sigma\alpha}\htilde u_{R}+\textrm{h.c.}\right)\left(B^{\mu\rho}\partial^{\alpha}_{\ \rho}B^{\nu\sigma}\right)$} & \\
\hline
 
 \end{tabular}
\begin{minipage}{5.7in}
\medskip
\caption{\label{tab:qqaa1} \footnotesize Primary operators for $\bar{q}q\gamma\gamma$ interactions. Under the assumption that $\bar{q}$ and $q$ are each other's anti-particles, the operators are Hermitean and have the listed CP properties. If they are not, each of these operators has a Hermitean conjugate, which can be used to create a CP even and a CP odd operator. To simplify the expressions, we use the shorthand $\bfd_{\mu\nu}=\bfd_{\mu}\bfd_{\nu}$, and similarly, $\partial_{\mu\nu}=\partial_{\mu}\partial_{\nu}$. To get the descendant operators, once can add contracted derivatives to get arbitrary Mandelstam factors of $s, (t-u)^2$. At dimension 11, $s\mathcal{O}_{7}$ and $s\mathcal{O}_{8}$ become redundant to other operators. For these two, one only needs their $(t-u)^{2n}\mathcal{O}_{7}$ and $(t-u)^{2n}\mathcal{O}_{8}$ descendants.}
\end{minipage}
\end{center}
\end{adjustwidth}
\end{table}

In Tables~\ref{tab:qqgg1} and~\ref{tab:qqgg2}, we list all of the primary operators for $\bar{q}qgg$ interactions. The Hilbert series says that we should expect 10 operators at dimension 7, 10 operators at dimension 8, 14 operators at dimension 9, 14 operators at dimension 10, and 6 operators at dimension 11. Additionally, we find that there are 2 redundancies at dimension 9---$s\mathcal{O}_{9}$ and $s\mathcal{O}_{10}$---and 4 redundancies at dimension 11---$s\mathcal{O}_{21}$, $s\mathcal{O}_{22}$, $s\mathcal{O}_{23}$, and $s\mathcal{O}_{24}$. As noted in Sec.~\ref{sec:HilbertSeries}, there are three ways we can contract the $SU(3)$ indices, two symmetric and one antisymmetric. For example, $\mathcal{O}_{1}$ and $\mathcal{O}_{2}$ in Table~\ref{tab:qqgg1} should be read as $\left(\bar{q}\delta_{AB}q\right)\left(G^{A\mu\nu}G^{B}_{\mu\nu}\right)$ and $d_{ABC}\left(\bar{q}T^{A}q\right)\left(G^{B\mu\nu}G^{C}_{\mu\nu}\right)$, respectively, where $T^{A}$ are the generators of $SU(3)$. $\mathcal{O}_{7}$ in Table~\ref{tab:qqgg1} should be ready as $f_{ABC}\left(\bar{q}T^{A}q\right)\left(G^{B\mu\nu}G^{C}_{\mu\nu}\right)$.  Thus, for $\mathcal{O}_{9,10,21,22,23,24}$, we only need to add their descendants with factors of $(t-u)^2$.

\begin{table}[p]
\begin{adjustwidth}{-.75in}{-.75in}  
\begin{center}
\footnotesize
\renewcommand{\arraystretch}{0.9}
\setlength{\tabcolsep}{6pt}
\begin{tabular}{|c|c|c|c|c|c|c|}

\hline
\multirow{2}{*}{$i$} & \multirow{2}{*}{$\mathcal{O}_i^{\bar{q}qgg}$}   & \multirow{2}{*}{CP} & \multirow{2}{*}{$d_{\mathcal{O}_i}$}& \multirow{2}{*}{$SU(3)$} & SMEFT & $c$ Unitarity  \\
 & & & & & Operator & Bound \\
 \hline
 
1, 2 & \scriptsize{$\left(\bar{q}q\right)\left(G^{\mu\nu}G_{\mu\nu}\right)$} & $+$ & \multirow{4}{*}{7} & \multirow{4}{*}{$\delta_{AB}$, $d_{ABC}$} & \scriptsize{$\left(\bar{Q}_{L}\htilde u_{R}+\textrm{h.c.}\right)\left(G^{\mu\nu}G_{\mu\nu}\right)$} & \multirow{4}{*}{  $\frac{0.4}{E_\text{TeV}^3}, \frac{1.2}{E_\text{TeV}^4}$ } \\

3, 4 & \scriptsize{$\left(i\bar{q}\gamma_{5}q\right)\left(G^{\mu\nu}G_{\mu\nu}\right)$} & $-$ & & & \scriptsize{$\left(i\bar{Q}_{L}\htilde u_{R}+\textrm{h.c.}\right)\left(G^{\mu\nu}G_{\mu\nu}\right)$} & \\

5, 6 & \scriptsize{$\left(\bar{q}q\right)\left(G^{\mu\nu}\tilde{G}_{\mu\nu}\right)$} & $-$ & & & \scriptsize{$\left(\bar{Q}_{L}\htilde u_{R}+\textrm{h.c.}\right)\left(G^{\mu\nu}\tilde{G}_{\mu\nu}\right)$} & \\

7, 8& \scriptsize{$\left(i\bar{q}\gamma_{5}q\right)\left(G^{\mu\nu}\tilde{G}_{\mu\nu}\right)$} & $+$ & & & \scriptsize{$\left(i\bar{Q}_{L}\htilde u_{R}+\textrm{h.c.}\right)\left(G^{\mu\nu}\tilde{G}_{\mu\nu}\right)$} & \\
\hline

9 & \scriptsize{$\left(\bar{q}\sigma_{\mu\nu}q\right)\left(G^{\mu\rho}G^{\nu}_{\ \rho}\right)$} & $+$ & \multirow{2}{*}{7} & \multirow{2}{*}{$f_{ABC}$} & \scriptsize{$\left(\bar{Q}_{L}\sigma_{\mu\nu}\htilde u_{R}+\textrm{h.c.}\right)\left(G^{\mu\rho}G^{\nu}_{\ \rho}\right)$} & \multirow{2}{*}{  $\frac{0.4}{E_\text{TeV}^3}, \frac{1.2}{E_\text{TeV}^4}$ } \\

10 & \scriptsize{$\left(i\bar{q}\sigma_{\mu\nu}\gamma_{5}q\right)\left(G^{\mu\rho}G^{\nu}_{\ \rho}\right)$} & $-$ & & & \scriptsize{$\left(i\bar{Q}_{L}\sigma_{\mu\nu}\htilde u_{R}+\textrm{h.c.}\right)\left(G^{\mu\rho}G^{\nu}_{\ \rho}\right)$} & \\
\hline

11, 12 & \scriptsize{$\left(i\bar{q}\gamma^{\nu}\bfd_{\mu}q\right)\left(G^{\mu\rho}G_{\nu\rho}\right)$} & $+$ & \multirow{2}{*}{8} & \multirow{2}{*}{$\delta_{AB}$, $d_{ABC}$} & \scriptsize{$\left(i\bar{Q}_{L}\gamma^{\nu}\bfd_{\mu}Q_{L}+i\bar{u}_{R}\gamma^{\nu}\bfd_{\mu}u_{R}\right)\left(G^{\mu\rho}G_{\nu\rho}\right)$} & \multirow{2}{*}{  $\frac{0.09}{E_\text{TeV}^4}$ } \\

13, 14 & \scriptsize{$\left(i\bar{q}\gamma^{\nu}\gamma_{5}\bfd_{\mu}q\right)\left(G^{\mu\rho}G_{\nu\rho}\right)$} & $+$ & & & \scriptsize{$\left(i\bar{Q}_{L}\gamma^{\nu}\bfd_{\mu}Q_{L}-i\bar{u}_{R}\gamma^{\nu}\bfd_{\mu}u_{R}\right)\left(G^{\mu\rho}G_{\nu\rho}\right)$} & \\
\hline

15 & \scriptsize{$\left(\bar{q}\gamma^{\nu}q\right)\left(\left[D_{\nu}G^{\mu\rho}\right]G_{\mu\rho}\right)$} & $+$ & \multirow{6}{*}{8} & \multirow{6}{*}{$f_{ABC}$} & \scriptsize{$\left(\bar{Q}_{L}\gamma^{\nu}Q_{L}+\bar{u}_{R}\gamma^{\nu}u_{R}\right)\left(\left[D_{\nu}G^{\mu\rho}\right]G_{\mu\rho}\right)$} & \multirow{6}{*}{  $\frac{0.09}{E_\text{TeV}^4}$ } \\

16 & \scriptsize{$\left(\bar{q}\gamma^{\nu}\gamma_{5}q\right)\left(\left[D_{\nu}G^{\mu\rho}\right]G_{\mu\rho}\right)$} & $+$ & & & \scriptsize{$\left(\bar{Q}_{L}\gamma^{\nu}Q_{L}-\bar{u}_{R}\gamma^{\nu}u_{R}\right)\left(\left[D_{\nu}G^{\mu\rho}\right]G_{\mu\rho}\right)$} & \\

17 & \scriptsize{$\left(i\bar{q}\gamma^{\nu}\bfd_{\mu}q\right)\left(G^{\mu\rho}G_{\nu\rho}\right)$} & $-$ & & & \scriptsize{$\left(i\bar{Q}_{L}\gamma^{\nu}\bfd_{\mu}Q_{L}+i\bar{u}_{R}\gamma^{\nu}\bfd_{\mu}u_{R}\right)\left(G^{\mu\rho}G_{\nu\rho}\right)$} & \\

18 & \scriptsize{$\left(i\bar{q}\gamma^{\nu}\gamma_{5}\bfd_{\mu}q\right)\left(G^{\mu\rho}G_{\nu\rho}\right)$} & $-$ & & & \scriptsize{$\left(i\bar{Q}_{L}\gamma^{\nu}\bfd_{\mu}Q_{L}-i\bar{u}_{R}\gamma^{\nu}\bfd_{\mu}u_{R}\right)\left(G^{\mu\rho}G_{\nu\rho}\right)$} & \\

19 & \scriptsize{$\left(i\bar{q}\gamma^{\mu}\bfd_{\rho}q\right)\left(G^{\nu\rho}\tilde{G}_{\mu\nu}\right)$} & $+$ & & & \scriptsize{$\left(i\bar{Q}_{L}\gamma^{\mu}\bfd_{\rho}q+i\bar{u}_{R}\gamma^{\mu}\bfd_{\rho}u_{R}\right)\left(G^{\nu\rho}\tilde{G}_{\mu\nu}\right)$} & \\

20 & \scriptsize{$\left(i\bar{q}\gamma^{\mu}\gamma_{5}\bfd_{\rho}q\right)\left(G^{\nu\rho}\tilde{G}_{\mu\nu}\right)$} & $+$ & & & \scriptsize{$\left(i\bar{Q}_{L}\gamma^{\mu}\bfd_{\rho}q-i\bar{u}_{R}\gamma^{\mu}\bfd_{\rho}u_{R}\right)\left(G^{\nu\rho}\tilde{G}_{\mu\nu}\right)$} & \\
\hline

21, 22 & \scriptsize{$\left(i\bar{q}\sigma_{\mu\nu}\bfd_{\sigma}q\right)\left(G^{\mu\rho}D^{\sigma}G^{\nu}_{\ \rho}\right)$} & $+$ & \multirow{4}{*}{9} & \multirow{4}{*}{$\delta_{AB}$, $d_{ABC}$} & \scriptsize{$\left(i\bar{Q}_{L}\sigma_{\mu\nu}\bfd_{\sigma}\htilde u_{R}+\textrm{h.c.}\right)\left(G^{\mu\rho}D^{\sigma}G^{\nu}_{\ \rho}\right)$} & \multirow{4}{*}{  $\frac{0.02}{E_\text{TeV}^5}, \frac{0.07}{E_\text{TeV}^6}$ } \\

23, 24 & \scriptsize{$\left(\bar{q}\sigma_{\mu\nu}\gamma_{5}\bfd_{\sigma}q\right)\left(G^{\mu\rho}D^{\sigma}G^{\nu}_{\ \rho}\right)$} & $-$ & & & \scriptsize{$\left(\bar{Q}_{L}\sigma_{\mu\nu}\bfd_{\sigma}\htilde u_{R}+\textrm{h.c.}\right)\left(G^{\mu\rho}D^{\sigma}G^{\nu}_{\ \rho}\right)$} & \\

25, 26 & \scriptsize{$\left(\bar{q}\bfd_{\mu\nu}q\right)\left(G^{\mu\rho}G^{\nu}_{\ \rho}\right)$} & $+$ & & & \scriptsize{$\left(\bar{Q}_{L}\bfd_{\mu\nu}\htilde u_{R}+\textrm{h.c.}\right)\left(G^{\mu\rho}G^{\nu}_{\ \rho}\right)$} & \\

27, 28 & \scriptsize{$\left(i\bar{q}\gamma_{5}\bfd_{\mu\nu}q\right)\left(G^{\mu\rho}G^{\nu}_{\ \rho}\right)$} & $-$ & & & \scriptsize{$\left(i\bar{Q}_{L}\bfd_{\mu\nu}\htilde u_{R}+\textrm{h.c.}\right)\left(G^{\mu\rho}G^{\nu}_{\ \rho}\right)$} & \\
\hline

29 & \scriptsize{$\left(\bar{q}\bfd_{\rho}q\right)\left(G^{\mu\nu}D^{\rho}G_{\mu\nu}\right)$} & $+$ & \multirow{6}{*}{9} & \multirow{6}{*}{$f_{ABC}$} & \scriptsize{$\left(\bar{Q}_{L}\bfd_{\rho}\htilde u_{R}+\textrm{h.c.}\right)\left(G^{\mu\nu}D^{\rho}G_{\mu\nu}\right)$} & \multirow{6}{*}{  $\frac{0.02}{E_\text{TeV}^5}, \frac{0.07}{E_\text{TeV}^6}$ } \\

30 & \scriptsize{$\left(i\bar{q}\gamma_{5}\bfd_{\rho}q\right)\left(G^{\mu\nu}D^{\rho}G_{\mu\nu}\right)$} & $-$ & & & \scriptsize{$\left(i\bar{Q}_{L}\bfd_{\rho}\htilde u_{R}+\textrm{h.c.}\right)\left(G^{\mu\nu}D^{\rho}G_{\mu\nu}\right)$} & \\

31 & \scriptsize{$\left(\bar{q}\bfd_{\rho}q\right)\left(G^{\mu\nu}D^{\rho}\tilde{G}_{\mu\nu}\right)$} & $-$ & & & \scriptsize{$\left(\bar{Q}_{L}\bfd_{\rho}\htilde u_{R}+\textrm{h.c.}\right)\left(G^{\mu\nu}D^{\rho}\tilde{G}_{\mu\nu}\right)$} & \\

32 & \scriptsize{$\left(i\bar{q}\gamma_{5}\bfd_{\rho}q\right)\left(G^{\mu\nu}D^{\rho} \tilde{G}_{\mu\nu}\right)$} & $+$ & & & \scriptsize{$\left(i\bar{Q}_{L}\bfd_{\rho}\htilde u_{R}+\textrm{h.c.}\right)\left(G^{\mu\nu}D^{\rho}\tilde{G}_{\mu\nu}\right)$} & \\

33 & \scriptsize{$\left(i\bar{q}\sigma_{\mu\nu}\bfd_{\sigma}q\right)\left(G^{\mu\rho}D_{\rho}G^{\nu\sigma}\right)$} & $-$ & & & \scriptsize{$\left(i\bar{Q}_{L}\sigma_{\mu\nu}\bfd_{\sigma}u_{R}+\textrm{h.c.}\right)\left(G^{\mu\rho}D_{\rho}G^{\nu\sigma}\right)$} & \\

34 & \scriptsize{$\left(\bar{q}\sigma_{\mu\nu}\gamma_{5}\bfd_{\sigma}q\right)\left(G^{\mu\rho}D_{\rho}G^{\nu\sigma}\right)$} & $+$ & & & \scriptsize{$\left(\bar{Q}_{L}\sigma_{\mu\nu}\bfd_{\sigma}u_{R}+\textrm{h.c.}\right)\left(G^{\mu\rho}D_{\rho}G^{\nu\sigma}\right)$} & \\
\hline

 \end{tabular}
\begin{minipage}{6.5in}
\medskip
\caption{\label{tab:qqgg1} \scriptsize Primary 7-, 8-, and 9-dimension operators for $\bar{q}qgg$ interactions. There are three allowed $SU(3)$ contractions, 2 symmetric ones---$\delta_{AB}$ and $d_{ABC}$---and one antisymmetric one---$f_{ABC}$. For example, $\mathcal{O}^{\bar{q}qgg}_{1}=\left(\bar{q}\delta_{AB}q\right)\left(G^{A\mu\nu}G^{B}_{\mu\nu}\right)$, $\mathcal{O}^{\bar{q}qgg}_{2}=d_{ABC}\left(\bar{q}T^{A}q\right)\left(G^{B\mu\nu}G^{C}_{\mu\nu}\right)$, and $\mathcal{O}^{\bar{q}qgg}_{9}=f_{ABC}\left(\bar{q}T^{A}\sigma_{\mu\nu}q\right)\left(G^{B\mu\rho}G^{C\nu}_{\ \rho}\right)$. Under the assumption that $\bar{q}$ and $q$ are each other's anti-particles, the operators are Hermitean and have the listed CP properties. If they are not, each of these operators has a Hermitean conjugate, which can be used to create a CP even and a CP odd operator. To simplify the expressions, we use the shorthand $\bfd_{\mu\nu}=\bfd_{\mu}\bfd_{\nu}$. To get the descendant operators, once can add contracted derivatives to get arbitrary Mandelstam factors of $s, (t-u)^2$. At dimension 9, $s\mathcal{O}_{9}$ and $s\mathcal{O}_{10}$ become redundant to other operators and at dimension 11, $s\mathcal{O}_{21}$, $s\mathcal{O}_{22}$, $s\mathcal{O}_{23}$ and $s\mathcal{O}_{24}$ become redundant to other operators. For the $\mathcal{O}_{9,10,21,22,23,24}$ operators, one only needs  descendants with factors of $(t-u)^2$.}
\end{minipage}
\end{center}
\end{adjustwidth}
\end{table}

\begin{table}[p]
\begin{adjustwidth}{-.85in}{-.85in}  
\begin{center}
\footnotesize
\renewcommand{\arraystretch}{0.9}
\setlength{\tabcolsep}{6pt}
\begin{tabular}{|c|c|c|c|c|c|c|}

\hline
\multirow{2}{*}{$i$} & \multirow{2}{*}{$\mathcal{O}_i^{\bar{q}qgg}$}   & \multirow{2}{*}{CP} & \multirow{2}{*}{$d_{\mathcal{O}_i}$}& \multirow{2}{*}{$SU(3)$} & SMEFT & $c$ Unitarity  \\
 & & & & & Operator & Bound \\
 \hline

35, 36 & \scriptsize{$\left(i\bar{q}\gamma^{\nu}\bfd_{\sigma}q\right)\left(\left[D_{\nu}G^{\mu\rho}\right]D^{\sigma}G_{\mu\rho}\right)$} & $+$ & \multirow{6}{*}{10} & \multirow{6}{*}{$\delta_{AB}$, $d_{ABC}$} & \scriptsize{$\left(i\bar{Q}_{L}\gamma^{\nu}\bfd_{\sigma}Q_{L}+i\bar{u}_{R}\gamma^{\nu}\bfd_{\sigma}u_{R}\right)\left(\left[D_{\nu}G^{\mu\rho}\right]D^{\sigma}G_{\mu\rho}\right)$} & \multirow{6}{*}{  $\frac{0.006}{E_\text{TeV}^6}$ } \\

37, 38 & \scriptsize{$\left(i\bar{q}\gamma^{\nu}\gamma_{5}\bfd_{\sigma}q\right)\left(\left[D_{\nu}G^{\mu\rho}\right]D^{\sigma}G_{\mu\rho}\right)$} & $+$ & & & \scriptsize{$\left(i\bar{Q}_{L}\gamma^{\nu}\bfd_{\sigma}Q_{L}-i\bar{u}_{R}\gamma^{\nu}\bfd_{\sigma}u_{R}\right)\left(\left[D_{\nu}G^{\mu\rho}\right]D^{\sigma}G_{\mu\rho}\right)$} & \\

39, 40 & \scriptsize{$\left(\bar{q}\gamma^{\nu}\bfd_{\mu\sigma}q\right)\left(G^{\mu\rho}D^{\sigma}G_{\nu\rho}\right)$} & $-$ & & & \scriptsize{$\left(\bar{Q}_{L}\gamma^{\nu}\bfd_{\mu\sigma}Q_{L}+\bar{u}_{R}\gamma^{\nu}\bfd_{\mu}u_{R}\right)\left(G^{\mu\rho}D^{\sigma}G_{\nu\rho}\right)$} & \\

41, 42 & \scriptsize{$\left(\bar{q}\gamma^{\nu}\gamma_{5}\bfd_{\mu\sigma}q\right)\left(G^{\mu\rho}D^{\sigma}G_{\nu\rho}\right)$} & $-$ & & & \scriptsize{$\left(\bar{Q}_{L}\gamma^{\nu}\bfd_{\mu\sigma}Q_{L}-\bar{u}_{R}\gamma^{\nu}\bfd_{\mu}u_{R}\right)\left(G^{\mu\rho}D^{\sigma}G_{\nu\rho}\right)$} & \\

43, 44 & \scriptsize{$\left(\bar{q}\gamma^{\mu}\bfd_{\rho\sigma}q\right)\left(G^{\nu\rho}D^{\sigma}\tilde{G}_{\mu\nu}\right)$} & $+$ & & & \scriptsize{$\left(\bar{Q}_{L}\gamma^{\mu}\bfd_{\rho\sigma}Q_{L}+\bar{u}_{R}\gamma^{\mu}\bfd_{\rho\sigma}u_{R}\right)\left(G^{\nu\rho}D^{\sigma}\tilde{G}_{\mu\nu}\right)$} & \\

45, 46 & \scriptsize{$\left(\bar{q}\gamma^{\mu}\gamma_{5}\bfd_{\rho\sigma}q\right)\left(G^{\nu\rho}D^{\sigma}\tilde{G}_{\mu\nu}\right)$} & $+$ & & & \scriptsize{$\left(\bar{Q}_{L}\gamma^{\mu}\bfd_{\rho\sigma}Q_{L}-\bar{u}_{R}\gamma^{\mu}\bfd_{\rho\sigma}u_{R}\right)\left(G^{\nu\rho}D^{\sigma}\tilde{G}_{\mu\nu}\right)$} & \\
\hline

47 & \scriptsize{$\left(\bar{q}\gamma^{\nu}\bfd_{\mu\sigma}q\right)\left(G^{\mu\rho}D^{\sigma}G_{\nu\rho}\right)$} & $+$ & \multirow{2}{*}{10} & \multirow{2}{*}{$f_{ABC}$} & \scriptsize{$\left(\bar{Q}_{L}\gamma^{\nu}\bfd_{\mu\sigma}Q_{L}+\bar{u}_{R}\gamma^{\nu}\bfd_{\mu\sigma}u_{R}\right)\left(G^{\mu\rho}D^{\sigma}G_{\nu\rho}\right)$} & \multirow{2}{*}{  $\frac{0.006}{E_\text{TeV}^6}$ } \\

48 & \scriptsize{$\left(\bar{q}\gamma^{\nu}\gamma_{5}\bfd_{\mu\sigma}q\right)\left(G^{\mu\rho}D^{\sigma}G_{\nu\rho}\right)$} & $+$ & & & \scriptsize{$\left(\bar{Q}_{L}\gamma^{\nu}\bfd_{\mu\sigma}Q_{L}-\bar{u}_{R}\gamma^{\nu}\bfd_{\mu\sigma}u_{R}\right)\left(G^{\mu\rho}D^{\sigma}G_{\nu\rho}\right)$} & \\
\hline

49, 50 & \scriptsize{$\left(\bar{q}\sigma_{\mu\nu}\bfd_{\sigma\alpha}q\right)\left(G^{\mu\rho}D^{\alpha}_{\ \rho}G^{\nu\sigma}\right)$} & $-$ & \multirow{2}{*}{11} & \multirow{2}{*}{$\delta_{AB}$, $d_{ABC}$} & \scriptsize{$\left(\bar{Q}_{L}\sigma_{\mu\nu}\bfd_{\sigma\alpha}\htilde u_{R}+\textrm{h.c.}\right)\left(G^{\mu\rho}D^{\alpha}_{\ \rho}G^{\nu\sigma}\right)$} & \multirow{2}{*}{  $\frac{0.001}{E_\text{TeV}^7}, \frac{0.004}{E_\text{TeV}^8}$ } \\

51, 52 & \scriptsize{$\left(i\bar{q}\sigma_{\mu\nu}\gamma_{5}\bfd_{\sigma\alpha}q\right)\left(G^{\mu\rho}D^{\alpha}_{\ \rho}G^{\nu\sigma}\right)$} & $+$ & & & \scriptsize{$\left(i\bar{Q}_{L}\sigma_{\mu\nu}\bfd_{\sigma\alpha}\htilde u_{R}+\textrm{h.c.}\right)\left(G^{\mu\rho}D^{\alpha}_{\ \rho}G^{\nu\sigma}\right)$} & \\
\hline

53 & \scriptsize{$\left(i\bar{q}\bfd_{\mu\nu\sigma}q\right)\left(G^{\mu\rho}D^{\sigma}G^{\nu}_{\ \rho}\right)$} & $+$ & \multirow{2}{*}{11} & \multirow{2}{*}{$f_{ABC}$} & \scriptsize{$\left(i\bar{Q}_{L}\bfd_{\mu\nu\sigma}\htilde u_{R}+\textrm{h.c.}\right)\left(G^{\mu\rho}\overset\leftrightarrow{D^{\sigma}}G^{\nu}_{\ \rho}\right)$} & \multirow{2}{*}{  $\frac{0.001}{E_\text{TeV}^7}, \frac{0.004}{E_\text{TeV}^8}$ } \\

54 & \scriptsize{$\left(\bar{q}\gamma_{5}\bfd_{\mu\nu\sigma}q\right)\left(G^{\mu\rho}D^{\sigma}G^{\nu}_{\ \rho}\right)$} & $-$ & & & \scriptsize{$\left(\bar{Q}_{L}\bfd_{\mu\nu\sigma}\htilde u_{R}+\textrm{h.c.}\right)\left(G^{\mu\rho}D^{\sigma}G^{\nu}_{\ \rho}\right)$} & \\
\hline
 
\end{tabular}
\begin{minipage}{5.7in}
\medskip
\caption{\label{tab:qqgg2}  \footnotesize Primary 10- and 11-dimension operators for $\bar{q}qgg$ interactions. There are three allowed $SU(3)$ contractions, 2 symmetric ones---$\delta_{AB}$ and $d_{ABC}$---and one antisymmetric one---$f_{ABC}$. Under the assumption that $\bar{q}$ and $q$ are each other's anti-particles, the operators are Hermitean and have the listed CP properties. If they are not, each of these operators has a Hermitean conjugate, which can be used to create a CP even and a CP odd operator. To simplify the expressions, we use the shorthand $\bfd_{\mu\nu}=\bfd_{\mu}\bfd_{\nu}$, and similarly $D_{\mu\nu}=D_{\mu}D_{\nu}$. To get the descendant operators, once can add contracted derivatives to get arbitrary Mandelstam factors of $s, (t-u)^2$.}
\end{minipage}
\end{center}
\end{adjustwidth}
\end{table}

\subsection{$ffff$ Amplitudes}

\begin{table}[p]
\begin{center}
\footnotesize
\centering
\renewcommand{\arraystretch}{0.9}
\tabcolsep6pt\begin{tabular}{|c|c|c|c|c|c|c|}
\hline
\multirow{2}{*}{$i$} & \multirow{2}{*}{$\mathcal{O}_i^{\bar{q}q\bar{\ell}\ell}$}   & \multirow{2}{*}{CP} & \multirow{2}{*}{$d_{\mathcal{O}_i}$}& SMEFT & $c$ Unitarity  \\ [-5pt]
&   && & Operator & Bound \\
\hline
1 &  $(\bar{q} q) (\bar{\ell}\ell) $ & +&    \multirow{4}{*}{ 6}&   $(\bar{Q}_L \tilde{H} u_R + \text{h.c.})(\bar{L}_L H e_R + \text{h.c.})$ &  \multirow{4}{*}{  $\frac{1.5}{E_\text{TeV}^2}, \frac{15}{E_\text{TeV}^4}$ }  \\
2 &  $(\bar{q} i \gamma_5 q) (\bar{\ell}\ell) $ &$-$&    &   $(i \bar{Q}_L \tilde{H} u_R + \text{h.c.})(\bar{L}_L H e_R + \text{h.c.})$ &  \\
3 &  $(\bar{q}  q) (\bar{\ell}i\gamma_5 \ell) $ & $-$&   &   $(\bar{Q}_L \tilde{H} u_R + \text{h.c.})(i \bar{L}_L H e_R + \text{h.c.})$ &   \\
4 &  $(\bar{q} i\gamma_5 q) (\bar{\ell}i\gamma_5 \ell) $ & $+$&   &   $(i\bar{Q}_L \tilde{H} u_R + \text{h.c.})(i \bar{L}_L H e_R + \text{h.c.})$&    \\ \hline 
5 &  $(\bar{q} \gamma^\mu q) (\bar{\ell}\gamma_\mu \ell) $ & +&  \multirow{4}{*}{ 6}  &   $(\bar{Q}_L \gamma^\mu Q_L + \bar{u}_R \gamma^\mu u_R)(\bar{L}_L \gamma_\mu L_L + \bar{e}_R \gamma_\mu e_R)$ &  \multirow{4}{*}{  $\frac{1.5}{E_\text{TeV}^2}$ } \\
6 &  $(\bar{q} \gamma^\mu \gamma_5 q) (\bar{\ell}\gamma_\mu \ell) $ & +&   &   $(\bar{Q}_L \gamma^\mu Q_L - \bar{u}_R \gamma^\mu u_R)(\bar{L}_L \gamma_\mu L_L + \bar{e}_R \gamma_\mu e_R)$ &    \\
7 &  $(\bar{q} \gamma^\mu  q) (\bar{\ell}\gamma_\mu \gamma_5 \ell) $ & +&   &   $(\bar{Q}_L \gamma^\mu Q_L + \bar{u}_R \gamma^\mu u_R)(\bar{L}_L \gamma_\mu L_L - \bar{e}_R \gamma_\mu e_R)$ &   \\
8 &  $(\bar{q} \gamma^\mu \gamma_5 q) (\bar{\ell}\gamma_\mu \gamma_5 \ell) $ & +&   &   $(\bar{Q}_L \gamma^\mu Q_L - \bar{u}_R \gamma^\mu u_R)(\bar{L}_L \gamma_\mu L_L - \bar{e}_R \gamma_\mu e_R)$ &   \\ \hline 
9 &  $(\bar{q} \sigma^{\mu\nu} q) (\bar{\ell}\sigma_{\mu\nu} \ell) $ & +&  \multirow{2}{*}{ 6}   &   $(\bar{Q}_L \sigma^{\mu\nu}\tilde{H} u_R + \text{h.c.})(\bar{L}_L \sigma_{\mu\nu}H e_R + \text{h.c.})$&  \multirow{2}{*}{  $\frac{1.5}{E_\text{TeV}^2}, \frac{15}{E_\text{TeV}^4}$ }   \\
10 &  $\epsilon_{\mu\nu\rho\sigma}(\bar{q} \sigma^{\mu\nu} q) (\bar{\ell}\sigma^{\rho\sigma} \ell) $ & $-$ &   &  $\epsilon_{\mu\nu\rho\sigma}(\bar{Q}_L \sigma^{\mu\nu}\tilde{H} u_R + \text{h.c.})(\bar{L}_L \sigma^{\rho\sigma}H e_R + \text{h.c.})$ &   \\
\hline 
11 &  $(\bar{q} \gamma^\mu q) (i \bar{\ell}\overset\leftrightarrow{D_\mu} \ell) $ & $+$ & \multirow{8}{*}{7}  &   $(\bar{Q}_L \gamma^\mu Q_L + \bar{u}_R \gamma^\mu u_R)(i \bar{L}_L  H \overset\leftrightarrow{D_\mu} e_R + \text{h.c.})$ &  \multirow{8}{*}{  $\frac{0.4}{E_\text{TeV}^3}, \frac{1.2}{E_\text{TeV}^4}$ }  \\
12 &  $(\bar{q} \gamma^\mu q)  ( \bar{\ell}  \gamma_5 \overset\leftrightarrow{D_\mu}  \ell) $ & $-$&   &   $(\bar{Q}_L \gamma^\mu Q_L + \bar{u}_R \gamma^\mu u_R)(\bar{L}_L H \overset\leftrightarrow{D_\mu}  e_R + \text{h.c.})$ &   \\
13 &  $(\bar{q} \gamma^\mu \gamma_5 q)  (i \bar{\ell} \overset\leftrightarrow{D_\mu}  \ell) $ & $+$ &   &   $(\bar{Q}_L \gamma^\mu Q_L - \bar{u}_R \gamma^\mu u_R)(i \bar{L}_L H \overset\leftrightarrow{D_\mu}  e_R + \text{h.c.})$ &   \\
14 &  $(\bar{q} \gamma^\mu \gamma_5 q)  (\bar{\ell} \gamma_5 \overset\leftrightarrow{D_\mu}  \ell) $ & $-$&   &   $(\bar{Q}_L \gamma^\mu Q_L - \bar{u}_R \gamma^\mu u_R)(\bar{L}_L H \overset\leftrightarrow{D_\mu}  e_R + \text{h.c.})$ &   \\

15 &  $ (i \bar{q}  \overset\leftrightarrow{D^\mu}   q) (\bar{\ell} \gamma_\mu \ell) $ & $+$ &   &   $(i\bar{Q}_L \tilde{H} \overset\leftrightarrow{D^\mu}  u_R + \text{h.c.})(\bar{L}_L \gamma_\mu L_L + \bar{e}_R \gamma_\mu e_R)$ &   \\
16 &  $ (\bar{q}  \gamma_5 \overset\leftrightarrow{D^\mu}  q) (\bar{\ell} \gamma_\mu \ell) $ & $-$&   &   $(\bar{Q}_L  \tilde{H}\overset\leftrightarrow{D^\mu}  u_R + \text{h.c.})(\bar{L}_L \gamma_\mu L_L + \bar{e}_R \gamma_\mu e_R)$ &   \\
17 &  $ (i \bar{q} \overset\leftrightarrow{D^\mu}  q) (\bar{\ell} \gamma_\mu \gamma_5 \ell) $ & $+$ &   &   $(i\bar{Q}_L  \tilde{H}\overset\leftrightarrow{D^\mu}  u_R + \text{h.c.})(\bar{L}_L \gamma_\mu L_L - \bar{e}_R \gamma_\mu e_R)$ &   \\
18 &  $ (\bar{q}  \gamma_5 \overset\leftrightarrow{D^\mu} q) (\bar{\ell} \gamma_\mu \gamma_5 \ell) $ & $-$&   &   $(\bar{Q}_L  \tilde{H}\overset\leftrightarrow{D^\mu} u_R + \text{h.c.})(\bar{L}_L \gamma_\mu L_L - \bar{e}_R \gamma_\mu e_R)$ &   \\
\hline
\end{tabular}
\medskip
\begin{minipage}{5.7in}
\caption{\label{tab:qqll} \footnotesize Primary operators for $\bar{q}q\bar{\ell}\ell$ interactions (As described in the text, these operators can be modified to yield the operators for baryon-lepton interactions $u u' d e$ and $ u d d' \nu$.).  Under the assumption the $\bar{q}, q$ and $\bar{\ell}, \ell$ are each other's anti-particles, the operators are Hermitean and have the listed CP properties.  If they are not, each of these operators has a Hermitean conjugate, which can be used to create a CP even and a CP odd operator.  To get the descendant operators, one can add contracted derivatives to get arbitrary Mandelstam factors of $s, t$.  At dimension 8, $s \mathcal{O}_9$ and $s\mathcal{O}_{10}$ become redundant and thus, one only needs to consider $\mathcal{O}_9$ and $\mathcal{O}_{10}$ with arbitrary factors of $t$. }
\end{minipage}
\end{center}
\end{table}

In Table~\ref{tab:qqll}, we've listed the primary operators for $\bar{q}q\bar{\ell}\ell$ interactions.  As the numerators of the Hilbert series suggests, there should be 10 primaries at dimension 6, 8 primaries at dimension 7, and at least two redundancies at dimension 8.  This is precisely what we find with the listed 18 operators, where at dimension 8, $s \mathcal{O}_9$ and $s\mathcal{O}_{10}$ are redundant to the other operators, where $s=(p_q+p_{\bar{q}})^2$.  Thus, for those two operators, one only needs their descendants $t^n \mathcal{O}_9$ and $t^n \mathcal{O}_{10}$.  We've listed a potential SMEFT operator to realize this interaction.  In some cases, a linear combination of the amplitudes may have a lower dimension SMEFT operator.  For example, $\bar{q}q\bar{\ell}\ell -\bar{q}i\gamma_5 q\bar{\ell}i\gamma_5 \ell$ can be realized by the SMEFT operator $(\epsilon^{ab} \bar{Q}_{L\, a} u_R \bar{L}_{L\, b} e_R + \text{h.c.})$.  This would affect the unitarity bound by removing the higher multiplicity bound of $15/E_\text{TeV}^4.$   We can also convert these operators to account for baryon-lepton interactions between $u u' d \bar{e}$ and $ u d d' \nu$.  The primes indicate different flavors and thus, we do not need to consider any issues with indistinguishable particles.  For example, $t c d e$ interactions can be found by replacing $\bar{q} \to \bar{t^c}, q\to c, \bar{\ell}\to \bar{e^c}, \ell \to d$ where $t^c$ and $e^c$ are the charge conjugated 4-component spinor for the top quark and the electron and the $SU(3)$ indices are contracted with an epsilon tensor.   For the baryon-neutrino coupling, the number of operators would depend on whether the neutrino is Majorana or Dirac, where the Dirac case has twice the operators, since one can use either $\bar{\nu}$ or $\bar{\nu^c}$.   

\begin{table}[p]
\begin{adjustwidth}{-.5in}{-.5in}  
\begin{center}
\scriptsize
\centering
\renewcommand{\arraystretch}{0.9}
\tabcolsep6pt\begin{tabular}{|c|c|c|c|c|c|}
\hline
\multirow{2}{*}{$i$} & \multirow{2}{*}{$\mathcal{O}_i^{uude}$}    & \multirow{2}{*}{$d_{\mathcal{O}_i}$}& SMEFT & $c$ Unitarity  \\[-5pt]
&   &&  Operator & Bound \\
\hline
1 &  $(\bar{d^c} u) (\bar{e^c}u) $ &    \multirow{4}{*}{6}&   $(\bar{d}^c_R \tilde{H}^\dag Q_L+\bar{Q}^c_L  H u_R)(\bar{e}^c_R \tilde{H}^\dag Q_L+\bar{L}^c_L H u_R)$ & \multirow{4}{*}{  $\frac{1.5}{E_\text{TeV}^2}, \frac{15}{E_\text{TeV}^4}$ } \\
2 &  $(\bar{d^c} i \gamma_5 u) (\bar{e^c}u) $ &    &   $(i\bar{d}^c_R \tilde{H}^\dag Q_L-i \bar{Q}^c_L  H u_R)(\bar{e}^c_R \tilde{H}^\dag Q_L+\bar{L}^c_L H u_R)$  &   \\
3 &  $(\bar{d^c}  u) (\bar{e^c}i \gamma_5 u) $ &    &    $(\bar{d}^c_R \tilde{H}^\dag Q_L+\bar{Q}^c_L  H u_R)(i\bar{e}^c_R \tilde{H}^\dag Q_L-i\bar{L}^c_L H u_R)$ &   \\
4 &  $(\bar{d^c}  i\gamma_5 u) (\bar{e^c}i \gamma_5 u) $ &    &    $(i\bar{d}^c_R \tilde{H}^\dag Q_L-i\bar{Q}^c_L  H u_R)(i\bar{e}^c_R \tilde{H}^\dag Q_L-i\bar{L}^c_L H u_R)$ &   \\
\hline 
5 &  $(\bar{d^c} \gamma^\mu u) (\bar{e^c}D_\mu u) $ &    \multirow{6}{*}{7}&  $(\bar{Q}^c_L \gamma^\mu Q_L+\bar{d}^c_R  \gamma^\mu u_R)(\bar{e}^c_R D_\mu \tilde{H}^\dag Q_L+\bar{L}^c_L H D_\mu  u_R)$  &  \multirow{6}{*}{  $\frac{0.4}{E_\text{TeV}^3}, \frac{1.2}{E_\text{TeV}^4}$ }    \\
6 &  $(\bar{d^c} \gamma^\mu u) (\bar{e^c}i\gamma_5 D_\mu u) $ &    &   $(\bar{Q}^c_L \gamma^\mu Q_L+\bar{d}^c_R  \gamma^\mu u_R)(i \bar{e}^c_R D_\mu \tilde{H}^\dag Q_L-i\bar{L}^c_L H D_\mu  u_R)$ &   \\
7 &  $(\bar{d^c} \gamma^\mu\gamma_5 u) (\bar{e^c}D_\mu u)$  &    &   $(\bar{Q}^c_L \gamma^\mu Q_L-\bar{d}^c_R  \gamma^\mu u_R)(\bar{e}^c_R D_\mu \tilde{H}^\dag Q_L+\bar{L}^c_L H D_\mu  u_R)$ &   \\
8 &  $(\bar{d^c} \gamma^\mu\gamma_5 u) (\bar{e^c}i\gamma_5 D_\mu u)$  &    &   $(\bar{Q}^c_L \gamma^\mu Q_L-\bar{d}^c_R  \gamma^\mu u_R)(i \bar{e}^c_R D_\mu \tilde{H}^\dag Q_L-i\bar{L}^c_L H D_\mu  u_R)$  &   \\
9 &  $(\bar{d^c} D_\mu u) (\bar{e^c}\gamma^\mu u)$  &    &   $(\bar{d}^c_R D_\mu \tilde{H}^\dag Q_L+\bar{Q}^c_L  H D_\mu  u_R)(\bar{L}^c_L \gamma^\mu Q_L + \bar{e}^c_R \gamma^\mu u_R)$ &   \\
10 &  $(\bar{d^c}i \gamma_5  D_\mu u) (\bar{e^c}\gamma^\mu u)$  &    &   $(i\bar{d}^c_R D_\mu \tilde{H}^\dag Q_L-i\bar{Q}^c_L  H D_\mu  u_R)(\bar{L}^c_L \gamma^\mu Q_L + \bar{e}^c_R \gamma^\mu u_R)$  &   \\
\hline 

11 &  $(\bar{d^c} u) ([D^\mu \bar{e^c}]\overset\leftrightarrow{D^u_\mu}u) $ &    \multirow{4}{*}{8}&  $(\bar{d}^c_R \tilde{H}^\dag Q_L+\bar{Q}^c_L  H u_R)([D^\mu \bar{e}^c_R]  \overset\leftrightarrow{D}\vphantom{D}^{Q,u}_\mu \tilde{H}^\dag Q_L+[D^\mu \bar{L}^c_L H]\overset\leftrightarrow{D}\vphantom{D}^{Q,u}_\mu u_R)$  &  \multirow{4}{*}{  $\frac{0.09}{E_\text{TeV}^4}, \frac{0.9}{E_\text{TeV}^6}$ }  \\
12 &  $(\bar{d^c} i \gamma_5 u) ([D^\mu\bar{e^c}]\overset\leftrightarrow{D^u_\mu}u) $ &    &   $(i\bar{d}^c_R \tilde{H}^\dag Q_L-i\bar{Q}^c_L  H u_R)([D^\mu \bar{e}^c_R] \overset\leftrightarrow{D}\vphantom{D}^{Q,u}_\mu  \tilde{H}^\dag Q_L+[D^\mu \bar{L}^c_L H ]\overset\leftrightarrow{D}\vphantom{D}^{Q,u}_\mu u_R)$ &   \\
13 &  $(\bar{d^c}  u) ([D^\mu\bar{e^c}]i \gamma_5 \overset\leftrightarrow{D^u_\mu}u) $ &    &   $(\bar{d}^c_R \tilde{H}^\dag Q_L+\bar{Q}^c_L  H u_R)(i [D^\mu \bar{e}^c_R]  \overset\leftrightarrow{D}\vphantom{D}^{Q,u}_\mu  \tilde{H}^\dag Q_L-i [D^\mu \bar{L}^c_L H ]\overset\leftrightarrow{D}\vphantom{D}^{Q,u}_\mu u_R)$ &   \\
14 &  $(\bar{d^c}  i\gamma_5 u) ([D^\mu\bar{e^c}]i \gamma_5 \overset\leftrightarrow{D^u_\mu}u) $ &    &  $(i \bar{d}^c_R \tilde{H}^\dag Q_L-i \bar{Q}^c_L  H u_R)(i [D^\mu \bar{e}^c_R]  \overset\leftrightarrow{D}\vphantom{D}^{Q,u}_\mu \tilde{H}^\dag Q_L-i [D^\mu \bar{L}^c_L H] \overset\leftrightarrow{D}\vphantom{D}^{Q,u}_\mu u_R)$   &   \\ \hline

15 &  $(\bar{d^c}\gamma^\mu u) ([D^\nu\bar{e^c}]\gamma_\mu \overset\leftrightarrow{D^u_\nu}u) $ &  \multirow{2}{*}{8}   &   $(\bar{Q}^c_L \gamma^\mu Q_L+\bar{d}^c_R  \gamma^\mu u_R) ([D^\nu \bar{L}^c_L] \gamma_\mu \overset\leftrightarrow{D}\vphantom{D}^{Q,u}_\nu Q_L + [D^\nu\bar{e}^c_R] \gamma_\mu \overset\leftrightarrow{D}\vphantom{D}^{Q,u}_\nu u_R)$ &   \multirow{2}{*}{  $\frac{0.09}{E_\text{TeV}^4}$}  \\

16 &  $(\bar{d^c}\gamma^\mu \gamma_5 u) ([D^\nu\bar{e^c}]\gamma_\mu \overset\leftrightarrow{D^u_\nu}u) $ &   &   $(\bar{Q}^c_L \gamma^\mu Q_L-\bar{d}^c_R  \gamma^\mu u_R) ([D^\nu \bar{L}^c_L] \gamma_\mu \overset\leftrightarrow{D}\vphantom{D}^{Q,u}_\nu Q_L + [D^\nu\bar{e}^c_R] \gamma_\mu \overset\leftrightarrow{D}\vphantom{D}^{Q,u}_\nu u_R) $ &  \\
\hline 

17 &  $(\bar{d^c} \gamma^\mu u) ([D^\nu \bar{e^c}]\overset\leftrightarrow{D^u_\nu}D_\mu u)$ &    \multirow{2}{*}{9}&   $(\bar{Q}^c_L \gamma^\mu Q_L+\bar{d}^c_R  \gamma^\mu u_R) ([D^\nu \bar{e}^c_R] \overset\leftrightarrow{D}\vphantom{D}^{Q,u}_\nu D_\mu \tilde{H}^\dag  Q_L+[D^\nu \bar{L}^c_L  H] \overset\leftrightarrow{D}\vphantom{D}^{Q,u}_\nu D_\mu u_R)$ &  \multirow{2}{*}{  $\frac{0.02}{E_\text{TeV}^5}, \frac{0.07}{E_\text{TeV}^6}$ }  \\

18 &  $(\bar{d^c} \gamma^\mu u) ([D^\nu \bar{e^c}]i \gamma_5 \overset\leftrightarrow{D^u_\nu}D_\mu u)$ &    &   $(\bar{Q}^c_L \gamma^\mu Q_L+\bar{d}^c_R  \gamma^\mu u_R)(i [D^\nu \bar{e}^c_R] \overset\leftrightarrow{D}\vphantom{D}^{Q,u}_\nu   D_\mu \tilde{H}^\dag Q_L-i [D^\nu \bar{L}^c_L H] \overset\leftrightarrow{D}\vphantom{D}^{Q,u}_\nu D_\mu  u_R) $ &   \\
\hline
\end{tabular}
\medskip
\begin{minipage}{6in}
\caption{\label{tab:uude}  \footnotesize Primary operators for $uude$ interactions, where $d^c$ and $e^c$ are the charge conjugated down-type quark and charged lepton 4-component spinor and $SU(3)$ indices are contracted with an epsilon tensor  (These operators can be modified to yield the operators for $udd\nu$ interactions by simply taking $u\to d, \bar{d^c} \to \bar{u^c}, \bar{e^c} \to \bar{\nu}/\bar{\nu^c}$.).  To simplify the expressions, we've defined a back-forth derivative $\overset\leftrightarrow{D^u_\mu}$, which only acts on the $u$ fields, and similarly $\overset\leftrightarrow{D}\vphantom{D}^{u,Q}_\mu$ which acts on $u_R$ and $Q_{L}$ (but not $\bar{Q}^c_L$).  To get the descendant operators, one can add contracted derivatives to get arbitrary Mandelstam factors that respect the exchange symmetry between the two up-type quarks, i.e.~$s, (t-u)^2$.  At dimension 8, $s \mathcal{O}_3$ and $s\mathcal{O}_{4}$ become redundant and thus, one only needs to consider $\mathcal{O}_3$ and $\mathcal{O}_{4}$ descendants with arbitrary factors of $(t-u)^2$. }
\end{minipage}
\end{center}
\end{adjustwidth}
\end{table}

In Table~\ref{tab:uude}, we've listed the primary operators for $uude$ interactions, where all $SU(3)$ indices are contracted by an epsilon tensor.  As the Hilbert series suggests, there should be 4 primaries at dimension 6, 6 primaries at dimension 7, 6 primaries with 2 redundancies at dimension 8, and 2 primaries at dimension 9.  The table shows the stated number of independent primaries and we find that  at dimension 8, $s \mathcal{O}_3$ and $s\mathcal{O}_{4}$ are redundant to the other operators, where $s=(p_u+p_{\bar{u}})^2$.  Thus, for those two, one only needs their descendants $(t-u)^{2n} \mathcal{O}_3$ and $(t-u)^{2n} \mathcal{O}_{4}$.  To account for $ u d d\nu$ interactions, one replaces $u\to d, \bar{d^c}\to \bar{u^c}, \bar{e^c}\to \bar{\nu}/\bar{\nu^c},$ where again the case of Dirac neutrinos allows twice as many operators.

\begin{table}[p]
\begin{center}
\footnotesize
\centering
\renewcommand{\arraystretch}{0.9}
\tabcolsep6pt\begin{tabular}{|c|c|c|c|c|c|c|}
\hline
\multirow{2}{*}{$i$} & \multirow{2}{*}{$\mathcal{O}_{i, S/A}^{\bar{q}q\bar{q}'q'}$}   & \multirow{2}{*}{CP} & \multirow{2}{*}{$d_{\mathcal{O}_i}$}& SMEFT & $c$ Unitarity  \\[-5pt]
&   && & Operator & Bound \\
\hline
1 &  $(\bar{q} q) (\bar{q}' q') $ & +&    \multirow{4}{*}{ 6}&   $(\bar{Q}_L \tilde{H} u_R + \text{h.c.})(\bar{Q}'_L \tilde H u'_R + \text{h.c.})$ &   \multirow{4}{*}{  $\frac{1.5}{E_\text{TeV}^2}, \frac{15}{E_\text{TeV}^4}$ }    \\
2 &  $(\bar{q} i \gamma_5 q) (\bar{q}'q') $ &$-$&    &   $(i\bar{Q}_L \tilde{H} u_R + \text{h.c.})(\bar{Q}'_L \tilde H u'_R + \text{h.c.})$ &   \\
3 &  $(\bar{q}  q) (\bar{q}'i\gamma_5  q') $ & $-$&   &   $(\bar{Q}_L \tilde{H} u_R + \text{h.c.})(i\bar{Q}'_L \tilde H u'_R + \text{h.c.})$ &   \\
4 &  $(\bar{q} i\gamma_5 q) (\bar{q}'i\gamma_5 q') $ & $+$&   &   $(i\bar{Q}_L \tilde{H} u_R + \text{h.c.})(i\bar{Q}'_L \tilde H u'_R + \text{h.c.})$ &   \\ \hline
5 &  $(\bar{q} \gamma^\mu q) (\bar{q}'\gamma_\mu q') $ & +& \multirow{4}{*}{ 6}  &   $(\bar{Q}_L \gamma^\mu Q_L + \bar{u}_R \gamma^\mu u_R)(\bar{Q}'_L \gamma_\mu Q'_L + \bar{u}'_R \gamma_\mu u'_R)$ &   \multirow{4}{*}{  $\frac{1.5}{E_\text{TeV}^2}$ }   \\
6 &  $(\bar{q} \gamma^\mu \gamma_5 q) (\bar{q}'\gamma_\mu q') $ & +&   &   $(\bar{Q}_L \gamma^\mu Q_L - \bar{u}_R \gamma^\mu u_R)(\bar{Q}'_L \gamma_\mu Q'_L + \bar{u}'_R \gamma_\mu u'_R)$ &   \\
7 &  $(\bar{q} \gamma^\mu  q) (\bar{q}'\gamma_\mu \gamma_5 q') $ & +&   &   $(\bar{Q}_L \gamma^\mu Q_L + \bar{u}_R \gamma^\mu u_R)(\bar{Q}'_L \gamma_\mu Q'_L - \bar{u}'_R \gamma_\mu u'_R)$ &   \\
8 &  $(\bar{q} \gamma^\mu \gamma_5 q) (\bar{q}'\gamma_\mu \gamma_5 q') $ & +&   &   $(\bar{Q}_L \gamma^\mu Q_L - \bar{u}_R \gamma^\mu u_R)(\bar{Q}'_L \gamma_\mu Q'_L - \bar{u}'_R \gamma_\mu u'_R)$ &   \\ \hline
9 &  $(\bar{q} \sigma^{\mu\nu} q) (\bar{q}'\sigma_{\mu\nu} q') $ & +& \multirow{2}{*}{ 6}  &   $(\bar{Q}_L \sigma^{\mu\nu}\tilde{H} u_R + \text{h.c.})(\bar{Q}'_L \sigma_{\mu\nu}\tilde{H} u'_R + \text{h.c.})$ & \multirow{2}{*}{  $\frac{1.5}{E_\text{TeV}^2}, \frac{15}{E_\text{TeV}^4}$ }   \\
10 &  $\epsilon_{\mu\nu\rho\sigma}(\bar{q} \sigma^{\mu\nu} q) (\bar{q}'\sigma^{\rho\sigma} q') $ & $-$ &   &    $\epsilon_{\mu\nu\rho\sigma}(\bar{Q}_L \sigma^{\mu\nu}\tilde{H} u_R + \text{h.c.})(\bar{Q}'_L \sigma^{\rho\sigma}\tilde{H} u'_R + \text{h.c.})$ &   \\
\hline 
11 &  $(\bar{q} \gamma^\mu q) (i \bar{q}' \overset\leftrightarrow{D_\mu}   q') $ & $+$& \multirow{8}{*}{7}  &    $(\bar{Q}_L \gamma^\mu Q_L + \bar{u}_R \gamma^\mu u_R)(i \bar{Q}'_L  \tilde{H} \overset\leftrightarrow{D_\mu} u'_R + \text{h.c.})$ &   \multirow{8}{*}{  $\frac{0.4}{E_\text{TeV}^3}, \frac{1.2}{E_\text{TeV}^4}$ }  \\
12 &  $(\bar{q} \gamma^\mu q) ( \bar{q}'\gamma_5 \overset\leftrightarrow{D_\mu}  q') $ & $-$&   &  $(\bar{Q}_L \gamma^\mu Q_L + \bar{u}_R \gamma^\mu u_R)(\bar{Q}'_L  \tilde{H} \overset\leftrightarrow{D_\mu}  u'_R + \text{h.c.})$ &   \\
13 &  $(\bar{q} \gamma^\mu \gamma_5 q) (i \bar{q}' \overset\leftrightarrow{D_\mu} q') $ & $+$&   &   $(\bar{Q}_L \gamma^\mu Q_L - \bar{u}_R \gamma^\mu u_R)(i \bar{Q}'_L  \tilde{H} \overset\leftrightarrow{D_\mu}  u'_R + \text{h.c.})$ &   \\
14 &  $(\bar{q} \gamma^\mu \gamma_5 q) (\bar{q}' \gamma_5 \overset\leftrightarrow{D_\mu}   q') $ & $-$&   &   $(\bar{Q}_L \gamma^\mu Q_L - \bar{u}_R \gamma^\mu u_R)( \bar{Q}'_L  \tilde{H} \overset\leftrightarrow{D_\mu}  u'_R + \text{h.c.})$ &   \\
15 &  $(i \bar{q}\overset\leftrightarrow{D^\mu}  q) (\bar{q}' \gamma_\mu q') $ & $+$&   &   $(i\bar{Q}_L   \tilde{H} \overset\leftrightarrow{D^\mu}u_R + \text{h.c.})(\bar{Q}'_L \gamma_\mu Q'_L + \bar{u}'_R \gamma_\mu u'_R)$ &   \\
16 &  $(\bar{q} \gamma_5 \overset\leftrightarrow{D^\mu}  q) (\bar{q}' \gamma_\mu q') $ & $-$&   &   $(\bar{Q}_L  \tilde{H}  \overset\leftrightarrow{D^\mu}  u_R + \text{h.c.})(\bar{Q}'_L \gamma_\mu Q'_L + \bar{u}'_R \gamma_\mu u'_R)$  &   \\
17 &  $(i \bar{q} \overset\leftrightarrow{D^\mu} q) (\bar{q}' \gamma_\mu \gamma_5 q') $ & $+$&   &    $(i\bar{Q}_L \tilde{H} \overset\leftrightarrow{D^\mu}  u_R + \text{h.c.})(\bar{Q}'_L \gamma_\mu Q'_L - \bar{u}'_R \gamma_\mu u'_R)$  &   \\
18 &  $(\bar{q} \gamma_5 \overset\leftrightarrow{D^\mu}  q) (\bar{q}'\gamma_\mu \gamma_5 q') $ & $-$&   &    $(\bar{Q}_L  \tilde{H}\overset\leftrightarrow{D^\mu}  u_R + \text{h.c.})(\bar{Q}'_L \gamma_\mu Q'_L - \bar{u}'_R \gamma_\mu u'_R)$  &   \\
\hline
\end{tabular}
\medskip
\caption{\label{tab:qqqpqp}  \footnotesize Primary operators for $\bar{q}q\bar{q}'q'$ interactions.  There are two allowed $SU(3)$ contractions, the $S$ indicates where $q, q'$ form a symmetric 6 representation under $SU(3)$, while $A$ has the antisymmetric $\bar{3}$ representation.  For example, with explicit indices we have $\mathcal{O}_{1, S}^{\bar{q}q\bar{q}'q'}= (\bar{q}^{\{\alpha} q_{\{\alpha}) (\bar{q}\,'^{\beta\}}q'_{\beta\}}) $ and $\mathcal{O}_{1, A}^{\bar{q}q\bar{q}'q'}= (\bar{q}^{[\alpha} q_{[\alpha}) (\bar{q}\,'^{\beta]}q'_{\beta]}) $, where $q_{\{\alpha} q_{\beta\}}=q_{\alpha} q_{\beta}+q_{\beta} q_{\alpha}$ and  $q_{[\alpha} q_{\beta]}=q_{\alpha} q_{\beta}-q_{\beta} q_{\alpha}$.  Under the assumption the $\bar{q}, q$ and $\bar{q}', q'$ are resprectively each other's anti-particles, the operators are Hermitean and have the listed CP properties.  If they are not, each of these operators has a Hermitean conjugate, which can be used to create a CP even and a CP odd operator.  To get the descendant operators, one can add contracted derivatives to get arbitrary Mandelstam factors of $s, t$.  At dimension 8, $s \mathcal{O}_9$ and $s\mathcal{O}_{10}$ become redundant and thus, one only needs to consider $\mathcal{O}_9$ and $\mathcal{O}_{10}$ with arbitrary factors of $t$.            }
\end{center}
\end{table}

In Table~\ref{tab:qqqpqp}, we've listed the primary operators for $\bar{q}q\bar{q'}q'$ interactions.  Notably the Hilbert series for this has a numerator that is twice the $\bar{q}q\bar{\ell}\ell$ Hilbert series.  This factor of two is simply for the two allowed $SU(3)$ contractions, one where the $qq'$ are either in the $6$ or $\bar{3}$ representation, leading to the symmetric ($S$) and antisymmetric ($A$) operators.   Again,  at dimension 8, $s \mathcal{O}_9$ and $s\mathcal{O}_{10}$ are redundant to the other operators, where $s=(p_q+p_{\bar{q}})^2$.  Thus one only needs to add their descendants $t^n \mathcal{O}_9$ and $t^n \mathcal{O}_{10}$.    

\begin{table}[p]
\begin{adjustwidth}{-.5in}{-.5in}  
\begin{center}
\scriptsize
\centering
\renewcommand{\arraystretch}{0.85}
\tabcolsep6pt\begin{tabular}{|c|c|c|c|c|c|c|}
\hline
\multirow{2}{*}{$i$} & \multirow{2}{*}{$\mathcal{O}_i^{uu\bar{t}\bar{c}}$}  &  \multirow{2}{*}{$d_{\mathcal{O}_i}$}&  \multirow{2}{*}{$SU(3)$}  & SMEFT & $c$ Unitarity  \\[-5pt]
&   & &  &  Operator & Bound \\
\hline
1 &  $(\bar{t} u) (\bar{c}u) $ &    \multirow{4}{*}{6}&  \multirow{4}{*}{A} &  $(\bar{Q}_{3L} \tilde{H} u_R + \bar{t}_{R} \tilde{H}^\dag Q_{1L})(\bar{Q}_{2L} \tilde{H} u_R + \bar{c}_{R} \tilde{H}^\dag Q_{1L})$ &   \multirow{4}{*}{  $\frac{1.5}{E_\text{TeV}^2}, \frac{15}{E_\text{TeV}^4}$ }    \\
2 &  $(\bar{t} i \gamma_5 u) (\bar{c}u) $ &  &    &   $(i\bar{Q}_{3L} \tilde{H} u_R -i \bar{t}_{R} \tilde{H}^\dag Q_{1L})(\bar{Q}_{2L} \tilde{H} u_R + \bar{c}_{R} \tilde{H}^\dag Q_{1L})$ &   \\
3 &  $(\bar{t}  u) (\bar{c}i \gamma_5 u) $ &  &    &   $(\bar{Q}_{3L} \tilde{H} u_R + \bar{t}_{R} \tilde{H}^\dag Q_{1L})(i\bar{Q}_{2L} \tilde{H} u_R -i \bar{c}_{R} \tilde{H}^\dag Q_{1L})$ &   \\
4 &  $(\bar{t}  i\gamma_5 u) (\bar{c}i \gamma_5 u) $ &  &    &   $(i\bar{Q}_{3L} \tilde{H} u_R - i\bar{t}_{R} \tilde{H}^\dag Q_{1L})(i\bar{Q}_{2L} \tilde{H} u_R +i \bar{c}_{R} \tilde{H}^\dag Q_{1L})$ &   \\
\hline 
5 &  $(\bar{t} u) (\bar{c}u) $ &    \multirow{4}{*}{6}&  \multirow{4}{*}{S} &  $(\bar{Q}_{3L} \tilde{H} u_R + \bar{t}_{R} \tilde{H}^\dag Q_{1L})(\bar{Q}_{2L} \tilde{H} u_R + \bar{c}_{R} \tilde{H}^\dag Q_{1L})$ &  \multirow{4}{*}{  $\frac{1.5}{E_\text{TeV}^2}, \frac{15}{E_\text{TeV}^4}$ }    \\
6 &  $(\bar{t} i \gamma_5 u) (\bar{c}u) $ &  &    &   $(i\bar{Q}_{3L} \tilde{H} u_R -i \bar{t}_{R} \tilde{H}^\dag Q_{1L})(\bar{Q}_{2L} \tilde{H} u_R + \bar{c}_{R} \tilde{H}^\dag Q_{1L})$  &   \\
7 &  $(\bar{t}  u) (\bar{c}i \gamma_5 u) $ &  &    &   $(\bar{Q}_{3L} \tilde{H} u_R + \bar{t}_{R} \tilde{H}^\dag Q_{1L})(i\bar{Q}_{2L} \tilde{H} u_R -i \bar{c}_{R} \tilde{H}^\dag Q_{1L})$ &   \\
8 &  $(\bar{t}  i\gamma_5 u) (\bar{c}i \gamma_5 u) $ &  &    &   $(i\bar{Q}_{3L} \tilde{H} u_R - i\bar{t}_{R} \tilde{H}^\dag Q_{1L})(i\bar{Q}_{2L} \tilde{H} u_R + i\bar{c}_{R} \tilde{H}^\dag Q_{1L})$  &   \\ \hline
9 &  $(\bar{t}  \gamma^\mu u) (\bar{c} \gamma_\mu u) $ & \multirow{2}{*}{6} & \multirow{2}{*}{S}   &   $(\bar{Q}_{3L} \gamma^\mu Q_{1L} + \bar{t}_R \gamma^\mu u_R)(\bar{Q}_{2L} \gamma_\mu Q_{1L} + \bar{c}_R \gamma_\mu u_R)$ &   \multirow{2}{*}{  $\frac{1.5}{E_\text{TeV}^2}$}    \\
10 &  $(\bar{t}  \gamma^\mu \gamma_5 u) (\bar{c} \gamma_\mu u) $ &  &    &   $(\bar{Q}_{3L} \gamma^\mu Q_{1L} - \bar{t}_R \gamma^\mu u_R)(\bar{Q}_{2L} \gamma_\mu Q_{1L} + \bar{c}_R \gamma_\mu u_R)$ &   \\
\hline 
11 &  $(\bar{t} \gamma^\mu u) (\bar{c}D_\mu u) $ &    \multirow{6}{*}{7}&  \multirow{6}{*}{A} &   $(\bar{Q}_{3L} \gamma^\mu Q_{1L} + \bar{t}_R \gamma^\mu u_R)(\bar{Q}_{2L} \tilde{H} D_\mu  u_R + \bar{c}_{R} D_\mu \tilde{H}^\dag Q_{1L})$ &  \multirow{6}{*}{  $\frac{0.4}{E_\text{TeV}^3}, \frac{1.2}{E_\text{TeV}^4}$ }  \\
12 &  $(\bar{t} \gamma^\mu u) (\bar{c}i\gamma_5 D_\mu u) $ & &   &   $(\bar{Q}_{3L} \gamma^\mu Q_{1L} + \bar{t}_R \gamma^\mu u_R)(i \bar{Q}_{2L}  \tilde{H} D_\mu u_R -i \bar{c}_{R} D_\mu \tilde{H}^\dag Q_{1L})$ &   \\
13 &  $(\bar{t} \gamma^\mu\gamma_5 u) (\bar{c}D_\mu u)$  &   & &   $(\bar{Q}_{3L} \gamma^\mu Q_{1L} - \bar{t}_R \gamma^\mu u_R)(\bar{Q}_{2L} \tilde{H} D_\mu  u_R + \bar{c}_{R} D_\mu \tilde{H}^\dag Q_{1L})$ &   \\
14 &  $(\bar{t} \gamma^\mu\gamma_5 u) (\bar{c}i\gamma_5 D_\mu u)$  & &   &   $(\bar{Q}_{3L} \gamma^\mu Q_{1L} - \bar{t}_R \gamma^\mu u_R)(i \bar{Q}_{2L}  \tilde{H} D_\mu u_R -i \bar{c}_{R} D_\mu \tilde{H}^\dag Q_{1L})$ &   \\
15 &  $(\bar{t} D_\mu u) (\bar{c}\gamma^\mu u)$  &  &  &   $(\bar{Q}_{3L} \tilde{H} D_\mu  u_R + \bar{t}_{R} D_\mu \tilde{H}^\dag Q_{1L})(\bar{Q}_{2L} \gamma^\mu Q_{1L} + \bar{c}_R \gamma^\mu u_R)$ &   \\
16 &  $(\bar{t}i \gamma_5 D_\mu u) (\bar{c}\gamma^\mu u)$  & &   &  $(i\bar{Q}_{3L}  \tilde{H} D_\mu u_R -i \bar{t}_{R} D_\mu \tilde{H}^\dag Q_{1L})(\bar{Q}_{2L} \gamma^\mu Q_{1L} + \bar{c}_R \gamma^\mu u_R)$ &   \\
\hline 
17 &  $(\bar{t} \gamma^\mu u) (\bar{c}D_\mu u) $ &    \multirow{2}{*}{7}&  \multirow{2}{*}{S} &   $(\bar{Q}_{3L} \gamma^\mu Q_{1L} + \bar{t}_R \gamma^\mu u_R)(\bar{Q}_{2L} \tilde{H} D_\mu  u_R + \bar{c}_{R} D_\mu \tilde{H}^\dag Q_{1L})$ & \multirow{2}{*}{  $\frac{0.4}{E_\text{TeV}^3}, \frac{1.2}{E_\text{TeV}^4}$ }  \\
18 &  $(\bar{t}\gamma^\mu u) (\bar{c}i \gamma_5  D_\mu u)$  & &   &    $(\bar{Q}_{3L} \gamma^\mu Q_{1L} + \bar{t}_R \gamma^\mu u_R)(i \bar{Q}_{2L} \tilde{H} D_\mu  u_R -i \bar{c}_{R} D_\mu \tilde{H}^\dag Q_{1L})$ &   \\
\hline
19 &  $(\bar{t} u) ([D^\mu \bar{c}]\overset\leftrightarrow{D^u_\mu}u) $ &    \multirow{4}{*}{8}&    \multirow{4}{*}{A} &   $(\bar{Q}_{3L} \tilde{H} u_R + \bar{t}_{R} \tilde{H}^\dag Q_{1L})([D^\mu \bar{Q}_{2L} \tilde{H}] \overset\leftrightarrow{D}\vphantom{D}^{u,Q_1}_\mu u_R + [D^\mu \bar{c}_{R}]  \overset\leftrightarrow{D}\vphantom{D}^{u,Q_1}_\mu \tilde{H}^\dag  Q_{1L})$ &  \multirow{4}{*}{  $\frac{0.09}{E_\text{TeV}^4}, \frac{0.9}{E_\text{TeV}^6}$ }   \\
20 &  $(\bar{t} i \gamma_5 u) ([D^\mu\bar{c}]\overset\leftrightarrow{D^u_\mu}u) $ & &    &  $(i\bar{Q}_{3L} \tilde{H} u_R -i \bar{t}_{R} \tilde{H}^\dag Q_{1L})([D^\mu \bar{Q}_{2L} \tilde{H}] \overset\leftrightarrow{D}\vphantom{D}^{u,Q_1}_\mu u_R + [D^\mu \bar{c}_{R}]   \overset\leftrightarrow{D}\vphantom{D}^{u,Q_1}_\mu \tilde{H}^\dag Q_{1L})$ &   \\
21 &  $(\bar{t}  u) ([D^\mu\bar{c}]i \gamma_5 \overset\leftrightarrow{D^u_\mu}u) $ &  &   &   $(\bar{Q}_{3L} \tilde{H} u_R + \bar{t}_{R} \tilde{H}^\dag Q_{1L})(i [D^\mu \bar{Q}_{2L} \tilde{H}] \overset\leftrightarrow{D}\vphantom{D}^{u,Q_1}_\mu u_R -i [D^\mu \bar{c}_{R}]  \overset\leftrightarrow{D}\vphantom{D}^{u,Q_1}_\mu \tilde{H}^\dag  Q_{1L})$ &   \\
22 &  $(\bar{t}  i\gamma_5 u) ([D^\mu\bar{c}]i \gamma_5 \overset\leftrightarrow{D^u_\mu}u) $ & &    &   $(i\bar{Q}_{3L} \tilde{H} u_R -i \bar{t}_{R} \tilde{H}^\dag Q_{1L})(i [D^\mu \bar{Q}_{2L} \tilde{H}] \overset\leftrightarrow{D}\vphantom{D}^{u,Q_1}_\mu u_R -i [D^\mu \bar{c}_{R}]   \overset\leftrightarrow{D}\vphantom{D}^{u,Q_1}_\mu \tilde{H}^\dag Q_{1L})$ &   \\ \hline
23 &  $(\bar{t}\gamma^\mu u) ([D^\nu\bar{c}]\gamma_\mu \overset\leftrightarrow{D^u_\nu}u) $ & \multirow{2}{*}{8} &  \multirow{2}{*}{A}  &   $(\bar{Q}_{3L} \gamma^\mu Q_{1L} + \bar{t}_R \gamma^\mu u_R)([D^\nu \bar{Q}_{2L}] \gamma_\mu \overset\leftrightarrow{D}\vphantom{D}^{u,Q_1}_\nu Q_{1L} + [D^\nu \bar{c}_R] \gamma_\mu \overset\leftrightarrow{D}\vphantom{D}^{u,Q_1}_\nu u_R)$ &  \multirow{2}{*}{  $\frac{0.09}{E_\text{TeV}^4}$ }  \\
24 &  $(\bar{t}\gamma^\mu \gamma_5 u) ([D^\nu\bar{c}]\gamma_\mu \overset\leftrightarrow{D^u_\nu}u) $ & &   &   $(\bar{Q}_{3L} \gamma^\mu Q_{1L} - \bar{t}_R \gamma^\mu u_R)([D^\nu \bar{Q}_{2L}] \gamma_\mu \overset\leftrightarrow{D}\vphantom{D}^{u,Q_1}_\nu Q_{1L} + [D^\nu \bar{c}_R] \gamma_\mu \overset\leftrightarrow{D}\vphantom{D}^{u,Q_1}_\nu u_R)$ &  \\
\hline 
25 &  $(\bar{t} u) ([D^\mu \bar{c}]\overset\leftrightarrow{D^u_\mu}u) $ &    \multirow{4}{*}{8}&  \multirow{4}{*}{S} &  $(\bar{Q}_{3L} \tilde{H} u_R + \bar{t}_{R} \tilde{H}^\dag Q_{1L})([D^\mu \bar{Q}_{2L} \tilde{H}] \overset\leftrightarrow{D}\vphantom{D}^{u,Q_1}_\mu u_R + [D^\mu \bar{c}_{R}] ^\dag  \overset\leftrightarrow{D}\vphantom{D}^{u,Q_1}_\mu \tilde{H} Q_{1L})$ &  \multirow{4}{*}{  $\frac{0.09}{E_\text{TeV}^4}, \frac{0.9}{E_\text{TeV}^6}$ }   \\
26 &  $(\bar{t} i \gamma_5 u) ([D^\mu\bar{c}]\overset\leftrightarrow{D^u_\mu}u)  $ &  &    &   $(i\bar{Q}_{3L} \tilde{H} u_R -i \bar{t}_{R} \tilde{H}^\dag Q_{1L})([D^\mu \bar{Q}_{2L} \tilde{H}] \overset\leftrightarrow{D}\vphantom{D}^{u,Q_1}_\mu u_R + [D^\mu \bar{c}_{R}]   \overset\leftrightarrow{D}\vphantom{D}^{u,Q_1}_\mu \tilde{H}^\dag Q_{1L})$ &   \\
27 &  $(\bar{t}  u) ([D^\mu\bar{c}]i \gamma_5 \overset\leftrightarrow{D^u_\mu}u) $ &  &    &   $(\bar{Q}_{3L} \tilde{H} u_R + \bar{t}_{R} \tilde{H}^\dag Q_{1L})(i [D^\mu \bar{Q}_{2L} \tilde{H}] \overset\leftrightarrow{D}\vphantom{D}^{u,Q_1}_\mu u_R -i [D^\mu \bar{c}_{R}]   \overset\leftrightarrow{D}\vphantom{D}^{u,Q_1}_\mu \tilde{H}^\dag Q_{1L})$ &   \\
28 &  $(\bar{t}  i\gamma_5 u) ([D^\mu\bar{c}]i \gamma_5 \overset\leftrightarrow{D^u_\mu}u)$ &  &    &   $(i\bar{Q}_{3L} \tilde{H} u_R -i \bar{t}_{R} \tilde{H}^\dag Q_{1L})(i [D^\mu \bar{Q}_{2L} \tilde{H}] \overset\leftrightarrow{D}\vphantom{D}^{u,Q_1}_\mu u_R -i [D^\mu \bar{c}_{R}]  \overset\leftrightarrow{D}\vphantom{D}^{u,Q_1}_\mu \tilde{H}^\dag Q_{1L})$ &   \\
\hline
29 &  $(\bar{t} \gamma^\mu u) ([D^\nu \bar{c}]\overset\leftrightarrow{D^u_\nu}D_\mu u)$ &    \multirow{2}{*}{9}&    \multirow{2}{*}{A} &   $(\bar{Q}_{3L} \gamma^\mu Q_{1L} + \bar{t}_R \gamma^\mu u_R)([D^\nu \bar{Q}_{2L} \tilde{H}] \overset\leftrightarrow{D}\vphantom{D}^{u,Q_1}_\nu D_\mu u_R + [D^\nu \bar{c}_{R}]   \overset\leftrightarrow{D}\vphantom{D}^{u,Q_1}_\nu  D_\mu \tilde{H}^\dag Q_{1L})$ &  \multirow{2}{*}{  $\frac{0.02}{E_\text{TeV}^5}, \frac{0.07}{E_\text{TeV}^6}$ }  \\
30 &  $(\bar{t} \gamma^\mu u) ([D^\nu \bar{c}]i \gamma_5 \overset\leftrightarrow{D^u_\nu}D_\mu u)$ &  &   &   $(\bar{Q}_{3L} \gamma^\mu Q_{1L} + \bar{t}_R \gamma^\mu u_R)(i[D^\nu \bar{Q}_{2L} \tilde{H}] \overset\leftrightarrow{D}\vphantom{D}^{u,Q_1}_\nu D_\mu u_R -i [D^\nu \bar{c}_{R}]   \overset\leftrightarrow{D}\vphantom{D}^{u,Q_1}_\nu D_\mu \tilde{H}^\dag  Q_{1L})$ &   \\
\hline

31 &  $(\bar{t} \gamma^\mu u) ([D^\nu \bar{c}]\overset\leftrightarrow{D^u_\nu}D_\mu u)$ &    \multirow{7}{*}{9}&  \multirow{7}{*}{S} &   $(\bar{Q}_{3L} \gamma^\mu Q_{1L} + \bar{t}_R \gamma^\mu u_R)([D^\nu \bar{Q}_{2L} \tilde{H}] \overset\leftrightarrow{D}\vphantom{D}^{u,Q_1}_\nu D_\mu u_R + [D^\nu \bar{c}_{R}]   \overset\leftrightarrow{D}\vphantom{D}^{u,Q_1}_\nu D_\mu \tilde{H}^\dag Q_{1L})$ &  \multirow{7}{*}{  $\frac{0.02}{E_\text{TeV}^5}, \frac{0.07}{E_\text{TeV}^6}$ }  \\
32 &  $(\bar{t} \gamma^\mu u) ([D^\nu \bar{c}]i \gamma_5 \overset\leftrightarrow{D^u_\nu}D_\mu u)$ & &   &   $(\bar{Q}_{3L} \gamma^\mu Q_{1L} + \bar{t}_R \gamma^\mu u_R)(i [D^\nu \bar{Q}_{2L} \tilde{H}] \overset\leftrightarrow{D}\vphantom{D}^{u,Q_1}_\nu D_\mu u_R -i  [D^\nu \bar{c}_{R}]   \overset\leftrightarrow{D}\vphantom{D}^{u,Q_1}_\nu  D_\mu \tilde{H}^\dag Q_{1L})$ &   \\
33 &  $(\bar{t} \gamma^\mu\gamma_5 u) ([D^\nu\bar{c}]\overset\leftrightarrow{D^u_\nu}D_\mu u)$  &   & &   $(\bar{Q}_{3L} \gamma^\mu Q_{1L} - \bar{t}_R \gamma^\mu u_R)([D^\nu \bar{Q}_{2L} \tilde{H}] \overset\leftrightarrow{D}\vphantom{D}^{u,Q_1}_\nu D_\mu u_R + [D^\nu \bar{c}_{R}]  \overset\leftrightarrow{D}\vphantom{D}^{u,Q_1}_\nu  D_\mu \tilde{H}^\dag Q_{1L})$ &   \\
34 &  $(\bar{t} \gamma^\mu\gamma_5 u) ([D^\nu\bar{c}]i\gamma_5\overset\leftrightarrow{D^u_\nu} D_\mu u)$  & &   &   $(\bar{Q}_{3L} \gamma^\mu Q_{1L} - \bar{t}_R \gamma^\mu u_R)(i [D^\nu \bar{Q}_{2L} \tilde{H}] \overset\leftrightarrow{D}\vphantom{D}^{u,Q_1}_\nu \!\!\! D_\mu u_R -i  [D^\nu \bar{c}_{R}]   \overset\leftrightarrow{D}\vphantom{D}^{u,Q_1}_\nu \!\!\! D_\mu \tilde{H}^\dag Q_{1L})$ &   \\

35 &  $(\bar{t} D_\mu u) ([D^\nu\bar{c}]\gamma^\mu \overset\leftrightarrow{D^u_\nu}u)$  &  &  &   $(\bar{Q}_{3L} \tilde{H}  D_\mu u_R + \bar{t}_{R} D_\mu \tilde{H}^\dag Q_{1L})([D^\nu \bar{Q}_{2L}] \gamma_\mu \overset\leftrightarrow{D}\vphantom{D}^{u,Q_1}_\nu Q_{1L} + [D^\nu \bar{c}_R] \gamma_\mu \overset\leftrightarrow{D}\vphantom{D}^{u,Q_1}_\nu u_R)$ &   \\
36 &  $(\bar{t}i \gamma_5 D_\mu u) ([D^\nu\bar{c}]\gamma^\mu \overset\leftrightarrow{D^u_\nu}u)$  & &   &   $(i\bar{Q}_{3L} \tilde{H}  D_\mu  u_R -i \bar{t}_{R} D_\mu \tilde{H}^\dag Q_{1L})([D^\nu \bar{Q}_{2L}] \gamma_\mu \overset\leftrightarrow{D}\vphantom{D}^{u,Q_1}_\nu Q_{1L} + [D^\nu \bar{c}_R] \gamma_\mu \overset\leftrightarrow{D}\vphantom{D}^{u,Q_1}_\nu u_R)$ &   \\
\hline
\end{tabular}
\caption{\label{tab:qqqbarqbarp}  \footnotesize Primary operators for $qq\bar{q}\bar{q}$ interactions with two indistinguishable quarks, for the specific case of $uu\bar{t}\bar{c}$ interactions  (Hermitean conjugate yields $tc\bar{u}\bar{u}$ and down-type interactions can be found by exchange for down quarks.).  The $SU(3)$ contractions are determined by $S (A)$ to be symmetric (antisymmetric) in the $uu$ indices.   We've defined a back-forth derivative $\overset\leftrightarrow{D^u_\mu}$, which only acts on the $u$ fields, and similarly $\overset\leftrightarrow{D}\vphantom{D}^{u,Q_1}_\mu$ which acts on $u_R$ and $Q_{1L}$.  For descendant operators, one adds contracted derivatives to get arbitrary Mandelstam factors that respect the exchange symmetry, i.e.~$s, (t-u)^2$.  At dimension 8, $s \mathcal{O}_3$ and $s\mathcal{O}_{4}$ become redundant, while at dimension 10, $s \mathcal{O}_{27}$ and $s\mathcal{O}_{28}$ become redundant.  Thus  one only needs to consider $\mathcal{O}_{3,4,27,28}$ descendants with arbitrary factors of $(t-u)^2$. }
\end{center}
\end{adjustwidth}
\end{table}

In Table~\ref{tab:qqqbarqbarp}, we've listed the primary operators for $\bar{q}q\bar{q}q$ interactions when two of the quarks are identical for the specific case of $uu\bar{t}\bar{c}$. There are again two allowed $SU(3)$ contractions, specified by whether the $uu$ are in symmetric $(S)$ or antisymmetric $(A)$ combination.  Since we're suppressing the $SU(3)$ indices, this makes some of the expressions look identical, which occurs in the blocks (1-4) and (5-8), (11-12) and (17-18),  (19-22) and (25-28), and  (29-30) and (31-32).  At dimension 8, $s \mathcal{O}_3$ and $s\mathcal{O}_{4}$ become redundant and at dimension 10, $s \mathcal{O}_{27}$ and $s\mathcal{O}_{28}$ become redundant, where $s=(p_u+p_{\bar{u}})^2$.  Thus one only needs to add descendants for $\mathcal{O}_{3,4,27,28}$ with factors of $(t-u)^2$.  These four redundancies explain the two $-2$ terms in the Hilbert series numerator.  

\begin{table}[p]
\begin{adjustwidth}{-.5in}{-.5in}  
\begin{center}
\scriptsize
\centering
\renewcommand{\arraystretch}{0.85}
\tabcolsep6pt\begin{tabular}{|c|c|c|c|c|c|c|}
\hline
\multirow{2}{*}{$i$} & \multirow{2}{*}{$\mathcal{O}_i^{uu\bar{t}\bar{t}}$}  &  \multirow{2}{*}{$d_{\mathcal{O}_i}$}& \multirow{2}{*}{$SU(3)$} & SMEFT & $c$ Unitarity  \\[-5pt]
&   & &  &  Operator & Bound \\
\hline
1 &  $(\bar{t} u) (\bar{t}u) $ &    \multirow{3}{*}{6}&  \multirow{3}{*}{A} &  $(\bar{Q}_{3L} \tilde{H} u_R + \bar{t}_{R} \tilde{H}^\dag Q_{1L})(\bar{Q}_{3L} \tilde{H} u_R + \bar{t}_{R} \tilde{H}^\dag Q_{1L})$ &  \multirow{3}{*}{  $\frac{1.5}{E_\text{TeV}^2}, \frac{15}{E_\text{TeV}^4}$ } \\
2 &  $(\bar{t} i \gamma_5 u) (\bar{t}u) $ &  &    &   $(i \bar{Q}_{3L} \tilde{H} u_R -i \bar{t}_{R} \tilde{H}^\dag Q_{1L})(\bar{Q}_{3L} \tilde{H} u_R + \bar{t}_{R} \tilde{H}^\dag Q_{1L})$ &   \\
3 &  $(\bar{t} i \gamma_5 u) (\bar{t}i \gamma_5 u) $ &  &    &   $(\bar{Q}_{3L} \tilde{H} u_R - \bar{t}_{R} \tilde{H}^\dag Q_{1L})(\bar{Q}_{3L} \tilde{H} u_R - \bar{t}_{R} \tilde{H}^\dag Q_{1L})$ &   \\
\hline 
4 &  $(\bar{t} u) (\bar{t}u) $ &    \multirow{3}{*}{6}&  \multirow{3}{*}{S} &  $(\bar{Q}_{3L} \tilde{H} u_R + \bar{t}_{R} \tilde{H}^\dag Q_{1L})(\bar{Q}_{3L} \tilde{H} u_R + \bar{t}_{R} \tilde{H}^\dag Q_{1L})$ & \multirow{3}{*}{  $\frac{1.5}{E_\text{TeV}^2}, \frac{15}{E_\text{TeV}^4}$ } \\
5 &  $(\bar{t} i \gamma_5 u) (\bar{t}u) $ &  &    &   $(i \bar{Q}_{3L} \tilde{H} u_R -i \bar{t}_{R} \tilde{H}^\dag Q_{1L})(\bar{Q}_{3L} \tilde{H} u_R + \bar{t}_{R} \tilde{H}^\dag Q_{1L})$ &   \\
6 &  $(\bar{t} i \gamma_5 u) (\bar{t}i \gamma_5 u) $ &  &    &   $(\bar{Q}_{3L} \tilde{H} u_R - \bar{t}_{R} \tilde{H}^\dag Q_{1L})(\bar{Q}_{3L} \tilde{H} u_R - \bar{t}_{R} \tilde{H}^\dag Q_{1L})$ &   \\ \hline

7 &  $(\bar{t} \gamma^\mu u) (\bar{t}\gamma_\mu u) $ &  \multirow{2}{*}{6} &  \multirow{2}{*}{S}   &   $(\bar{Q}_{3L} \gamma^\mu Q_{1L} + \bar{t}_R \gamma^\mu u_R)(\bar{Q}_{3L} \gamma_\mu Q_{1L} + \bar{t}_R \gamma_\mu u_R)$ &  \multirow{2}{*}{  $\frac{1.5}{E_\text{TeV}^2}$} \\
8 &  $(\bar{t} \gamma^\mu\gamma_5 u) (\bar{t}\gamma_\mu u) $ &  &    &   $(\bar{Q}_{3L} \gamma^\mu Q_{1L} - \bar{t}_R \gamma^\mu u_R)(\bar{Q}_{3L} \gamma_\mu Q_{1L} + \bar{t}_R \gamma_\mu u_R)$ &   \\
\hline
9 &  $(\bar{t} \gamma^\mu u) (\bar{t}D_\mu u) $ &    \multirow{4}{*}{7}&  \multirow{4}{*}{A} &  $(\bar{Q}_{3L} \gamma^\mu Q_{1L} + \bar{t}_R \gamma^\mu u_R)(\bar{Q}_{3L} \tilde{H} D_\mu  u_R + \bar{t}_{R} D_\mu \tilde{H}^\dag Q_{1L})$ &  \multirow{4}{*}{  $\frac{0.4}{E_\text{TeV}^3}, \frac{1.2}{E_\text{TeV}^4}$ }  \\
10 &  $(\bar{t} \gamma^\mu u) (\bar{t}i\gamma_5 D_\mu u) $ &  &    &   $(\bar{Q}_{3L} \gamma^\mu Q_{1L} + \bar{t}_R \gamma^\mu u_R)(i \bar{Q}_{3L} \tilde{H} D_\mu  u_R -i \bar{t}_{R} D_\mu \tilde{H}^\dag Q_{1L})$ &   \\
11 &  $(\bar{t} \gamma^\mu\gamma_5 u) (\bar{t}D_\mu u) $ &    &   &  $(\bar{Q}_{3L} \gamma^\mu Q_{1L} - \bar{t}_R \gamma^\mu u_R)(\bar{Q}_{3L} \tilde{H} D_\mu   u_R + \bar{t}_{R} D_\mu \tilde{H}^\dag Q_{1L})$ &   \\
12 &  $(\bar{t} \gamma^\mu \gamma_5 u) (\bar{t}i\gamma_5 D_\mu u) $ &  &    &   $(\bar{Q}_{3L} \gamma^\mu Q_{1L} - \bar{t}_R \gamma^\mu u_R)(i\bar{Q}_{3L}\tilde{H} D_\mu u_R -i \bar{t}_{R} D_\mu \tilde{H}^\dag Q_{1L})$ &   \\
\hline

13 &  $(\bar{t} u) ([D^\mu\bar{t}\, ]\overset\leftrightarrow{D^u_\mu}u) $ &    \multirow{3}{*}{8}&  \multirow{3}{*}{A} &  $(\bar{Q}_{3L} \tilde{H} u_R + \bar{t}_{R} \tilde{H}^\dag Q_{1L})([D^\mu \bar{Q}_{3L} \tilde{H}] \overset\leftrightarrow{D}\vphantom{D}^{u,Q_1}_\mu u_R + [D^\mu \bar{t}_{R}]  \overset\leftrightarrow{D}\vphantom{D}^{u,Q_1}_\mu \tilde{H}^\dag Q_{1L})$ &  \multirow{3}{*}{  $\frac{0.09}{E_\text{TeV}^4}, \frac{0.9}{E_\text{TeV}^6}$ } \\
14 &  $(\bar{t} i \gamma_5 u) ([D^\mu\bar{t}\, ]\overset\leftrightarrow{D^u_\mu}u) $ &  &    &   $(i \bar{Q}_{3L} \tilde{H} u_R -i \bar{t}_{R} \tilde{H}^\dag Q_{1L})([D^\mu \bar{Q}_{3L} \tilde{H}] \overset\leftrightarrow{D}\vphantom{D}^{u,Q_1}_\mu u_R + [D^\mu \bar{t}_{R}]   \overset\leftrightarrow{D}\vphantom{D}^{u,Q_1}_\mu \tilde{H}^\dag Q_{1L})$&   \\
15 &  $(\bar{t} i \gamma_5 u) ([D^\mu\bar{t}\, ]i \gamma_5 \overset\leftrightarrow{D^u_\mu} u) $ &  &    &   $(i \bar{Q}_{3L} \tilde{H} u_R -i \bar{t}_{R} \tilde{H}^\dag Q_{1L})(i [D^\mu \bar{Q}_{3L} \tilde{H}] \overset\leftrightarrow{D}\vphantom{D}^{u,Q_1}_\mu u_R -i [D^\mu \bar{t}_{R}]   \overset\leftrightarrow{D}\vphantom{D}^{u,Q_1}_\mu \tilde{H}^\dag Q_{1L})$ &   \\ \hline

16 &  $(\bar{t} \gamma^\mu u) ([D^\nu\bar{t}\, ]\gamma_\mu \overset\leftrightarrow{D^u_\nu} u) $ & \multirow{2}{*}{8} &  \multirow{2}{*}{A}  &   $(\bar{Q}_{3L} \gamma^\mu Q_{1L} + \bar{t}_R \gamma^\mu u_R)([D^\nu \bar{Q}_{3L}] \gamma_\mu \overset\leftrightarrow{D}\vphantom{D}^{u,Q_1}_\nu  Q_{1L} + [D^\nu \bar{t}_R] \gamma_\mu \overset\leftrightarrow{D}\vphantom{D}^{u,Q_1}_\nu  u_R)$ &  \multirow{2}{*}{  $\frac{0.09}{E_\text{TeV}^4}$ }  \\

17 &  $(\bar{t} \gamma^\mu\gamma_5 u) ([D^\nu\bar{t}\, ]\gamma_\mu \overset\leftrightarrow{D^u_\nu}u) $ &  &    &  $(\bar{Q}_{3L} \gamma^\mu Q_{1L} - \bar{t}_R \gamma^\mu u_R)([D^\nu \bar{Q}_{3L}] \gamma_\mu \overset\leftrightarrow{D}\vphantom{D}^{u,Q_1}_\nu Q_{1L} + [D^\nu \bar{t}_R] \gamma_\mu \overset\leftrightarrow{D}\vphantom{D}^{u,Q_1}_\nu u_R)$ &   \\
\hline

18 &  $(\bar{t} u) ([D^\mu\bar{t}\, ]\overset\leftrightarrow{D^u_\mu}u) $ &    \multirow{3}{*}{8}&  \multirow{3}{*}{S} &  $(\bar{Q}_{3L} \tilde{H} u_R + \bar{t}_{R} \tilde{H}^\dag Q_{1L})([D^\mu \bar{Q}_{3L} \tilde{H}] \overset\leftrightarrow{D}\vphantom{D}^{u,Q_1}_\mu u_R + [D^\mu \bar{t}_{R}]   \overset\leftrightarrow{D}\vphantom{D}^{u,Q_1}_\mu \tilde{H}^\dag Q_{1L})$ &  \multirow{3}{*}{  $\frac{0.09}{E_\text{TeV}^4}, \frac{0.9}{E_\text{TeV}^6}$ } \\
19 &  $(\bar{t} i \gamma_5 u) ([D^\mu\bar{t}\, ] \overset\leftrightarrow{D^u_\mu}u) $ &  &    &   $(i \bar{Q}_{3L} \tilde{H} u_R -i \bar{t}_{R} \tilde{H}^\dag Q_{1L})([D^\mu \bar{Q}_{3L} \tilde{H}] \overset\leftrightarrow{D}\vphantom{D}^{u,Q_1}_\mu u_R + [D^\mu \bar{t}_{R}]  \overset\leftrightarrow{D}\vphantom{D}^{u,Q_1}_\mu \tilde{H}^\dag  Q_{1L})$ &   \\
20 &  $(\bar{t} i \gamma_5 u) ([D^\mu\bar{t}\, ]i \gamma_5 \overset\leftrightarrow{D^u_\mu}u) $ &  &    &   $(i \bar{Q}_{3L} \tilde{H} u_R -i \bar{t}_{R} \tilde{H}^\dag Q_{1L})(i [D^\mu \bar{Q}_{3L} \tilde{H}] \overset\leftrightarrow{D}\vphantom{D}^{u,Q_1}_\mu u_R -i [D^\mu \bar{t}_{R}]   \overset\leftrightarrow{D}\vphantom{D}^{u,Q_1}_\mu \tilde{H}^\dag Q_{1L})$ &   \\
\hline

21 &  $(\bar{t} \gamma^\mu u) ([D^\nu\bar{t}\, ]D_\mu \overset\leftrightarrow{D^u_\nu}u) $ &    \multirow{4}{*}{9}&  \multirow{4}{*}{S} &  $(\bar{Q}_{3L} \gamma^\mu Q_{1L} + \bar{t}_R \gamma^\mu u_R)([D^\nu \bar{Q}_{3L} \tilde{H}] \overset\leftrightarrow{D}\vphantom{D}^{u,Q_1}_\nu D_\mu u_R + [D^\nu \bar{t}_{R}]  \overset\leftrightarrow{D}\vphantom{D}^{u,Q_1}_\nu   D_\mu \tilde{H}^\dag Q_{1L})$ &  \multirow{4}{*}{  $\frac{0.02}{E_\text{TeV}^5}, \frac{0.07}{E_\text{TeV}^6}$ } \\
22 &  $(\bar{t} \gamma^\mu u) ([D^\nu\bar{t}\, ]i\gamma_5 D_\mu \overset\leftrightarrow{D^u_\nu}u) $ &  &    &   $(\bar{Q}_{3L} \gamma^\mu Q_{1L} + \bar{t}_R \gamma^\mu u_R)(i [D^\nu \bar{Q}_{3L} \tilde{H}] \overset\leftrightarrow{D}\vphantom{D}^{u,Q_1}_\nu  D_\mu u_R -i [D^\nu \bar{t}_{R}]   \overset\leftrightarrow{D}\vphantom{D}^{u,Q_1}_\nu D_\mu \tilde{H}^\dag Q_{1L})$ &   \\
23 &  $(\bar{t} \gamma^\mu\gamma_5 u) ([D^\nu\bar{t}\, ]D_\mu \overset\leftrightarrow{D^u_\nu}u) $ &    &   &  $(\bar{Q}_{3L} \gamma^\mu Q_{1L} - \bar{t}_R \gamma^\mu u_R)([D^\nu \bar{Q}_{3L} \tilde{H}] \overset\leftrightarrow{D}\vphantom{D}^{u,Q_1}_\nu  D_\mu u_R + [D^\nu \bar{t}_{R}]   \overset\leftrightarrow{D}\vphantom{D}^{u,Q_1}_\nu D_\mu \tilde{H}^\dag Q_{1L})$ &   \\
24 &  $(\bar{t} \gamma^\mu \gamma_5 u) ([D^\nu\bar{t}\, ]i\gamma_5 D_\mu \overset\leftrightarrow{D^u_\nu}u) $ &  &    &   $(\bar{Q}_{3L} \gamma^\mu Q_{1L} - \bar{t}_R \gamma^\mu u_R)(i [D^\nu \bar{Q}_{3L} \tilde{H}] \overset\leftrightarrow{D}\vphantom{D}^{u,Q_1}_\nu \!\!\! D_\mu u_R -i [D^\nu \bar{t}_{R}]   \overset\leftrightarrow{D}\vphantom{D}^{u,Q_1}_\nu D_\mu \tilde{H}^\dag Q_{1L})$ &   \\

\hline
\end{tabular}
\medskip
\begin{minipage}{6in}
\caption{\label{tab:qqqq2id}  \footnotesize Primary operators for $qq\bar{q}\bar{q}$ interactions with two indistinguishable quarks and two indistinguishable antiquarks, for the specific case of $uu\bar{t}\bar{t}$ interactions 
  (The Hermitean conjugate yields the $tt\bar{u}\bar{u}$ interactions and the down-type interactions can be found by exchange for down quarks.).  The $SU(3)$ contractions are determined by $S$ to be symmetric in the $uu$ indices and $A$ to be antisymmetric.   To simplify the expressions, we've defined a back-forth derivative $\overset\leftrightarrow{D^u_\mu}$, which only acts on the $u$ fields, and similarly $\overset\leftrightarrow{D}\vphantom{D}^{u,Q_1}_\mu$ which acts on $u_R$ and $Q_{1L}$.  To get the descendant operators, one can add contracted derivatives to get arbitrary Mandelstam factors that respect the exchange symmetries, i.e.~$s, (t-u)^2$.  At dimension 8, $s \mathcal{O}_2$ and $s\mathcal{O}_{3}$ become redundant, while at dimension 10, $s \mathcal{O}_{19}$ and $s\mathcal{O}_{20}$ become redundant.  Thus, one only needs to consider $\mathcal{O}_{2}, \mathcal{O}_{3}, \mathcal{O}_{19}, \mathcal{O}_{20}$ with arbitrary factors of $(t-u)^2$. }
\end{minipage}
\end{center}
\end{adjustwidth}
\end{table}

In Table~\ref{tab:qqqq2id}, we've listed the primary operators for $\bar{q}q\bar{q}q$ interactions when the two quarks are identical and the two anti-quarks are identical,  for the specific case of $uu\bar{t}\bar{t}$. There are again two allowed $SU(3)$ contractions, specified by whether the $uu$ are in symmetric $(S)$ or antisymmetric $(A)$ combination.  Since we're suppressing the $SU(3)$ indices, this makes some of the expressions look identical, with (1-3) and (4-6) being the same, as well as (13-15) and (18-20).  At dimension 8, $s \mathcal{O}_2$ and $s\mathcal{O}_{3}$ become redundant and at dimension 10, $s \mathcal{O}_{19}$ and $s\mathcal{O}_{20}$ become redundant.  Thus one only needs the descendants of $\mathcal{O}_{2,3,19,20}$ with factors of $(t-u)^2$.  These four redundancies explain the two $-2$ terms in the Hilbert series. 

\section{Interesting Top Decay Amplitudes for the HL-LHC}
\label{sec:interestingdecays}
Now that we have all of the results, we can  compare our unitarity upper bounds on the coupling strengths with our estimate of the couplings needed for HL-LHC sensitivity to the new top quark decays in Eqn.~\ref{eqn:cdecaybound}, to highlight which top decay amplitudes are worth studying in more detail  at the HL-LHC.   In the following, we will assume we have top quark pair production, where one top quark decays into a $b$ quark and a leptonic $W$, with a $b$-tagging efficiency of 0.5, a lepton tagging efficiency of 0.8, and a $W$ leptonic branching ratio of 0.2.  For the Higgs modes, we will assume it decays to photons with a branching ratio of $\sim 2\times 10^{-3}$.  

First, let's consider two body decays of the top quark.  For the charged current decays, we have $t\to W (b,s,d)$, which have left and right handed vector and tensor couplings, which can be distinguished by the lepton angular distributions \cite{TopQuarkWorkingGroup:2013hxj}.  In addition, the tensor operators can be constrained by top quark production \cite{Schulze:2016qas}.   For flavor changing neutral current decays, we have $t \to (u,c)(h,Z, \gamma, g)$, which are all actively being searched for at the LHC \cite{ATLAS:2021amo, ATLAS:2022per,CMS-PAS-TOP-21-013,ATLAS:2022gzn, ATLAS:2023qzr, CMS:2021gfa, CMS:2021hug}.  For all of these two body decays, there is a dimension 6 SMEFT operator that realizes the coupling, which explains why they are actively being studied.  Our constraints on the coupling strengths agree that these are interesting and could potentially probe unitarity violating scales up to several tens of TeV.

Now, let's consider three body decays.  We do not consider all hadronic decays of the top quark since those suffer from large combinatorial backgrounds at the LHC and our estimates would be entirely too optimistic.  The charged current contact interaction $t\to (b,s,d)(\bar{e}, \bar{\mu}, \bar{\tau}) \nu$ has a different lepton pair invariant mass, which could be interesting to look for in terms of the quark-charged lepton invariant mass distribution.  Here our estimates say that all of the dimension 6 CP even amplitudes could be interesting, even with unitarity violation occurring around 5 TeV, while the dimension 7 CP even amplitudes are interesting if unitarity violation occurs at about $\sim 3$ TeV.  Thus, these are worth exploring as there is room to increase the coupling for lower scales of unitarity violation.  The other three body decays with a charged current interaction are $t\to (b,s,d)W(\gamma, g)$, which are generated at higher order in the Standard Model (we do not consider $t\to d WZ$ since this is so close to being kinematically closed and thus, our assumptions about the phase space and matrix element would be wrong.).  Contact amplitudes, unlike the Standard Model processes,  are not enhanced in the collinear/soft limits so these might be distinguishable.  Here, we find that of the operators in Table~\ref{tab:qqza1} the operators 3-4, 5 and 8 could be interesting for unitarity violation occurring at $\sim 6$ TeV, operators 10 and 14-15 need unitarity violation by $\sim 3$ TeV, and operators 19-22 and 25 need unitarity violation just above a TeV.   However, since we should interpret our estimates carefully for these photon and gluon decays, the lowest dimension operators are probably the most realistic to explore.

Flavor changing decays are highly suppressed in the Standard Model, so these are very promising to search for.  To start with, four fermion contact terms $t\to (c,u)(e,\mu,\tau)(\bar{e},\bar{\mu},\bar{\tau})$ are being searched for at the LHC in the lepton flavor violating modes to $e\mu$ \cite{CMS:2022ztx}.  Here our estimates say that dimension 6 CP even and odd amplitudes are interesting for unitarity violation above 9 TeV, while dimension 7 CP even and odd amplitudes require unitarity violation by $\sim 4$ TeV.  The existing CMS search probes the dimension 6 amplitudes \cite{CMS:2022ztx}, but does not look for the dimension 7 amplitudes since they appear at dimension 8 in SMEFT.  We can also consider flavor changing neutral current decays involving gauge bosons, including $t\to (c,u)(h\gamma, hg, Z\gamma, Zg, \gamma\gamma, \gamma g, gg)$, but not $t\to (c,u) WW$ since it is also nearly kinematically closed.  Again, our estimates are too optimistic for the decay modes that are completely hadronic, so we will focus on the other cases.    For the decays with a Higgs and a photon or gluon, using the amplitudes and unitarity bounds in Table 3 of \cite{Chang:2022crb} and assuming the diphoton Higgs decay, we find that the dimension 6, 7, 8 operators require unitarity violation respectively by $\sim 5, 2, 1$ TeV, so the dimension 6 and 7 ones are the most promising.  For the decays into a $Z$ and a photon or gluon, assuming the $Z$ decays to $ee$ or $\mu\mu$, we find that the dimension 6, 7, 8, 9 operators in Table~\ref{tab:qqza1}, require unitarity violation respectively by $\sim 3.5, 2.5, 1.2, 0.8$ TeV so the dimension 6, 7, 8 ones should be explored more closely, but the dimension 9 operators are likely out of reach.    For the decays with two photons or a photon and gluon, we find that the dimension 7, 8, 9, 10, 11 operators in Tables~\ref{tab:qqga1}, \ref{tab:qqaa1} require unitarity violation respectively by $\sim 5, 2, 1.3,1, 0.7$ TeV and given that we should be careful with these estimates (especially for the $\gamma g$ case), the dimension 7 ones are likely the only relevant ones.  

There are also baryon number violating three body decays mediated by our amplitudes, $t\to (\bar{c}, \bar{u})(\bar{b},\bar{s},\bar{d})(\bar{e},\bar{\mu},\bar{\tau})$.  These  would have combinatorial backgrounds, but have been searched for in the past by CMS \cite{CMS:2013zol}.   Again, theory explorations of these have focused on the dimension 6 SMEFT operators \cite{Hou:2005iu,Dong:2011rh}, so it would be interesting if the ones parameterized by dimension 8 SMEFT operators give distinguishable signals.

To conclude, our unitarity bounds combined with our estimates for the interesting size of couplings for top quark decays has allowed us a quick survey of which of the decay amplitudes may be worth pursuing at the HL-LHC.  As the dimension of the amplitude gets larger, these two constraints become more challenging to satisfy without lowering the scale of unitarity to the TeV scale.  Since the SMEFT operator realization must be at the same or higher dimension, this motivates studying in more detail top decays from many dimension 8 and a few dimension 10 SMEFT operators to determine their sensitivity at HL-LHC and future colliders.

\section{Conclusions}
\label{sec:conclusions}

In this paper, we have extended an approach \cite{Chang:2022crb} to determine the on-shell 3 and 4 point amplitudes that are needed for modeling general top quark phenomenology at colliders.  These serve as an intermediary between the observables searched for by experimental analyses  and the operators in effective field theories for the Standard Model.  This involved characterizing the general amplitudes for processes involving four fermions or two fermions and two gauge bosons.  We were able to characterize these respectively to dimension 12 and 13, finding the structure of primary and descendant amplitudes, where descendants are primaries multiplied by Mandelstam factors.  Interestingly, we find two classes of interactions whose Hilbert series numerator has a complete cancellation in the numerator. This na\"{i}vely would suggest that there are no primary operators at a certain mass dimension, but in actuality there are an equal number of new primaries and redundancies that appear at that mass dimension.  This illustrates the importance of using the Hilbert series in conjunction with the amplitudes, as they complement each other in this process.  We also note that our approach is a complementary check to the existing results up to dimension 8 using spinor-helicity variables \cite{Dong:2022jru,Liu:2023jbq} and extends the amplitude structure to higher dimension.   

To provide an initial survey of the potential phenomenology, we've used perturbative unitarity to place upper bounds on the coupling strengths of these interactions.  These depend on the scale where unitarity is violated $E_\text{TeV} = E_\text{max}/\text{TeV},$ with more stringent constraints as one increases $E_\text{TeV}$.  Given the expected sample of top quarks at HL-LHC, we've estimated the coupling size needed for the top quark decays to be seen over irreducible backgrounds.  This allowed us to highlight the that top quark decays into both FCNC  modes, like $t\to c(\bar{\ell}\ell, h\gamma, hg, Z\gamma, Zg, \gamma\gamma, \gamma g)$,  and  non-FCNC modes, like $t\to b(W\gamma, Wg)$, could be interesting to search for at the HL-LHC.  Some of these highlighted modes occur at dimension 8 and 10 in SMEFT and thus would be interesting to explore how distinctive these new amplitudes are compared to existing searches.  We leave such detailed phenomenology to future work.

To conclude, the high energy program at colliders is entering the phase of testing whether the Standard Model is indeed the correct description of physics at the TeV scale.  To do so, we must look for new physics in the most general way, so that we can find such deviations or constrain them.  On-shell amplitudes are a useful intermediary between experimental analyses and the parameterization of new physics by effective field theories.  Finally, by determining the on-shell amplitude structure to high dimension and writing down a concrete basis for them, we hope this will allow the field to maximize its efforts to find what exists beyond the Standard Model.

\section*{Acknowledgements}
We would like to thank Markus Luty for discussions and Gauthier Durieux for extensive comments on the manuscript.
The work of LB and SC was supported in part by the U.S. Department of Energy under Grant Number DE-SC0011640.

\bibliographystyle{utphys}
\bibliography{references}

\providecommand{\href}[2]{#2}\begingroup\raggedright\begin{thebibliography}{10}

\bibitem{Buchmuller:1985jz}
W.~Buchmuller and D.~Wyler, ``{Effective Lagrangian Analysis of New
  Interactions and Flavor Conservation},''
  \href{http://dx.doi.org/10.1016/0550-3213(86)90262-2}{{\em Nucl. Phys. B}
  {\bfseries 268} (1986) 621--653}.

\bibitem{Grzadkowski:2010es}
B.~Grzadkowski, M.~Iskrzynski, M.~Misiak, and J.~Rosiek, ``{Dimension-Six Terms
  in the Standard Model Lagrangian},''
  \href{http://dx.doi.org/10.1007/JHEP10(2010)085}{{\em JHEP} {\bfseries 10}
  (2010) 085}, \href{http://arxiv.org/abs/1008.4884}{{\ttfamily arXiv:1008.4884
  [hep-ph]}}.

\bibitem{Feruglio:1992wf}
F.~Feruglio, ``{The Chiral approach to the electroweak interactions},''
  \href{http://dx.doi.org/10.1142/S0217751X93001946}{{\em Int. J. Mod. Phys. A}
  {\bfseries 8} (1993) 4937--4972},
  \href{http://arxiv.org/abs/hep-ph/9301281}{{\ttfamily arXiv:hep-ph/9301281}}.

\bibitem{Gupta:2014rxa}
R.~S. Gupta, A.~Pomarol, and F.~Riva, ``{BSM Primary Effects},''
  \href{http://dx.doi.org/10.1103/PhysRevD.91.035001}{{\em Phys. Rev. D}
  {\bfseries 91} no.~3, (2015) 035001},
  \href{http://arxiv.org/abs/1405.0181}{{\ttfamily arXiv:1405.0181 [hep-ph]}}.

\bibitem{Gonzalez-Alonso:2014eva}
M.~Gonzalez-Alonso, A.~Greljo, G.~Isidori, and D.~Marzocca,
  ``{Pseudo-observables in Higgs decays},''
  \href{http://dx.doi.org/10.1140/epjc/s10052-015-3345-5}{{\em Eur. Phys. J. C}
  {\bfseries 75} (2015) 128}, \href{http://arxiv.org/abs/1412.6038}{{\ttfamily
  arXiv:1412.6038 [hep-ph]}}.

\bibitem{Greljo:2015sla}
A.~Greljo, G.~Isidori, J.~M. Lindert, and D.~Marzocca, ``{Pseudo-observables in
  electroweak Higgs production},''
  \href{http://dx.doi.org/10.1140/epjc/s10052-016-4000-5}{{\em Eur. Phys. J. C}
  {\bfseries 76} no.~3, (2016) 158},
  \href{http://arxiv.org/abs/1512.06135}{{\ttfamily arXiv:1512.06135
  [hep-ph]}}.

\bibitem{Falkowski:2001958}
A.~Falkowski, ``{Higgs Basis: Proposal for an EFT basis choice for LHC
  HXSWG},''. \url{https://cds.cern.ch/record/2001958}.

\bibitem{Shadmi:2018xan}
Y.~Shadmi and Y.~Weiss, ``{Effective Field Theory Amplitudes the On-Shell Way:
  Scalar and Vector Couplings to Gluons},''
  \href{http://dx.doi.org/10.1007/JHEP02(2019)165}{{\em JHEP} {\bfseries 02}
  (2019) 165}, \href{http://arxiv.org/abs/1809.09644}{{\ttfamily
  arXiv:1809.09644 [hep-ph]}}.

\bibitem{Durieux:2019eor}
G.~Durieux, T.~Kitahara, Y.~Shadmi, and Y.~Weiss, ``{The electroweak effective
  field theory from on-shell amplitudes},''
  \href{http://dx.doi.org/10.1007/JHEP01(2020)119}{{\em JHEP} {\bfseries 01}
  (2020) 119}, \href{http://arxiv.org/abs/1909.10551}{{\ttfamily
  arXiv:1909.10551 [hep-ph]}}.

\bibitem{Durieux:2020gip}
G.~Durieux, T.~Kitahara, C.~S. Machado, Y.~Shadmi, and Y.~Weiss,
  ``{Constructing massive on-shell contact terms},''
  \href{http://dx.doi.org/10.1007/JHEP12(2020)175}{{\em JHEP} {\bfseries 12}
  (2020) 175}, \href{http://arxiv.org/abs/2008.09652}{{\ttfamily
  arXiv:2008.09652 [hep-ph]}}.

\bibitem{Dong:2022jru}
Z.-Y. Dong, T.~Ma, J.~Shu, and Z.-Z. Zhou, ``{The New Formulation of Higgs
  Effective Field Theory},'' \href{http://arxiv.org/abs/2211.16515}{{\ttfamily
  arXiv:2211.16515 [hep-ph]}}.

\bibitem{Liu:2023jbq}
H.~Liu, T.~Ma, Y.~Shadmi, and M.~Waterbury, ``{An EFT hunter's guide to
  two-to-two scattering: HEFT and SMEFT on-shell amplitudes},''
  \href{http://arxiv.org/abs/2301.11349}{{\ttfamily arXiv:2301.11349
  [hep-ph]}}.

\bibitem{Chang:2022crb}
S.~Chang, M.~Chen, D.~Liu, and M.~A. Luty, ``{Primary Observables for Indirect
  Searches at Colliders},'' \href{http://arxiv.org/abs/2212.06215}{{\ttfamily
  arXiv:2212.06215 [hep-ph]}}.

\bibitem{Lehman:2015via}
L.~Lehman and A.~Martin, ``{Hilbert Series for Constructing Lagrangians:
  expanding the phenomenologist's toolbox},''
  \href{http://dx.doi.org/10.1103/PhysRevD.91.105014}{{\em Phys. Rev. D}
  {\bfseries 91} (2015) 105014},
  \href{http://arxiv.org/abs/1503.07537}{{\ttfamily arXiv:1503.07537
  [hep-ph]}}.

\bibitem{Henning:2015daa}
B.~Henning, X.~Lu, T.~Melia, and H.~Murayama, ``{Hilbert series and operator
  bases with derivatives in effective field theories},''
  \href{http://dx.doi.org/10.1007/s00220-015-2518-2}{{\em Commun. Math. Phys.}
  {\bfseries 347} no.~2, (2016) 363--388},
  \href{http://arxiv.org/abs/1507.07240}{{\ttfamily arXiv:1507.07240
  [hep-th]}}.

\bibitem{Lehman:2015coa}
L.~Lehman and A.~Martin, ``{Low-derivative operators of the Standard Model
  effective field theory via Hilbert series methods},''
  \href{http://dx.doi.org/10.1007/JHEP02(2016)081}{{\em JHEP} {\bfseries 02}
  (2016) 081}, \href{http://arxiv.org/abs/1510.00372}{{\ttfamily
  arXiv:1510.00372 [hep-ph]}}.

\bibitem{Henning:2015alf}
B.~Henning, X.~Lu, T.~Melia, and H.~Murayama, ``{2, 84, 30, 993, 560, 15456,
  11962, 261485, ...: Higher dimension operators in the SM EFT},''
  \href{http://dx.doi.org/10.1007/JHEP08(2017)016}{{\em JHEP} {\bfseries 08}
  (2017) 016}, \href{http://arxiv.org/abs/1512.03433}{{\ttfamily
  arXiv:1512.03433 [hep-ph]}}. [Erratum: JHEP 09, 019 (2019)].

\bibitem{Henning:2017fpj}
B.~Henning, X.~Lu, T.~Melia, and H.~Murayama, ``{Operator bases, $S$-matrices,
  and their partition functions},''
  \href{http://dx.doi.org/10.1007/JHEP10(2017)199}{{\em JHEP} {\bfseries 10}
  (2017) 199}, \href{http://arxiv.org/abs/1706.08520}{{\ttfamily
  arXiv:1706.08520 [hep-th]}}.

\bibitem{Graf:2020yxt}
L.~Graf, B.~Henning, X.~Lu, T.~Melia, and H.~Murayama, ``{2, 12, 117, 1959,
  45171, 1170086, \textellipsis{}: a Hilbert series for the QCD chiral
  Lagrangian},'' \href{http://dx.doi.org/10.1007/JHEP01(2021)142}{{\em JHEP}
  {\bfseries 01} (2021) 142}, \href{http://arxiv.org/abs/2009.01239}{{\ttfamily
  arXiv:2009.01239 [hep-ph]}}.

\bibitem{Graf:2022rco}
L.~Gr\'af, B.~Henning, X.~Lu, T.~Melia, and H.~Murayama, ``{Hilbert Series, the
  Higgs Mechanism, and HEFT},''
  \href{http://arxiv.org/abs/2211.06275}{{\ttfamily arXiv:2211.06275
  [hep-ph]}}.

\bibitem{Note1}
One can certainly model build situations where high dimension operators can be
  the leading correction to the Standard Model. For example, a theory having
  multiple massive mediators, which all need to be integrated out to generate
  an operator with just Standard Model particles, could be a potential
  realization.

\bibitem{Note2}
How to treat massive gauge bosons has only recently been worked out and is best
  explained in \cite {Graf:2022rco}.

\bibitem{Schomerus:2016epl}
V.~Schomerus, E.~Sobko, and M.~Isachenkov, ``{Harmony of Spinning Conformal
  Blocks},'' \href{http://dx.doi.org/10.1007/JHEP03(2017)085}{{\em JHEP}
  {\bfseries 03} (2017) 085}, \href{http://arxiv.org/abs/1612.02479}{{\ttfamily
  arXiv:1612.02479 [hep-th]}}.

\bibitem{Kravchuk:2016qvl}
P.~Kravchuk and D.~Simmons-Duffin, ``{Counting Conformal Correlators},''
  \href{http://dx.doi.org/10.1007/JHEP02(2018)096}{{\em JHEP} {\bfseries 02}
  (2018) 096}, \href{http://arxiv.org/abs/1612.08987}{{\ttfamily
  arXiv:1612.08987 [hep-th]}}.

\bibitem{Chang:2019vez}
S.~Chang and M.~A. Luty, ``{The Higgs Trilinear Coupling and the Scale of New
  Physics},'' \href{http://dx.doi.org/10.1007/JHEP03(2020)140}{{\em JHEP}
  {\bfseries 03} (2020) 140}, \href{http://arxiv.org/abs/1902.05556}{{\ttfamily
  arXiv:1902.05556 [hep-ph]}}.

\bibitem{Abu-Ajamieh:2020yqi}
F.~Abu-Ajamieh, S.~Chang, M.~Chen, and M.~A. Luty, ``{Higgs coupling
  measurements and the scale of new physics},''
  \href{http://dx.doi.org/10.1007/JHEP07(2021)056}{{\em JHEP} {\bfseries 07}
  (2021) 056}, \href{http://arxiv.org/abs/2009.11293}{{\ttfamily
  arXiv:2009.11293 [hep-ph]}}.

\bibitem{Abu-Ajamieh:2021egq}
F.~Abu-Ajamieh, ``{The scale of new physics from the Higgs couplings to
  \ensuremath{\gamma}\ensuremath{\gamma} and \ensuremath{\gamma}Z},''
  \href{http://dx.doi.org/10.1007/JHEP06(2022)091}{{\em JHEP} {\bfseries 06}
  (2022) 091}, \href{http://arxiv.org/abs/2112.13529}{{\ttfamily
  arXiv:2112.13529 [hep-ph]}}.

\bibitem{Abu-Ajamieh:2022ppp}
F.~Abu-Ajamieh, ``{The scale of new physics from the Higgs couplings to gg},''
  \href{http://dx.doi.org/10.1016/j.physletb.2022.137389}{{\em Phys. Lett. B}
  {\bfseries 833} (2022) 137389},
  \href{http://arxiv.org/abs/2203.07410}{{\ttfamily arXiv:2203.07410
  [hep-ph]}}.

\bibitem{Maltoni:2001dc}
F.~Maltoni, J.~M. Niczyporuk, and S.~Willenbrock, ``{The Scale of fermion mass
  generation},'' \href{http://dx.doi.org/10.1103/PhysRevD.65.033004}{{\em Phys.
  Rev. D} {\bfseries 65} (2002) 033004},
  \href{http://arxiv.org/abs/hep-ph/0106281}{{\ttfamily arXiv:hep-ph/0106281}}.

\bibitem{Dicus:2004rg}
D.~A. Dicus and H.-J. He, ``{Scales of fermion mass generation and electroweak
  symmetry breaking},''
  \href{http://dx.doi.org/10.1103/PhysRevD.71.093009}{{\em Phys. Rev. D}
  {\bfseries 71} (2005) 093009},
  \href{http://arxiv.org/abs/hep-ph/0409131}{{\ttfamily arXiv:hep-ph/0409131}}.

\bibitem{Dror:2015nkp}
J.~A. Dror, M.~Farina, E.~Salvioni, and J.~Serra, ``{Strong tW Scattering at
  the LHC},'' \href{http://dx.doi.org/10.1007/JHEP01(2016)071}{{\em JHEP}
  {\bfseries 01} (2016) 071}, \href{http://arxiv.org/abs/1511.03674}{{\ttfamily
  arXiv:1511.03674 [hep-ph]}}.

\bibitem{Falkowski:2019tft}
A.~Falkowski and R.~Rattazzi, ``{Which EFT},''
  \href{http://dx.doi.org/10.1007/JHEP10(2019)255}{{\em JHEP} {\bfseries 10}
  (2019) 255}, \href{http://arxiv.org/abs/1902.05936}{{\ttfamily
  arXiv:1902.05936 [hep-ph]}}.

\bibitem{Maltoni:2019aot}
F.~Maltoni, L.~Mantani, and K.~Mimasu, ``{Top-quark electroweak interactions at
  high energy},'' \href{http://dx.doi.org/10.1007/JHEP10(2019)004}{{\em JHEP}
  {\bfseries 10} (2019) 004}, \href{http://arxiv.org/abs/1904.05637}{{\ttfamily
  arXiv:1904.05637 [hep-ph]}}.

\bibitem{Note3}
Note that the dimension 6 operator $\protect \frac {1}{\Lambda ^2}(\protect
  \bar {Q_L} (D^2 \protect \tilde {H}) u_R +\protect \text {h.c.})$ can be
  reduced by equations of motion and does not result in the correct high energy
  behavior of the $\protect \bar {q}qWW$ interaction.

\bibitem{CMS:2021hug}
{\bfseries CMS} Collaboration, A.~Tumasyan {\em et~al.}, ``{Search for
  Flavor-Changing Neutral Current Interactions of the Top Quark and Higgs Boson
  in Final States with Two Photons in Proton-Proton Collisions at
  $\sqrt{s}=13\text{ }\text{ }\mathrm{TeV}$},''
  \href{http://dx.doi.org/10.1103/PhysRevLett.129.032001}{{\em Phys. Rev.
  Lett.} {\bfseries 129} no.~3, (2022) 032001},
  \href{http://arxiv.org/abs/2111.02219}{{\ttfamily arXiv:2111.02219
  [hep-ex]}}.

\bibitem{CMS:2021gfa}
{\bfseries CMS} Collaboration, A.~Tumasyan {\em et~al.}, ``{Search for
  flavor-changing neutral current interactions of the top quark and the Higgs
  boson decaying to a bottom quark-antiquark pair at $ \sqrt{s} $ = 13 TeV},''
  \href{http://dx.doi.org/10.1007/JHEP02(2022)169}{{\em JHEP} {\bfseries 02}
  (2022) 169}, \href{http://arxiv.org/abs/2112.09734}{{\ttfamily
  arXiv:2112.09734 [hep-ex]}}.

\bibitem{ATLAS:2021amo}
{\bfseries ATLAS} Collaboration, G.~Aad {\em et~al.}, ``{Search for
  flavour-changing neutral-current interactions of a top quark and a gluon in
  pp collisions at $\sqrt{s}=13$~TeV with the ATLAS detector},''
  \href{http://dx.doi.org/10.1140/epjc/s10052-022-10182-7}{{\em Eur. Phys. J.
  C} {\bfseries 82} no.~4, (2022) 334},
  \href{http://arxiv.org/abs/2112.01302}{{\ttfamily arXiv:2112.01302
  [hep-ex]}}.

\bibitem{ATLAS:2022per}
{\bfseries ATLAS} Collaboration, ``{Search for flavour-changing neutral-current
  couplings between the top quark and the photon with the ATLAS detector at
  $\sqrt{s} = 13$ TeV},'' \href{http://arxiv.org/abs/2205.02537}{{\ttfamily
  arXiv:2205.02537 [hep-ex]}}.

\bibitem{ATLAS:2022gzn}
{\bfseries ATLAS} Collaboration, ``{Search for flavour-changing neutral current
  interactions of the top quark and the Higgs boson in events with a pair of
  $\tau$-leptons in pp collisions at $\sqrt{s}=13$ TeV with the ATLAS
  detector},'' \href{http://arxiv.org/abs/2208.11415}{{\ttfamily
  arXiv:2208.11415 [hep-ex]}}.

\bibitem{CMS-PAS-TOP-21-013}
{\bfseries CMS} Collaboration, ``{Search for flavor-changing neutral current
  interactions of the top quark in final states with a photon and additional
  jets in proton-proton collisions at $\sqrt{s}$=13 TeV},'' tech. rep., CERN,
  Geneva, 2023.
\newblock \url{https://cds.cern.ch/record/2851857}.

\bibitem{ATLAS:2023qzr}
{\bfseries ATLAS} Collaboration, ``{Search for flavor-changing neutral-current
  couplings between the top quark and the $Z$ boson with LHC Run 2
  proton-proton collisions at $\sqrt{s} = 13$ TeV with the ATLAS detector},''
  \href{http://arxiv.org/abs/2301.11605}{{\ttfamily arXiv:2301.11605
  [hep-ex]}}.

\bibitem{Aguilar-Saavedra:2008nuh}
J.~A. Aguilar-Saavedra, ``{A Minimal set of top anomalous couplings},''
  \href{http://dx.doi.org/10.1016/j.nuclphysb.2008.12.012}{{\em Nucl. Phys. B}
  {\bfseries 812} (2009) 181--204},
  \href{http://arxiv.org/abs/0811.3842}{{\ttfamily arXiv:0811.3842 [hep-ph]}}.

\bibitem{Durieux:2014xla}
G.~Durieux, F.~Maltoni, and C.~Zhang, ``{Global approach to top-quark
  flavor-changing interactions},''
  \href{http://dx.doi.org/10.1103/PhysRevD.91.074017}{{\em Phys. Rev. D}
  {\bfseries 91} no.~7, (2015) 074017},
  \href{http://arxiv.org/abs/1412.7166}{{\ttfamily arXiv:1412.7166 [hep-ph]}}.

\bibitem{Aguilar-Saavedra:2018ksv}
D.~Barducci {\em et~al.}, ``{Interpreting top-quark LHC measurements in the
  standard-model effective field theory},''
  \href{http://arxiv.org/abs/1802.07237}{{\ttfamily arXiv:1802.07237
  [hep-ph]}}.

\bibitem{Altmannshofer:2023bfk}
W.~Altmannshofer, S.~Gori, B.~V. Lehmann, and J.~Zuo, ``{UV physics from IR
  features: new prospects from top flavor violation},''
  \href{http://arxiv.org/abs/2303.00781}{{\ttfamily arXiv:2303.00781
  [hep-ph]}}.

\bibitem{Diaz-Cruz:1989tem}
J.~L. Diaz-Cruz, R.~Martinez, M.~A. Perez, and A.~Rosado, ``{Flavor Changing
  Radiative Decay of The T Quark},''
  \href{http://dx.doi.org/10.1103/PhysRevD.41.891}{{\em Phys. Rev. D}
  {\bfseries 41} (1990) 891--894}.

\bibitem{Eilam:1990zc}
G.~Eilam, J.~L. Hewett, and A.~Soni, ``{Rare decays of the top quark in the
  standard and two Higgs doublet models},''
  \href{http://dx.doi.org/10.1103/PhysRevD.44.1473}{{\em Phys. Rev. D}
  {\bfseries 44} (1991) 1473--1484}. [Erratum: Phys.Rev.D 59, 039901 (1999)].

\bibitem{Jenkins:1996zd}
E.~E. Jenkins, ``{The Rare top decays $t \to b W^{+} Z$ and $t \to c W^{+}
  W^{-}$},'' \href{http://dx.doi.org/10.1103/PhysRevD.56.458}{{\em Phys. Rev.
  D} {\bfseries 56} (1997) 458--466},
  \href{http://arxiv.org/abs/hep-ph/9612211}{{\ttfamily arXiv:hep-ph/9612211}}.

\bibitem{Mele:1998ag}
B.~Mele, S.~Petrarca, and A.~Soddu, ``{A New evaluation of the t
  ---\ensuremath{>} cH decay width in the standard model},''
  \href{http://dx.doi.org/10.1016/S0370-2693(98)00822-3}{{\em Phys. Lett. B}
  {\bfseries 435} (1998) 401--406},
  \href{http://arxiv.org/abs/hep-ph/9805498}{{\ttfamily arXiv:hep-ph/9805498}}.

\bibitem{Aguilar-Saavedra:2004mfd}
J.~A. Aguilar-Saavedra, ``{Top flavor-changing neutral interactions:
  Theoretical expectations and experimental detection},'' {\em Acta Phys.
  Polon. B} {\bfseries 35} (2004) 2695--2710,
  \href{http://arxiv.org/abs/hep-ph/0409342}{{\ttfamily arXiv:hep-ph/0409342}}.

\bibitem{TopQuarkWorkingGroup:2013hxj}
{\bfseries Top Quark Working Group} Collaboration, K.~Agashe {\em et~al.},
  ``{Working Group Report: Top Quark},'' in {\em {Community Summer Study 2013}:
  {Snowmass on the Mississippi}}.
\newblock 11, 2013.
\newblock \href{http://arxiv.org/abs/1311.2028}{{\ttfamily arXiv:1311.2028
  [hep-ph]}}.

\bibitem{Schulze:2016qas}
M.~Schulze and Y.~Soreq, ``{Pinning down electroweak dipole operators of the
  top quark},'' \href{http://dx.doi.org/10.1140/epjc/s10052-016-4263-x}{{\em
  Eur. Phys. J. C} {\bfseries 76} no.~8, (2016) 466},
  \href{http://arxiv.org/abs/1603.08911}{{\ttfamily arXiv:1603.08911
  [hep-ph]}}.

\bibitem{CMS:2022ztx}
{\bfseries CMS} Collaboration, A.~Tumasyan {\em et~al.}, ``{Search for
  charged-lepton flavor violation in top quark production and decay in pp
  collisions at $ \sqrt{s} $ = 13 TeV},''
  \href{http://dx.doi.org/10.1007/JHEP06(2022)082}{{\em JHEP} {\bfseries 06}
  (2022) 082}, \href{http://arxiv.org/abs/2201.07859}{{\ttfamily
  arXiv:2201.07859 [hep-ex]}}.

\bibitem{CMS:2013zol}
{\bfseries CMS} Collaboration, S.~Chatrchyan {\em et~al.}, ``{Search for Baryon
  Number Violation in Top-Quark Decays},''
  \href{http://dx.doi.org/10.1016/j.physletb.2014.02.033}{{\em Phys. Lett. B}
  {\bfseries 731} (2014) 173--196},
  \href{http://arxiv.org/abs/1310.1618}{{\ttfamily arXiv:1310.1618 [hep-ex]}}.

\bibitem{Hou:2005iu}
W.-S. Hou, M.~Nagashima, and A.~Soddu, ``{Baryon number violation involving
  higher generations},''
  \href{http://dx.doi.org/10.1103/PhysRevD.72.095001}{{\em Phys. Rev. D}
  {\bfseries 72} (2005) 095001},
  \href{http://arxiv.org/abs/hep-ph/0509006}{{\ttfamily arXiv:hep-ph/0509006}}.

\bibitem{Dong:2011rh}
Z.~Dong, G.~Durieux, J.-M. Gerard, T.~Han, and F.~Maltoni, ``{Baryon number
  violation at the LHC: the top option},''
  \href{http://dx.doi.org/10.1103/PhysRevD.85.016006}{{\em Phys. Rev. D}
  {\bfseries 85} (2012) 016006},
  \href{http://arxiv.org/abs/1107.3805}{{\ttfamily arXiv:1107.3805 [hep-ph]}}.

\end{thebibliography}\endgroup

\end{document}